%% file: Bures_geodesics_metrology_Quantum_vf.tex
\documentclass[aps,amsmath,amssymb,superscriptaddress,showpacs,floatfix,twocolumn, pra,reprint,a4paper,nopdfoutputerror,accepted=2025-04-08]{quantumarticle}
\usepackage{amssymb,amsthm,latexsym,amsmath,dsfont,pstricks,hyperref,multirow}
\usepackage{graphicx,epsfig,float}
\usepackage{caption}
\usepackage{epstopdf}
\usepackage{psfrag}
\usepackage{color} 
\usepackage{subfig}
\usepackage[T1]{fontenc}
\usepackage[latin2]{inputenc}
\usepackage{natbib}
\usepackage{etoolbox}
\apptocmd{\sloppy}{\hbadness 10000\relax}{}{}

\newcommand{\re}{\mathrm{Re}\, }

\newcommand{\I}{{\rm{i}}}
\newcommand{\D}{{\rm{d}}}
\newcommand{\E}{e}
\newcommand{\ket}[1]{| #1 \rangle}
\newcommand{\bra}[1]{\langle #1 |}
\newcommand{\braket}[2]{\langle #1 | #2 \rangle}
\newcommand{\ketbra}[2]{| #1 \rangle \langle #2 |}
\newcommand{\onehalf}{{\frac{1}{2}}}

\newcommand{\ie}{i.e.}

\newcommand{\mmax}{{\rm max}}
\newcommand{\mmin}{{\rm min}}

\newcommand{\RHS}{right-hand side\;\,}
\newcommand{\LHS}{left-hand side\;\,}
\newcommand{\ONB}{orthonormal basis\;\,}

\def\real{{\mathbb{R}}}

\def\complex{{\mathbb{C}}}

\def\proba{{\rm I\kern -.18em P}}

\newcommand{\Span}{\operatorname{span}}

\newcommand{\supp}{\operatorname{supp}}

\newcommand{\tensorproduct}{\operatornamewithlimits{\otimes}}
\newcommand{\identity}{{\mathds{1}}}

\newcommand{\tr}{\operatorname{tr}}

\newcommand{\rank}{\operatorname{rank}}

\newcommand{\be}{\begin{equation}}
\newcommand{\ee}{\end{equation}}
\newcommand{\ba}{\begin{eqnarray}}
\newcommand{\ea}{\end{eqnarray}}
\newcommand{\nn}{\nonumber}

\newcommand{\Proof}{\noindent {\it Proof. }}

\newcommand{\finpro}{\hfill $\Box$}

\newcommand{\eps}{\varepsilon}

\newtheorem{theorem}{Theorem}
\newtheorem{lemma}{Lemma}

\newcommand{\Ee}{{\cal E}}
\newcommand{\Ff}{{\cal F}}

\newcommand{\Hh}{{\cal H}}

\newcommand{\Kk}{{\cal K}}

\newcommand{\Mm}{{\cal M}}

\newcommand{\Oo}{{\cal O}}

\newcommand{\Ss}{{\cal S}}

\newcommand{\Xx}{{\cal X}}

\newcommand{\Hf}{{\mathfrak{h}}}
\newcommand{\Vf}{{\mathfrak{v}}}
\newcommand{\AAA}{{\sf A}}
\newcommand{\SSS}{{\sf S}}
\newcommand{\PP}{{\sf P}}
\newcommand{\EE}{{\sf E}}

\newcommand{\inp}{{\rm in}}
\newcommand{\clas}{{\rm clas}}
\newcommand{\meas}{{\rm meas}}

\newcommand{\opt}{{\rm{opt}}}
\newcommand{\ho}{{\rm{h}}}

\newcommand{\states}{{\cal E}_{\cal H}}
\newcommand{\geo}{{\mathrm{g}}}
\newcommand{\inv}{{\,\mathrm{inv}}}

\newcommand{\Mgeo}{M_{\rho\sigma}}
\newcommand{\MgeoV}{M_{\rho\sigma,V}}
\newcommand{\HhA}{{\cal H}_{\sf A}}
\newcommand{\dPsi}{\dot{\Psi}}
\newcommand{\dPhi}{\dot{\Phi}}
\newcommand{\drho}{\dot{\rho}}
\newcommand{\dsigma}{\dot{\sigma}}
\newcommand{\Hgeo}{H_{{\rm{g}},V}}
\newcommand{\Hg}{H_{{\rm{g}}}}
\newcommand{\est}{{\rm est}}
\newcommand{\SL}{L}

\newcommand{\gap}{\epsilon}

\begin{document}

\title{Bures geodesics and quantum metrology}

\author{Dominique Spehner}
\affiliation{Departamento de Ingenier\'{\i}a Matem\'atica, Universidad de Concepci\'on, Concepci\'on, Chile}
\affiliation{Univ. Grenoble Alpes, CNRS, Institut Fourier and LPMMC,  F-38000 Grenoble, France}


 \begin{abstract}
   We study the geodesics on the manifold of mixed quantum states for the Bures metric.
   It is shown that these geodesics correspond to physical non-Markovian evolutions of the system
   coupled to an ancilla.
   Furthermore, we argue that geodesics lead to optimal precision in single-parameter estimation
    in quantum metrology.
    More precisely, if the unknown parameter $x$ is a phase shift proportional to the time parametrizing the geodesic,  the estimation
   error obtained by processing the data of measurements on the system
   is equal to the smallest error that can be achieved from joint detections on the system and ancilla, meaning that there is no information loss
   on this parameter in the ancilla. 
 This error
 can saturate the Heisenberg bound.
 Reciprocally, assuming that the system-ancilla output and input states are related by a unitary 
   $\E^{-\I x H}$ with $H$ a
   $x$-independent Hamiltonian, we show that if the error obtained from measurements on the system is equal to the minimal error obtained from joint measurements on the system and ancilla then the system evolution is given by a geodesic. In such a case,  
   the measurement on the system bringing most information on $x$
   is $x$-independent and can be determined in terms of the intersections 
   of the geodesic with the boundary of quantum states.
   These results show that geodesic evolutions are of interest for high-precision detections
in systems coupled to an ancilla in the absence of measurements on the ancilla.
 \end{abstract}


\maketitle

 \section{Introduction.}

 Geodesics play a prominent role in  classical mechanics and general relativity as they describe the trajectories of free
 particles and light. In contrast, at first sight they are not relevant in quantum mechanics.
 In the quantum theory, the notion of trajectories in space or space-time has to be abandoned. Instead, quantum dynamics are described
 by time evolutions of quantum states. Such states are given by density matrices  $\rho$
 forming a manifold of dimension $n^2-1$, where $n$ is the dimension of the system Hilbert space $\Hh$
 (which we assume here to be finite).
 Different distances can be defined on this manifold.
 A distance appearing naturally in various contexts in quantum information theory
 is the Bures arccos distance $d_{\rm B}$~\cite{Nielsen}.
 This distance  is a good measure of  the distinguishability of quantum states, being a simple
 function of the fidelity~\cite{Uhlmann76}. Furthermore, it has a clear information content, in particular
 it satisfies the data-processing inequality~\cite{Nielsen} and it is closely related to the quantum Fisher information quantifying the maximal amount of
 information on a parameter in a quantum state~\cite{Caves96,myreview}.
 Unlike the trace distance,  $d_{\rm B}$ is a Riemannian distance, \ie, it has an associated
 metric $g$ giving the square infinitesimal distance 
 $\D s^2 = (g_\rho)_ {\alpha \beta} \,\partial_\alpha \rho \, \partial_\beta \rho$,
 where $\partial_\alpha\rho$ is the derivative of $\rho$ with respect to the coordinate $\alpha$
 and we make use of Einstein's summation convention. The manifold of quantum states  $\states$ equipped with the Bures metric is a 
 Riemannian manifold, on which one can define geodesics.

 In this work, we study these geodesics   and analyze their usefulness in quantum metrology. Metrology is the science of devising schemes that extract as precise as possible
 estimates of the parameters associated to the system.
 In quantum metrology, the  estimation of the unknown parameters (for instance, a phase shift in an interferometer)
 is obtained from the detection outcomes on quantum probes undergoing a parameter-dependent transformation process (for instance, the propagation in the
 two arms of the interferometer).
 It has been recognized that estimation errors 
 scaling like the inverse of the number $N$ of probes  (Heisenberg limit) can be achieved  using entangled probes,
 yielding an improvement by a factor of $1/\sqrt{N}$ with respect to the error for classical probes~\cite{Bollinger96,Kok02,Giovannetti06,Giovannetti2,Smerzi09,Hyllus12,Toth12,Demkowitz-Dobrzanski_review}.
 Quantum-enhanced precision have been observed experimentally in optical systems~\cite{Nagata07,Kacprowicz10,Daryanoosh18}, trapped ions~\cite{MeyerPRL2001,Leibfried_Science2004}, and
 Bose-Einstein condensates~\cite{Riedel10,Gross10,Pezze_review18}. 
 In these experiments, noise and losses lead to dephasing and entanglement losses, thus
 limiting the precision.
 In order to account for such limiting effects, several authors have studied parameter estimation in quantum systems
 coupled to their environment~\cite{Ji08,Kolodynski10,Knysh11,Escher11,Escher12,Demkowicz-Dobrzanski12,Huelga2016,Huelga2018,FSMH11,PSMG13,SPFG14}. It has been argued that  for dephasing and photon loss processes the $\sqrt{N}$-improvement is lost, the best precision having the classical scaling
 for large $N$ albeit with a better prefactor~\cite{Huelga97,Ji08,Kolodynski10,Knysh11,Escher11,Escher12,Demkowicz-Dobrzanski12}.
  On the other hand,
   intermediate scalings  $\sim N^{-\kappa}$ of the estimation error,  with $1/2 \leq  \kappa < 1$,
   have been obtained  for phase-covariant and spin-boson  models by optimizing not only the initial state but also the probe time~\cite{Huelga2016,Huelga2018}.  The results of the aforementioned references are, however, model-dependent.
  
On a general ground, one expects that the environment coupling should increase the estimation error since information on the
unknown parameter $x$ can be lost in the environment and measurements on the latter are not possible.
It is nevertheless known that, for any non-unitary transformation process on the probe, there exists an environment which does not deteriorate the precision of the estimation, \ie, a system-environment coupling such that no information on $x$ is lost in the environment~\cite{Escher11,Escher12}. Such a coupling depends in general on $x$. To determine it one must solve a non-trivial optimization problem. 


{  In this paper we show that, in contrast to the aforementioned general expectation,
  it is possible to engineer parameter-independent coupling Hamiltonians of the probe with its environment  such that
  there is no information losses   on the estimated parameter $x$ in the environment for any value of $x$, even though
  the coupling strongly entangles the probe and environment.
  Such a situation occurs when the transformation of the probe state is given by a Bures geodesic. As we shall see, the latter arises from
  a time evolution of the probe coupled to an ancilla with some specific coupling Hamiltonian,
  after tracing out the ancilla (i.e., the environment) degrees of freedom.
  The unknown parameter is a phase multiplying this Hamiltonian; it is proportional to the geodesic distance.
  For $N$ probes coupled to independent ancillas, the precision on this phase saturates the Heisenberg bound, with an error $\sim N^{-1}$.

  More precisely,  we establish that  (i) Bures geodesics 
 are not purely mathematical objects but correspond to physical
 non-Markovian evolutions of the system coupled to an ancilla; we find explicitly the system-ancilla
 coupling Hamiltonian such that a state on the geodesic 
 corresponds to the system state after an  interaction with the ancilla during a lapse of time $\tau$;
 (ii)~if the transformation process on a probe is given by a geodesic and 
the estimated parameter is a phase shift $x$ proportional to the time $\tau$ parametrizing this geodesic, then
the environment does not carry any information on $x$.
  This means that the estimation precision is equal to the best precision which
can be achieved from joint measurements on the probe and ancilla, even if one can measure only the probe. 
Conversely, we show that
  system-ancilla Hamiltonians $H$ with such a property always correspond to geodesics after tracing out the ancilla,
  if one moreover assumes that
  the input system-ancilla state achieves the best possible precision for $H$
  (this is in particular the case for $N$ probes when the Heisenberg bound is saturated).
We also show that 
(iii)~there is an optimal measurement on the probes yielding the smallest error
which is independent of the estimated parameter $x$ and given in terms of the intersection
states of the geodesic with  the boundary  $\partial \states$ of the manifold of  quantum states.

As a consequence of (i),  the geodesics are physical processes that can be simulated in quantum systems coupled to ancillas.
We exhibit below
examples of quantum circuits implementing some system-ancilla coupling Hamiltonians and corresponding geodesics.
The geodesic evolutions being periodic in time, they are strongly non-Markovian, in the sense that memory effects and back action of the ancilla are important.}  
Our results (ii) and (iii) show that geodesics  are of practical interest in quantum metrology.


Our analysis relies on an application to the manifold of
quantum states $\states$ of the concept of Riemannian submersions in Riemannian geometry~\cite{Gallot_book}. On the way,
previous results in the literature~\cite{Barnum_thesis,Ericson05} on the explicit form of the Bures geodesics
for arbitrary Hilbert space dimensions $n$ are revisited. We show that these results are incomplete as they
miss the geodesic curves joining two quantum states along paths which are not the shortest ones.
We derive the explicit forms of {all}  geodesics joining two invertible states $\rho$ and $\sigma$  at arbitrary dimensions $n$ 
and study the intersections of these geodesics with the boundary  $\partial \states$.
Moreover, we argue that the approach based on Riemannian geometry provides useful tools in quantum metrology.

The rest of the paper is organized as follows.
A summary of our main results is presented in Sec.~\ref{main-results} after a brief introduction to quantum parameter estimation.
The mathematical background on Riemannian geometry and submersions is given in Sec.~\ref{sec-preliminaries}.
In  Sec.~\ref{sec_study_of_Bures_geodesics}, the explicit form of the Bures geodesics is derived
and we study their intersections with the boundary $\partial \states$.
In Sec.~\ref{sec-geodesics_physcial_evol}, we show that the geodesics correspond to physical evolutions of the system coupled to an ancilla.
The optimality of geodesics in quantum metrology is investigated in Sec.~\ref{sec-Q_metrology}.
Finally, our main conclusions and perspectives are drawn in Sec.~\ref{sec-Conclusion}. 
Two appendices contain some technical properties and
proofs.

\begin{figure}[htbp]
    \scalebox{0.3}{\input{geGriemann_submersion.pstex_t}}
    \captionsetup{format=plain,justification=raggedright}
    \caption{\small
The manifold of quantum states $\Ee=\states$ is the projection $\pi(\Ss)$ of the manifold $\Ss$ of pure states on an enlarged Hilbert
  space $\Hh\otimes \Hh_\AAA$, where $\pi$ is the partial trace over $\Hh_\AAA$.
 The horizontal subspaces  at
  $\ket{\Psi}$ and $\ket{\Phi_V}$ are orthogonal to the orbits $\pi^{-1} (\rho)$  and $\pi^{-1} (\sigma)$ (red lines).
  A geodesic in $\Ss$  joining $\ket{\Psi}$ to $\ket{\Phi_V}$ (plain black curve) with
  a horizontal initial tangent vector $\ket{{\dPsi}^\ho}$ 
  projects out to a geodesic $\gamma_{\geo,V}$  on ${\mathcal{E}}$  (green plain curve). In contrast, if the
  geodesic in $\Ss$   (blue dashed  curve) has a non horizontal initial tangent
  vector, its projection (green dashed line)  is 
  not a geodesic on $\Ee$. The differential $\D \pi$ maps
  the horizontal tangent vector $\ket{{\dPsi}^\ho}$  to a tangent vector
  $\dot{\rho}$ of  $\gamma_{\geo,V}$ having the same length $\| \dot{\rho}\|=\| {\dPsi}^\ho\|$.
  A  non horizontal vector $\ket{ {\dPsi}}$ is mapped
  by $\D \pi$ to a vector $\dot{\rho}$ with a smaller length, given by $\| \dot{\rho}\|^2 = \| \dot{\Psi}\|^2 - \| {\dPsi}^{\rm{v}} \|^2$,
  where $\ket{ {\dPsi}^{\rm{v}}}$ is the vertical component of $\ket{ {\dPsi}}$ (Pythagorean theorem).
  }
\label{fig-geodesics_Riemannian_submersion}
\end{figure}

 \section{Main results} \label{main-results}

In this section we describe our main results and orient the reader to the subsequent sections, where
these results are presented with more mathematical details.

\subsection{Determination of the geodesics and their intersections with the boundary of quantum states}

The explicit form of the Bures geodesics
has been derived in Refs.~\cite{Barnum_thesis,Ericson05}  for arbitrary Hilbert space dimensions $n<\infty$. The geodesic joining two invertible states $\rho$ and $\sigma$ determined in these references
is given by
\begin{eqnarray} \label{eq-shortest_Bures_geodesics}
 && 
    \gamma_{\geo} ( \tau) =\frac{1}{\sin^2 \theta}
  \bigg( \sin^2 ( \theta - \tau)\,  \rho  + \sin^2 ( \tau )\, \sigma
  \\ \nonumber
  & & 
   +\sin ( \theta - \tau)  \sin ( \tau) \big(
   \rho^{-1/2} | \sqrt{\sigma} \sqrt{\rho} |  \rho^{1/2} + {\mathrm{h.c.}} \big)  \bigg)
\end{eqnarray}
with $0 \leq \tau \leq \theta= \arccos \sqrt{F(\rho,\sigma)}$, where
$F(\rho,\sigma)=(\tr |\sqrt{\sigma} \sqrt{\rho}|)^2$ is the fidelity between $\rho$ and $\sigma$, $|O|= \sqrt{O^\dagger O}$ stands for the
modulus of the operator $O$, and h.c. refers to the Hermitian conjugate. 
The geodesic (\ref{eq-shortest_Bures_geodesics}) has a length $\theta$
equal to the arccos Bures distance $d_{\rm B} (\rho, \sigma)$, it is thus the shortest geodesic arc joining  $\rho$ and $\sigma$.
Recall that a curve is a geodesic if it has constant velocity and minimizes {\it locally} the length of curves between two points.
In Riemannian manifolds, there exists in general geodesics joining two points which do not follow the shortest path from
one point to the other.  
For instance,  there are two geodesic arcs joining two non-diametrically opposite points on a sphere, namely
the two arcs of the great circle passing through them; the smallest arc is the shortest geodesic, which minimizes the length globally, and
the largest arc is another geodesic with a length strictly larger than the distance between its two extremities. On the other hand,
if the two points  on the sphere are diametrically opposite, there are infinitely many geodesics joining them, which have all the same length.  

In a similar way, we show in Sec.~\ref{sec_study_of_Bures_geodesics} that,  depending on the two invertible mixed states $\rho$ and $\sigma$,
there is either a finite or an infinite number of Bures geodesics joining $\rho$ and $\sigma$.
The explicit form of these geodesics is given  by a formula generalizing (\ref{eq-shortest_Bures_geodesics}) in Theorem~\ref{prop-Bures_geodesics} below.
For generic invertible states $\rho$ and $\sigma\in \states$, the number of geodesics is finite and equal to $2^n$ (recall that $n = \dim \Hh$).
The geodesics can be classified according to the number of times they bounce on the boundary of quantum states $\partial \states$ between $\rho$ and $\sigma$.
Recall that $\partial \states$ is 
  the set of non-invertible density matrices.
The shortest geodesic (\ref{eq-shortest_Bures_geodesics}) is the only geodesic 
starting at $\rho$ and ending at $\sigma$ without intersecting the boundary (but it does so if one extend it after
$\sigma$, as shown in~\cite{Ericson05}).

Although these results are not the most original contribution of the paper, they form the starting point of the subsequent analysis.  
The method to determine the Bures geodesics is similar, albeit technically more involved,
to textbook derivations of
the Fubini-Study geodesics on the complex projective space  ${\mathbf{CP}}^n$ (manifold of pure
quantum states) \cite{Gallot_book}. It relies on the notion of Riemannian submersions.
The main observation is that the manifold  of (mixed) quantum states
can be viewed as the projection of {  the manifold of pure states from an enlarged
  Hilbert space $\Hh \otimes \Hh_\AAA$ (purifications), where
$\Hh_\AAA$ is an ancilla Hilbert space and the projection is the partial trace over the ancilla.} The set of all
purifications $\ket{\Psi} \in \Hh \otimes \Hh_\AAA$ of $\rho$ projecting out to the same density matrix $\rho$ forms an orbit under the action of local unitaries on the ancilla. 
As noted by Uhlmann~\cite{Uhlmann76,Uhlmann86}, the Bures distance between $\rho$ and $\sigma$ is the norm distance between
the corresponding orbits, that is, $d_{\rm Bures} ( \rho,\sigma) = \min \|\ket{\Psi}-\ket{\Phi}\|$ where the minimum is over all purifications $\ket{\Psi}$
 and $\ket{\Phi}$ on the orbits of $\rho$ and $\sigma$, respectively.
For such a distance, the geodesics $\gamma_\geo(\tau)$ joining $\rho$ and $\sigma$ are obtained
by projecting  onto $\states$ geodesics on the purification manifold having horizontal tangent vectors, as illustrated in Fig.~\ref{fig-geodesics_Riemannian_submersion}.
The latter geodesics are easy to determine since the metric on this manifold is the euclidean metric (given by the scalar product)
restricted to a unit hypersphere (since purifications are normalized vectors).
More details on Riemannian submersions are given in Sec.~\ref{sec-preliminaries} below.

\subsection{Geodesics correspond to physical evolutions} \label{sec-result_geodesics_are_physical}

One of the  purposes of this paper is to show that the Bures geodesics are not only mathematical objects but correspond to physical dynamical evolutions that could in principle be realized in the laboratory.
Let $\gamma_\geo(\tau)$ be a geodesic on $\states$ starting at $\rho = \gamma_\geo (0)$.
Consider an ancilla system $\AAA$ with  Hilbert space $\Hh_\AAA$ of dimension $n_\AAA \geq n$.
Let $\ket{\Psi}$ be a purification of $\rho$ on $\Hh \otimes \Hh_\AAA$, \ie, $\rho = \tr_\AAA \ketbra{\Psi}{\Psi}$, where $\tr_\AAA$ is the partial trace over the ancilla.
We show in Sec.~\ref{sec-geodesics_physcial_evol} (see Theorem~\ref{prop-geodesics_as-Q_evol})
that there exists a system-ancilla Hamiltonian $\Hg$ such that  
\begin{equation} \label{eq-geo_as_physical_evol}
  \gamma_\geo (\tau) = \tr_\AAA  \E^{-\I \tau \Hg} \ketbra{\Psi}{\Psi} \,\E^{\I \tau \Hg} \;.
 \end{equation} 
In other words, $\gamma_\geo (\tau) $ is the system state at the (dimensionless) time $\tau$,
given that the system is coupled to the ancilla at time $0$ and
interacts with it up to time $\tau$ with the Hamiltonian $\Hg$. This Hamiltonian reads
\begin{equation} \label{eq-geodesic_Hamil}
  \Hg = -\I \big( \ketbra{\Psi}{\dPsi} - \ketbra{\dPsi}{\Psi} \bigr)\;,
\end{equation}
where $\ket{\dot{\Psi}}$ is a normalized vector satisfying the horizontality condition
\begin{equation} \label{eq-horizontality_cond}
 \ket{\dot{\Psi}} = {  L_\SSS} \otimes \identity_\AAA \ket{\Psi} 
\end{equation}
for some self-adjoint operator {  $L_\SSS$ acting on the system such that $ \langle L_\SSS \otimes \identity_\AAA \rangle_\Psi = 0$}.
Condition (\ref{eq-horizontality_cond}) can be interpreted geometrically as follows:  
$\ket{\dot{\Psi}}$ is a vector in the tangent space at $\ket{\Psi}$ which is orthogonal to the orbit 
$\{ \identity \otimes U_\AAA \ket{\Psi}\,;\,U_\AAA \text{ unitary on }\Hh_\AAA\}$ of $\rho$ under the unitary group on the ancilla.
Note that $\braket{\Psi}{\dot{\Psi}}=0$.
As will be shown in Sec.~\ref{sec-Q_metrology}, the operator $2 L_{\SSS}$ is equal to the symmetric logarithmic derivative of $\gamma_\geo(\tau)$ at $\tau=0$.

Since $\rho$ can be chosen arbitrarily on $\gamma_\geo$ and 
all geodesics extended over the time interval $[0,\pi]$ intersect the
boundary of quantum states $\partial \states$ (see Appendix~\ref{sec-intersections_boundary}), one can without loss of generality assume
that $\rho \in \partial \states$.
If $\gamma_\geo$ has an intersection  with $\partial \states$ given by a pure state $\rho_\psi= \ketbra{\psi}{\psi}$,
one can choose $\rho=\rho_\psi$. Then the purifications of $\rho$ are product states $\ket{\Psi}=\ket{\psi} \ket{\alpha}$ and,
as a consequence of (\ref{eq-geo_as_physical_evol}), there is a smooth family of  Completely Positive
Trace Preserving (CPTP) maps $\Mm_{\geo,\tau}$ (quantum channels) such that
\begin{equation}
  \gamma_\geo (\tau)= \Mm_{\geo,\tau} ( \rho)\;,
\end{equation}
\ie,   the geodesic evolution is obtained by applying $\Mm_{\geo,\tau}$ to $\rho$.
The quantum evolution $\{ \Mm_{\geo,\tau} \}_{\tau \geq 0}$ is  strongly non-Markovian.
Actually, we show in Sec.~\ref{sec-geodesics_physcial_evol} that this evolution is periodic in time,
\begin{equation} \label{eq-geo_evolutyion_is_periodic}
  \Mm_{\geo,\tau+ 2\pi} = \Mm_{\geo,\tau}\;.
\end{equation}

\begin{figure}[htbp]
  \begin{center}
    \includegraphics[height=4.5cm,angle=0]{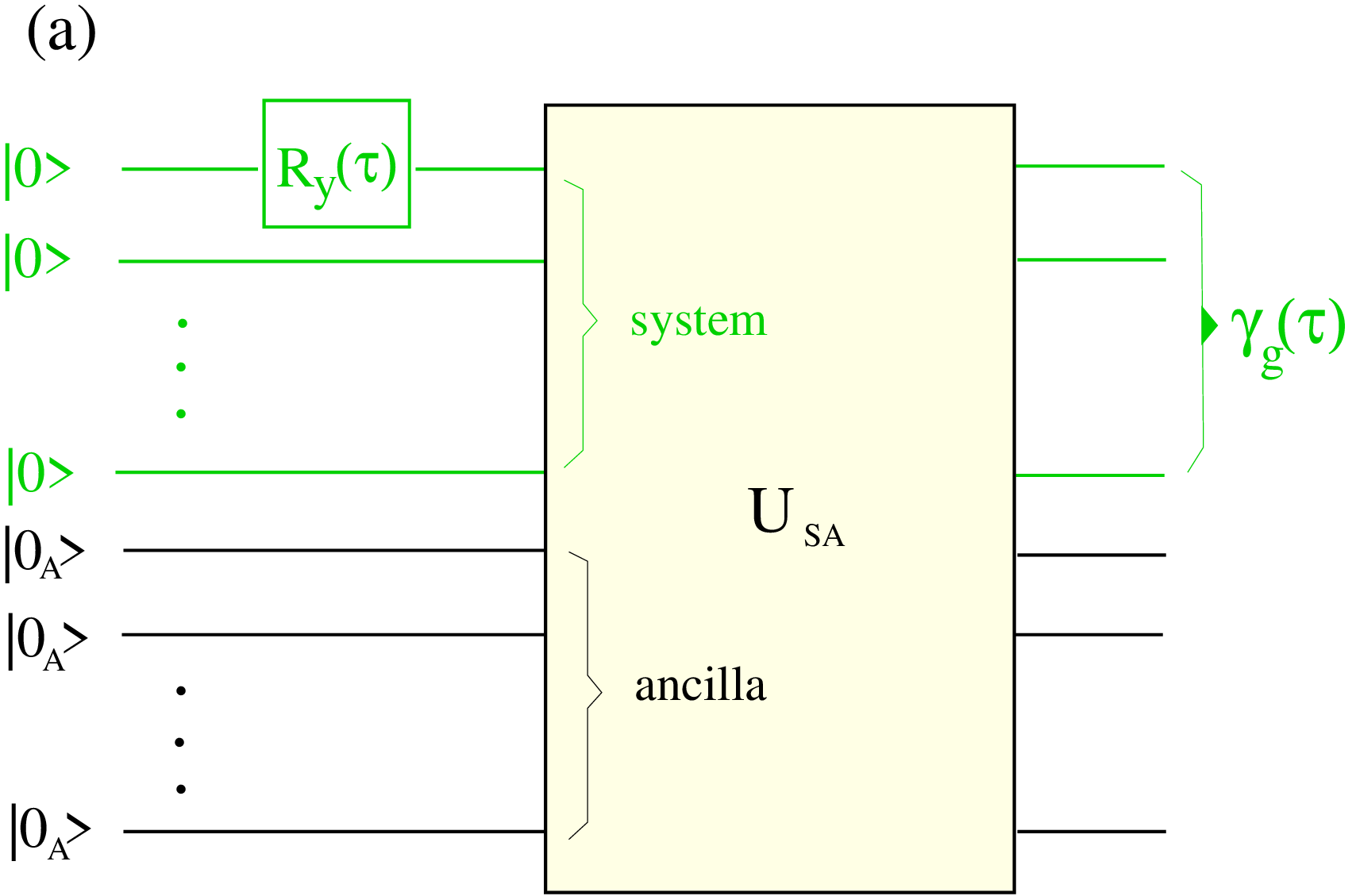}
\end{center}  
  \vspace{1mm}

  \includegraphics[height=3.1cm,angle=0]{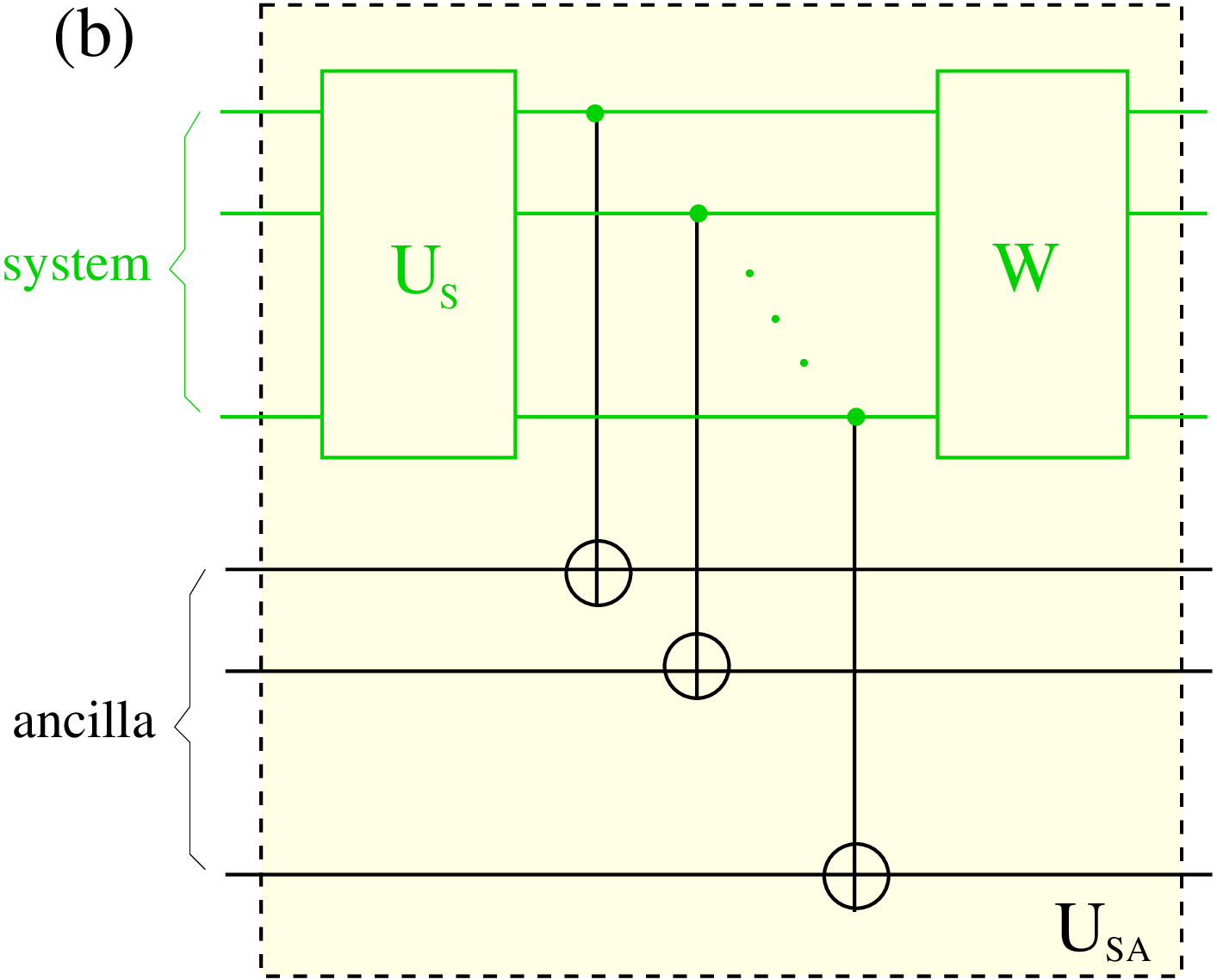}\hspace{2mm}
   \includegraphics[height=3.1cm,angle=0]{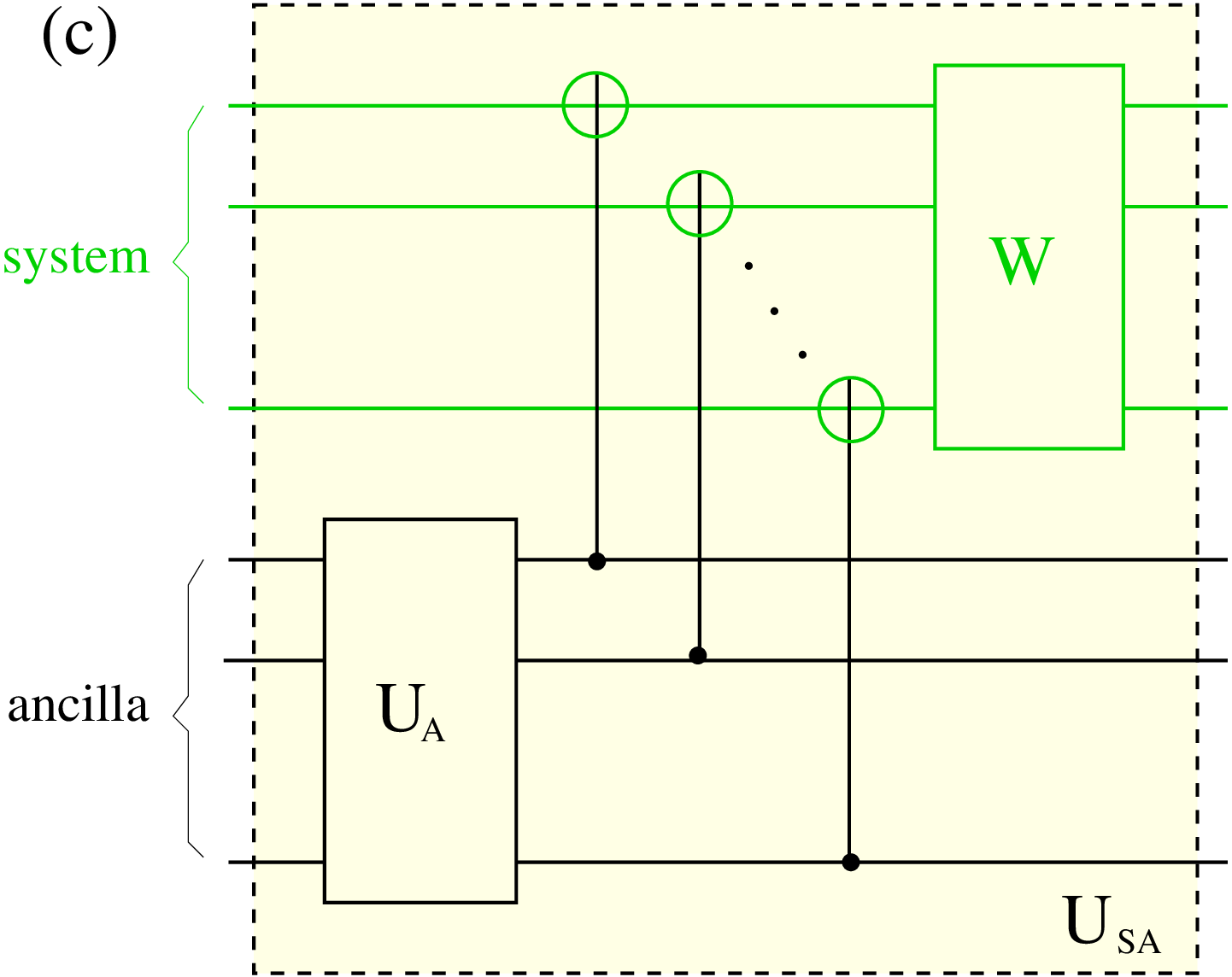}
   \captionsetup{format=plain,justification=raggedright}
   \caption{\small
     Quantum circuit implementing the geodesic evolution.
  {\bf (a):} General circuit for a $d$-qubit system; $R_y (\tau)= \E^{-\I \tau \sigma_y}$ is a rotation around the $y$-axis
  and $U_{\SSS \AAA}$ a system-ancilla entangling unitary satisfying (\ref{eq-def_entangling_unitary}).
  {\bf (b),~(c):}~examples of quantum circuits for $U_{\SSS \AAA}$;
  the unitaries $W$, $U_\SSS$, and $U_\AAA$ are such that
  $U_i \ket{0}_i  =   \sum_k \sqrt{p_k} \ket{k}_{i}$, $i= \SSS,\AAA$, $U_\SSS \ket{1}_\SSS = \sum_k \alpha_k \ket{k}$,
  and $W \ket{k}_\SSS  =  \ket{w_k}_\SSS$, $k=0,\ldots,2^d-1$,
  where  $\sqrt{p_k}$ and $\{ \ket{w_k}_\SSS\}$ are the coefficients and orthonormal basis
  in the Schmidt decomposition of $\ket{\Psi}$ and $\alpha_k \in \real\setminus\{0\}$, $\sum_k \alpha_k \sqrt{p_k}=0$. 
  }
\label{fig-Q_circuit_geodesic_evol}
\end{figure}

For a system formed by $d$ qubits coupled to $d$ ancilla qubits,
the geodesics can  be implemented by the quantum circuit of
Fig.~\ref{fig-Q_circuit_geodesic_evol}(a), where $U_{\SSS \AAA}$ is a unitary operator on $\Hh \otimes \Hh_\AAA$ such that
\begin{equation} \label{eq-def_entangling_unitary}
  \ket{\Psi} = U_{\SSS \AAA} \ket{0} \ket{0}_\AAA  
  \quad , \quad \ket{\dPsi}  = U_{\SSS \AAA} \ket{1} \ket{0}_\AAA \;.
\end{equation}
Here, $\{ \ket{k} \}_{k=0}^{n-1}$ and $\{ \ket{k}_\AAA \}_{k=0}^{n-1}$ denotes the computational bases of
$\Hh \simeq \complex^n$ and $\Hh_\AAA\simeq \complex^n$ (with $n=2^d$).
Indeed, denoting by  $\sigma_y^{(1)}$ the $y$-Pauli matrix acting on the first qubit and by $\identity^{(2 \ldots{  d})}$ the identity operator on
the other  qubits of the system, the Hamiltonian
\begin{equation}
  \widetilde{H}_\geo = U_{\SSS \AAA} \, \sigma_y^{(1)} \otimes \identity^{(2 \ldots {  d})} \otimes \identity_\AAA \, U^\dagger_{\SSS \AAA}
\end{equation}
leaves the subspace $\Span \{ \ket{\Psi}, \ket{\dPsi} \}$ invariant and coincides with the geodesic Hamiltonian {   (\ref{eq-geodesic_Hamil})} on this subspace.
Thus
\begin{equation}
 \E^{- i \tau  {H}_\geo} \ket{\Psi} = U_{\SSS \AAA} \,\E^{-\I \tau \sigma_y^{(1)}} \otimes \identity^{(2\ldots {  d})} \otimes \identity_\AAA \,\ket{0} \ket{0}_\AAA\;.
\end{equation}
By (\ref{eq-geo_as_physical_evol}), the state of the system at the output of the circuit is $\gamma_\geo (\tau)$.
The two circuits
 of Fig.~\ref{fig-Q_circuit_geodesic_evol}(b) and (c) give
examples of entangling unitaries $U_{\SSS \AAA}$ implementing a geodesic $\gamma_\geo(\tau)$
through  an arbitrary invertible state $\rho= \gamma_\geo (0)$.
Introducing the spectral decomposition $\rho= \sum_k p_k \ketbra{w_k}{w_k}$,
one checks that
(\ref{eq-def_entangling_unitary})  holds for both circuits, with
$\ket{\Psi} = \sum_{k} \sqrt{p_k} \ket{w_k} \ket{k}_\AAA$ a purification of $\rho$ and
$\ket{\dPsi}$  a horizontal tangent vector of the form
(\ref{eq-horizontality_cond}) for some self-adjoint operator $L_\SSS$.
Note that the system-ancilla entangling operation in these circuits is
obtained by means of $d$ C-NOT gates.
The geodesic implemented by the unitary $U_{\SSS \AAA}$ of Fig.~\ref{fig-Q_circuit_geodesic_evol}(b) is a geodesic joining two commuting states
(see Sec.~\ref{sec_study_of_Bures_geodesics}).

\subsection{Optimality of geodesics for parameter estimation in open quantum systems}
\label{sec_results_Q_metrology}

Before explaining our results, let us introduce some basic background on quantum metrology for readers not familiar with this field.
More information can be found e.g. in the review articles~\cite{Caves96,Giovannetti2,Demkowitz-Dobrzanski_review,Kok_review2020}.

\vspace{2mm}

\noindent {\it Parameter estimation in closed and open quantum systems.}

The goal of parameter estimation is to estimate an unknown real parameter
$x$ pertaining to an interval $X \subset \real$ by performing measurements on a quantum system (probe).
Before each measurement, the system undergoes a $x$-dependent process transforming the input state $\rho_{\inp}$ into an
output state $\rho_x$.
The  estimator $x_\est \in X$ is a function of the measurement outcomes.
The precision of the estimation is characterized by the
variance $(\Delta x )^2 = \langle (x_\est - x)^2 \rangle_x$, where
$\langle \cdot \rangle_x$ refers to the average over the outcomes conditioned to the parameter value $x$. Hereafter, 
we assume that the estimator is   locally unbiased, \ie, $\langle x_\est \rangle_x = x$ and 
$\partial_x \langle x_\est \rangle_x  = 1$. Furthermore, we assume that the map $ x \in X \mapsto \rho_x$ is injective and
that $\rho_x$ has a rank independent of $x$.
If one performs $N_\meas$ independent identical measurements on identical probes prepared in state $\rho_x$, then
the estimation error  satisfies the quantum Cr\'amer-Rao bound~\cite{Helstrom68,Braunstein94}
\begin{equation} \label{eq-Q_Cramer_Rao_bound}
  \Delta x \geq
   (\Delta x)_{\rm{QCRB}} = \frac{1}{\sqrt{N_\meas \,\Ff_Q (x, \{ \rho_x \}_{x \in X} )}}\;,
\end{equation}
where  $\Ff_Q (x,\{ \rho_x \}_{x \in X} )$ is the quantum Fisher information (QFI).
The QFI is obtained by maximizing  over all measurements
the classical Fisher information (CFI)
\begin{equation} \label{eq-classical_Fisher_info}
  \Ff_\clas ( x, \{ p_{j | x } \}_{x \in X}  ) = \sum_{j,p_{j|x}>0} \frac{( \partial_x p_{j| x} )^2}{p_{j| x}} \;,
\end{equation}
where  $p_{j | x }$ is the probability of the measurement outcome $j$ given that the system is in state $\rho_x$.
Note that the QFI and CFI depend on general on the parameter value $x$. For clarity we write  this dependence on $x$ explicitly.
The bound (\ref{eq-Q_Cramer_Rao_bound}) is saturated asymptotically (in the limit $N_\meas \gg 1$) by choosing:
(i)~the maximum likelihood estimator $x_\est$ (ii)~the optimal measurement maximizing the CFI, for which $\Ff_\clas =\Ff_Q$.
This means that the \RHS of  (\ref{eq-Q_Cramer_Rao_bound}) gives the smallest error
that can be achieved in the estimation.
 { Note that this result is not true if the map $x \in X \mapsto \rho_x$ is not injective, because different values of $x$ corresponding to the same $\rho_x$ can not be distinguished. To circumvent this problem, one can partition
the interval $X$ into smaller intervals $X_k$ such that $x \in X_k \mapsto \rho_x$ is injective for each $k$, provided one has some
 prior knowledge about the $X_k$ to which $x$ belongs~\cite{Demkowitz-Dobrzanski_review}.    
 The saturation of the bound  (\ref{eq-Q_Cramer_Rao_bound}) and the bound itself may also not be true when
 the rank of $\rho_x$ has a jump at the parameter value $x$, see e.g.~\cite{Seveso2019}.

 Using classical resources, the smallest estimation error $(\Delta x)_{\rm{QCRB}}$} scales with the number of probes $N$ like $1/\sqrt{N}$
(shot noise limit).
Multipartite entanglement among the quantum probes can enhance the precision by a factor $1/\sqrt{N}$,
  leading to errors $ (\Delta x)_{\rm{QCRB}}$ scaling like $1/N$ (Heisenberg limit)~\cite{Bollinger96,Kok02,Giovannetti06,Giovannetti2,Smerzi09,Hyllus12,Toth12}.

For a closed system in a pure state undergoing the $x$-dependent unitary transformation
$
  \ket{\Psi_x} = \E^{-\I x H} \ket{\Psi_\inp}
$,
where $H$ is a given observable, the QFI is given by~\cite{myreview,Braunstein94}
\begin{eqnarray} \label{eq-QFI_pure_states}
  \nonumber
  \Ff_Q ( \{ \ket{\Psi_x \}_{x \in X} )} & = & 4 \big( \| \dPsi_x \|^2 - \big| \braket{\Psi_x}{\dPsi_x} \big|^2 \big)  
\\
   & = & 4 \langle ( \Delta H )^2 \rangle_{\Psi_\inp} \;,
\end{eqnarray}
where $\ket{\dPsi_x}$ is the derivative of $\ket{\Psi_x}$ with respect to $x$ and
 $ \langle (\Delta H )^2 \rangle_{\Psi} = \bra{\Psi} H^2 \ket{\Psi} - \bra{\Psi} H \ket{\Psi}^2$ is the square quantum fluctuation of $H$
in state $\ket{\Psi}$.
The input states maximizing this fluctuation are the superpositions
\begin{equation} \label{eq-Schrodinger_cat_states_intro}
  \ket{\Psi_\inp} = \frac{1}{\sqrt{2}} \Big(   \ket{\epsilon_\mmax} + \E^{\I \varphi} \ket{\epsilon_\mmin} \Big)\;,
\end{equation}  
where $\varphi$ is a real phase and $\ket{\epsilon_\mmax}$ (respectively $\ket{\epsilon_\mmin}$) is an eigenstate of $H$ with
maximal (minimal) eigenvalue $\epsilon_\mmax$ ($\epsilon_\mmin$). (We assume here that these eigenvalues are non-degenerated.)
{  The corresponding maximal QFI is $ \Ff_Q ( \{ \ket{\Psi_x \}_{x \in X} )}= 4 \gap^2$ with 
  $\gap = (\epsilon_\mmax - \epsilon_\mmin )/2$.}
According to (\ref{eq-Q_Cramer_Rao_bound}) and (\ref{eq-QFI_pure_states}),
the states (\ref{eq-Schrodinger_cat_states_intro}) are the optimal input states
minimizing the estimation error  $(\Delta x)_{\rm{QCRB}}$.
Indeed,  by convexity of the QFI, using mixed input states $\rho_{\inp}$ can not lead to smaller errors.
Note that the QFI (\ref{eq-QFI_pure_states}) is
independent of $x$, as clear from the last expression. For this reason, we omit $x$ in the argument of  $\Ff_Q$.
  In contrast, for non-unitary evolutions the QFI depends in general upon
  the value $x$ of the estimated parameter.

For $N$  probes undergoing a unitary ``parallel'' transformation with the observable $H_N = \sum_{i=1}^N H_i$, where
$H_i$ stands for the action of $H$ on the $i$th probe, the error has the Heisenberg scaling
$(\Delta x)_{\rm{QCRB}} \propto 1/N$ (more precisely, 
{  $(\Delta x)_{\rm{QCRB}} =(2  N \gap\sqrt{N_\meas})^{-1}$).}
The optimal input states, given by replacing  $ \ket{\epsilon_\mmax} $ and
$\ket{\epsilon_\mmin}$ in (\ref{eq-Schrodinger_cat_states_intro}) by the eigenvectors of $H_N$ with
maximal and minimal eigenvalues $N \epsilon_\mmax$ and $N\epsilon_\mmin$,
show genuine multipartite  entanglement.

In experimental setups, the coupling of the probes  with their environment can not be neglected.
A general description of the state transformation process is given by a family $\{ \Mm_x \}_{x \in X}$ of
$x$-dependent quantum channels, 
 which accounts for the joint effects of the free evolutions of the probe $\PP$ and environment $\EE$ and 
the coupling between them.
The probe output state is related to the input state  $\rho_\inp$ by 
\begin{equation} \label{eq-output_states}
  \rho_x = \Mm_x ( \rho_\inp )\; .
\end{equation}
In a realistic scenario, measurements can be performed on the probe only,
\ie, one can not  extract information from the environment. 
{  The  error $(\Delta x)_{\rm{QCRB}}$ obtained by measuring the probe can clearly not be smaller than the error
$(\Delta x)_{\rm{QCRB,\PP\EE}}$
obtained from joint measurements on the probe and environment.}
A natural question is whether there exists a family of quantum channels {  $\{ \Mm_x \}_{x \in X}$ and input states $\ket{\Psi_\inp}$}
such that
$(\Delta x)_{\rm{QCRB}}= (\Delta x)_{\rm{QCRB,\PP\EE}}$ for {any value of $x$.}
This  means that the environment does not carry any information about the parameter $x$.

\vspace{2mm}

\begin{figure}[htbp]
  \includegraphics[height=5cm,angle=0]{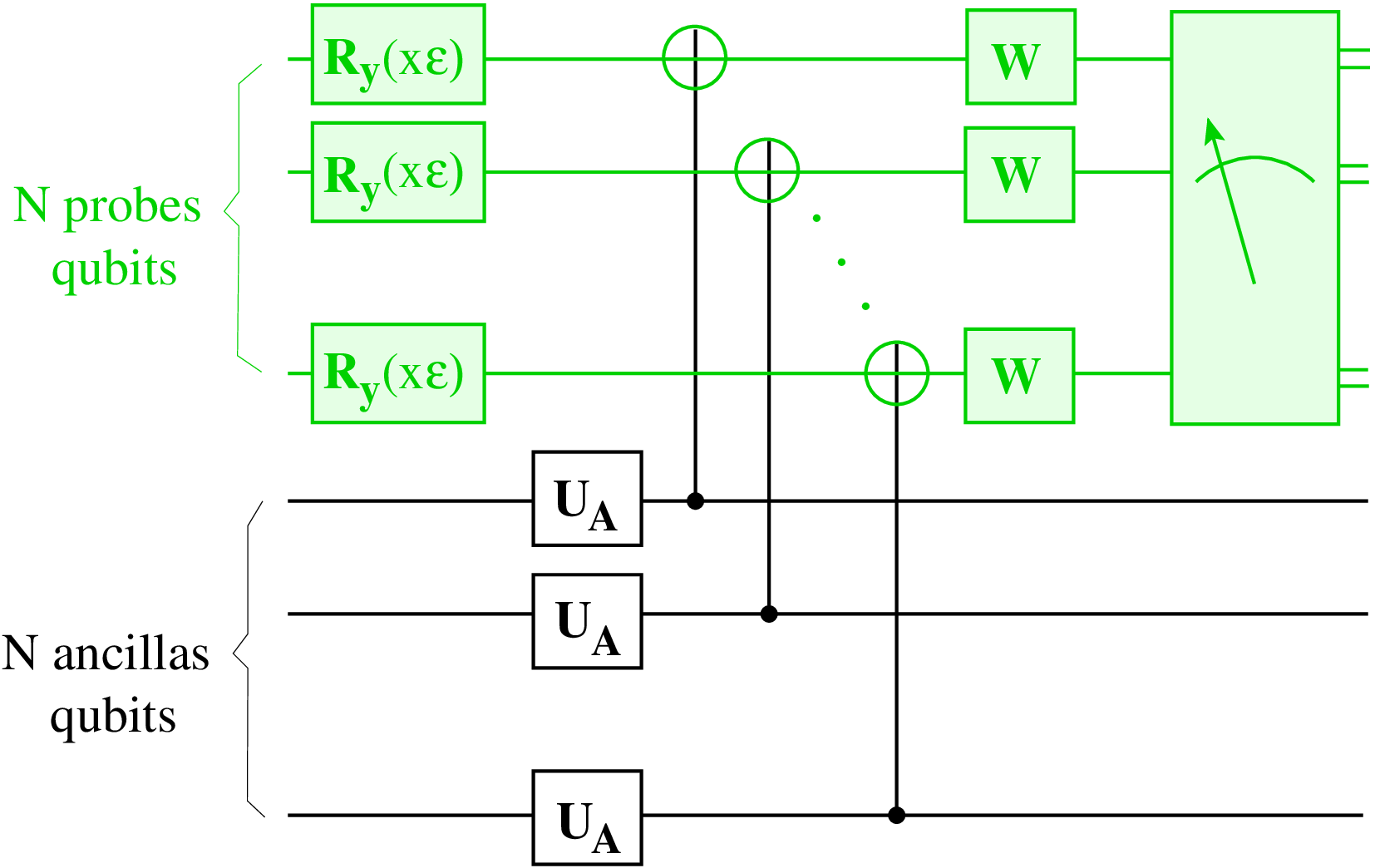}
  \captionsetup{format=plain,justification=raggedright}
  \caption{\small Quantum circuit implementing a geodesic transformation
    for the estimation of a phase shift $x$, with an  error reaching the Heisenberg scaling.
       The single qubit unitaries $R_y$, $U_\AAA$ and $W$ are as in Fig.~\ref{fig-Q_circuit_geodesic_evol} with $d=1$. 
  In spite of the presence of the C-NOT gates entangling the probe qubits with the ancilla qubits,
  the maximal information on $x$ can be recovered from measurements on the probe qubits only, with a minimal
  error $(\Delta x)_{\rm QCRB} = (2 N \gap\sqrt{N_{\rm meas}})^{-1}$. 
  }
\label{fig-Q_circuit_metrology}
\end{figure}

\newpage

\noindent {\it Optimality of geodesics in parameter estimation.}

We will show in Sec.~\ref{sec-Q_metrology}
below that it is possible to find such a family of quantum channels and input state,
which are moreover such that the error $(\Delta x)_{\rm{QCRB}}$ is equal to the smallest possible error
\begin{equation}
(\Delta x)_{\rm{QCRB}} =   
  (\Delta x)_{\rm{QCRB,\PP\EE}}=  ( 2 {  N} \gap\sqrt{N_\meas} )^{-1}\; ,
\end{equation}
where {  $\gap = (\epsilon_\mmax - \epsilon_\mmin )/2$} is as before
the maximal quantum fluctuation of the observable $H$. 
Such a family is given by the CPTP maps associated to
the Bures geodesics described in the previous subsection.
The corresponding state transformation  is
\begin{equation} \label{eq-ouput_states_geodesic}
  \rho_x = \gamma_{\geo} (x \gap) \;,
\end{equation}
where $\gamma_{\geo}(\tau)$ is a geodesic starting at 
\begin{equation} \label{eq-initial_state_geodesic}
  \gamma_{\geo}(0) = \rho_\inp = \tr_\AAA \ketbra{\Psi_\inp}{\Psi_\inp}\;.
\end{equation}
Note that the estimated parameter $x$ appears in (\ref{eq-ouput_states_geodesic}) as a phase shift proportional to the
dimensionless time $\tau$,
 and that $\gamma_\geo$ is an arbitrary {\it fixed} geodesic passing through $\rho_\inp$.

In other words, for the state transformation and phase shift $x$ given in (\ref{eq-ouput_states_geodesic}), one has:
(i)~the precision on the estimation of $x$ obtained from  measurements on the probe only is the same as that obtained
by performing joint measurements on the probe and
environment;
(ii)~the error $(\Delta x)_{\rm QCRB}= (\Delta x)_{\rm{QCRB,\PP\EE}}$  is the smallest achievable among all  probe-environment initial states.
Furthermore, the error scales like $N^{-1}$ with the number of probes, \ie, it 
reaches the Heisenberg bound. An example of quantum circuit implementing the corresponding state transformation is shown in Fig.~\ref{fig-Q_circuit_metrology}.

To justify properties (i)-(ii), let us look at the transformation (\ref{eq-output_states}) as resulting from the coupling of the
probe with an ancilla $\AAA$, the probe and ancilla being initially in a pure state  and undergoing a unitary transformation.
  Note that this is always possible according to Stinepring's theorem~\cite{Stinespring55}. Let us assume that the unitary evolution of the probe and ancilla
  is generated by some $x$-independent Hamiltonian $H$.
Thus
\begin{equation} \label{eq-purification_rho_x}
  \rho_x = \tr_\AAA \ketbra{\Psi_x}{\Psi_x}\quad ,  \quad   \ket{\Psi_x} = \E^{-\I x H } \ket{\Psi_\inp}\;.
\end{equation}
Using the physical irrelevance of phase factors in the quantum states $\ket{\Psi_x}$, we may assume that
$\langle H \rangle_{\inp} = \bra{\Psi_\inp} H \ket{\Psi_\inp}=0$ (in fact, 
this amounts to  multiply  $\ket{\Psi_x}$ by the phase factor $\E^{\I x \langle H \rangle_{\inp}}$, \ie, to replace $H$ by $\Delta H = H -  \langle H \rangle_{\inp}$ in (\ref{eq-purification_rho_x})).     
Let us decompose
the tangent vector $\ket{\dPsi_x}= \partial_x \ket{\Psi_x}$ into the sum of its horizontal part $\ket{\dPsi^\ho_x}$ and its vertical part
$\ket{\dPsi^{\rm v}_x}$, with $\braket{\dPsi^\ho_x}{\dPsi^{\rm v}_x}=0$, where horizontality means orthogonality to the orbit of $\rho_x$, see Fig.~\ref{fig-geodesics_Riemannian_submersion}.
According to the theory of Riemannian submersions, the square norm of the horizontal part coincides with the
square norm $(g_{\rm B})_{\rho_x} ( \dot{\rho}_x, \dot{\rho}_x)$ of $\dot{\rho}_x=\partial_x \rho_x$, the latter
norm being given by the Bures metric $g_{\rm B}$ at $\rho_x$ (see Sec.~\ref{sec-preliminaries}).
It is known that this square norm is equal to the QFI up to a factor of one fourth~\cite{Braunstein94,myreview}.
Thus, thanks to the Pythagorean theorem
$\| \dPsi_x \|^2 = \| \dPsi_x^\ho\|^2 + \| \dPsi_x^{\rm v}\|^2$, one obtains the following formula for the QFI of the probe~\cite{Escher11}
\begin{equation} \label{eq-formula_QFI_Pythagore}
  \Ff_Q (x, \{ \rho_x\}_{x \in X})
  = \Ff_Q( \{ \ket{\Psi_x} \}_{x \in X}) - 4 \| \dPsi^{\rm v}_x\|^2\;,
\end{equation}
where the {  QFI of the total system (probe and ancilla) is given by
 $\Ff( \{ \ket{\Psi_x} \}_{x \in X})= 4 \| \dPsi_x \|^2$ thanks to (\ref{eq-QFI_pure_states}) and our assumption $\langle H \rangle_\inp=0$, and
$\| \dPsi^{\rm v}_x\|^2$ quantifies the amount of information on $x$ in the ancilla.  
It follows that the equality  $\Ff_Q (x, \{ \rho_x\}_{x \in X}) =  \Ff_Q ( \{ \ket{\Psi_x} \}_{x \in X}) $
holds whenever $\ket{\dPsi^{\rm v}_x}=0$, \ie, for horizontal tangent vectors $\ket{\dPsi_x}$.

Let  $\ket{\Psi_\geo (\tau)}$ be a pure--state geodesic for the norm distance on $\Kk= \Hh \otimes \Hh_\AAA$.
A general result on Riemannian submersions tells us that if the tangent  vector $\ket{\dPsi_\geo (\tau)}$ is horizontal at $\tau=0$
then it remains horizontal at all times $\tau$
and $\ket{\Psi_\geo (\tau)}$ projects out to a geodesic $\gamma_\geo (\tau)$ on $\states$,
where the projection corresponds  here to the partial trace over the ancilla (see Sec.~\ref{sec-Riemannian_submersions} for more details).
{  Reciprocally, any geodesic on $\states$ can be lifted locally to a pure--state geodesic
 $\ket{\Psi_\geo (\tau)}$ with horizontal tangent vectors.}  
Therefore, for the state transformation (\ref{eq-ouput_states_geodesic}) one has $\ket{\dPsi_x^{\rm v}} =0$
for all parameter values $x$.
Thanks to (\ref{eq-Q_Cramer_Rao_bound}) and (\ref{eq-formula_QFI_Pythagore}), $(\Delta x)_{\rm{QCRB}}$ is then equal to the best error
obtained from measurements on the probe and ancilla.
This justifies property (i) above.
Furthermore, it is easy to show that $\ket{\Psi_\inp}$ is the superposition (\ref{eq-Schrodinger_cat_states_intro}) for the Hamiltonian
$H = \gap \,\Hg$, $\Hg$ being given by (\ref{eq-geodesic_Hamil})
with $\ket{\Psi}= \ket{\Psi_\inp}$ and $\ket{\dPsi} = \ket{\dPsi_\inp}/\|\dPsi_\inp\|$
{  (note that our hypothesis $\langle H \rangle_\inp=0$ is fulfilled for this Hamiltonian).}
This justifies (ii).

Conversely, we show in Sec.~\ref{sec-Q_metrology} that if 
  the state transformation of the probe and ancilla is given by a unitary $U_x= \E^{-\I x H}$, where $x$ is the unknown parameter and $H$ 
  a parameter-independent Hamiltonian, and if properties (i) and (ii) hold, then the state transformation of the probe is necessarily a geodesic evolution. More precisely,
Theorem~\ref{prop-geodesics_and_parameter_estimation}
 characterizes all Hamiltonians $H$ and input states $\ket{\Psi_\inp}$ of the probe and ancilla satisfying the two following
conditions, which are equivalent to (i) and (ii):
(I)~$\ket{\dPsi_x}$ is horizontal for all $x\in X$ (\ie, it satisfies (\ref{eq-horizontality_cond}));
(II) the QFI of the probe and ancilla is maximal (\ie, $\Ff(\{ \ket{\Psi_x} \}_{x \in X})= 4 N^2 \gap^2$). 
We prove that these conditions  hold if and only if
$\ket{\Psi_x} = \ket{\Psi_\geo ( x \gap )} = \E^{-\I x \gap  H_\geo} \ket{\Psi_\inp}$ is a pure--state geodesic of the probe and ancilla
with a horizontal initial tangent vector $\ket{\dPsi_\inp}$,
\ie, if and only if  the state transformation has the form (\ref{eq-ouput_states_geodesic}) for some geodesic $\gamma_\geo$
  and initial state (\ref{eq-initial_state_geodesic}).
 By (\ref{eq-formula_QFI_Pythagore}), this implies that geodesic evolutions are optimal for quantum parameter estimation in the following sense: among all coupling Hamiltonians $H$ of the  probe and ancilla having a fixed energy gap $2 \gap$
  and all initial states $\ket{\Psi_\inp}$ satisfying (II),
  the highest QFI of the probe for arbitrary parameter values $x$ is obtained when $H$ generates
  a geodesic evolution of the probe.
  As we show in  Sec.~\ref{sec-Q_metrology}, this occurs when}  
the restriction of $H$ to the two-dimensional subspace spanned by
its eigenvectors associated to the maximal and minimal eigenvalues 
coincides with the geodesic Hamiltonian (\ref{eq-geodesic_Hamil}) {  multiplied by  $\epsilon$}, with $\ket{\Psi}= \ket{\Psi_\inp}$ and
$\ket{\dPsi}= \ket{\dPsi_\inp}/\|\dPsi_\inp\|$.
Theorem~\ref{prop-geodesics_and_parameter_estimationIII} shows that these eigenvectors must
be related by $\ket{\epsilon_\mmin} = U \otimes \identity \ket{\epsilon_\mmax}$ for some local unitary $U$ acting on the probe.

Another nice property of the state transformation (\ref{eq-ouput_states_geodesic}) is related to the
optimal measurement.
Recall that a measurement is given by a POVM $\{ M_j \}$, that is, a set of non-negative operators $M_j \geq 0$ on $\Hh$ such that $\sum_j M_j = \identity$.
The outcome probabilities when the system is in state $\rho_x$ are given by $p_{j|x}= \tr M_j \rho_x$.
An optimal measurement is a POVM $\{ M_j^\opt\}$ for which the CFI
(\ref{eq-classical_Fisher_info}) coincides with the QFI.
Such a measurement leads to the smallest error $\Delta x = (\Delta x)_{\rm QCRB}$ in  (\ref{eq-Q_Cramer_Rao_bound}).
In general, $\{ M_j^\opt \}$ depends on the parameter $x$~\cite{Caves96}.
Since  $x$ is unknown {\it a priori}, it is then  in practice impossible to implement directly an optimal measurement strategy.
A notable exception
is  a closed system  undergoing a unitary transformation  with an input state given by the superposition 
(\ref{eq-Schrodinger_cat_states_intro}) minimizing the error.  Then the optimal measurement is independent of $x$.
Theorem~\ref{prop-geodesics_and_parameter_estimationII} in 
Sec.~\ref{sec-Q_metrology} shows that the  geodesic transformation (\ref{eq-ouput_states_geodesic}) enjoys the same property.
More precisely, one has: (iii)~there is an optimal POVM  $\{ M_j^\opt\}$ maximizing the CFI which is independent of $x$ and is
given by the von Neumann measurement with projectors onto $\ker \rho_j$,
where $\rho_j$ are the states at which the geodesic $\gamma_\geo$ intersects
the boundary $\partial \states$ of quantum states. 
This property gives a practical way to determine an optimal measurement in numerical simulations or experiments:
first determine the smallest eigenvalue $p_n (x)= \min_{\| \psi\| = 1} \bra{\psi} \rho_x \ket{\psi}$ of  $\rho_x$ for different values of $x$
 until finding a parameter value $x_j$ such that $p_n(x_j)\simeq 0$  (if $x$ is random and can not be tuned, just repeat the transformation many times); then
 determine all  eigenvalues very close to zero  of $\rho_j=\rho_{x_j}$ and the associated eigenvectors (this can be done by minimization of
 $ \bra{\psi} \rho_j \ket{\psi}$ in orthogonal subspaces, or by using quantum state tomography);
repeat this procedure for different values of $x_j$ until
obtaining an orthonormal basis of $\Hh$ formed by eigenvectors of the states $\rho_j$ with vanishing eigenvalues.
Note that Theorem~\ref{prop-intersection_Bures_geo_boundary} in Appendix~\ref{sec-intersections_boundary} shows that these eigenvectors indeed form an orthonormal basis. 
Then such a  basis is an  optimal measurement basis for the estimation of $x$.

Let us comment that the geometric approach described in the following sections is not restricted to geodesic evolutions and
provides a new method for studying parameter estimation in open quantum systems.
Actually, the above arguments show that, {  for a fixed QFI of the probe and environment (i.e., a fixed value of the  square fluctuation $\langle ( \Delta H)^2\rangle_{\Psi_\inp}$),
  one can enhance the QFI of the probe and thus the precision of the estimation by
  reducing the vertical component of the tangent vector $\ket{\dPsi_x}= -\I \Delta H \ket{\Psi_x}$, see (\ref{eq-formula_QFI_Pythagore}). Given a Hamiltonian $H$ of the probe and environment,
  the direction of  $\ket{\dPsi_x}$ could be changed using 
  control techniques on the probe or its environment, to make $\ket{\dPsi_x}$ more horizontal while keeping its norm fixed.
  This}
may be useful for designing new engineering reservoir techniques in order to increase precision
in quantum metrology in the presence of losses and dephasing.

\section{Mathematical preliminaries} \label{sec-preliminaries}

In this section we describe the geometrical properties of the manifold of mixed states of a quantum system equipped with the Bures distance
and introduce the notion of Riemannian submersions. 

 \subsection{Riemannian geometry for quantum states and Bures distance.}

Let us first recall some basic notions of Riemannian geometry.
 A metric on a smooth manifold $\Ee$ is a smooth map $g$ 
associating to each point $\rho \in \Ee$ a scalar product $g_\rho$ on the
tangent space $T_\rho  \Ee$ at $\rho$. 
A curve $\gamma$ on $\Ee$ joining two points $\rho$ and $\sigma$ is parametrized by 
a piecewise $C^1$ map
$\gamma : t \in [ t_0,t_1]  \mapsto \gamma(t) \in \Ee$
such that $\gamma (t_0)= \rho$ and $\gamma (t_1) = \sigma$. 
Its length $\ell (\gamma) $ is 
\begin{equation}
\ell (\gamma) = \int_\gamma \D s = \int_{t_0}^{t_1} \D t \, \sqrt{ g_{\gamma(t)} ( \dot{\gamma}(t), \dot{\gamma}(t) )}\;,
\end{equation} 
where $\dot{\gamma}(t)$ stands for the time derivative $\D \gamma/\D t$.
A Riemannian distance $d$ on $\Ee$  can be associated to any metric $g$, defined as the infimum
$d (\rho, \sigma) = \inf_\gamma \ell (\gamma)$ of the lengths of  all curves $\gamma$ joining $\rho$ and $\sigma$.
Such a distance is called the geodesic distance on $(\Ee,g)$.
Curves $\gamma_\geo$ with constant velocity minimizing  the length {\it locally} are called geodesics.
More precisely, $\gamma_\geo : [t_0,t_1] \rightarrow \Ee  $ is a geodesic if 
(i)~$g_{\gamma_\geo(t)} ( \dot{\gamma_\geo }(t),\dot{\gamma_\geo} (t) ) = {\rm const.}$
and (ii)~$\forall\;\,t \in (t_0,t_1)$, $\exists\;\delta >0$ such that
$\ell ( \gamma_\geo |_{[ t,t+\delta]} ) = d( \gamma_\geo (t),\gamma_\geo (t+\delta))$.
In particular, if there is a geodesic
$\gamma_\geo$ with length $\ell (\gamma_\geo )=d(\rho,\sigma)$ minimizing the length globally, one says that $\gamma_\geo$ is
the shortest geodesic joining $\rho$ and $\sigma$.

Conversely, 
one can associate  to a distance $d$ on $\Ee$ a  metric $g$ if $d$ 
satisfies the following condition (we ignore here regularity assumptions):  
for any $\rho\in \Ee$ and $\dot{\rho} \in T_\rho \Ee$, the square distance between $\rho$ and $\rho + t \dot{\rho}$  behaves as $t \to 0$ as 
\begin{equation} \label{eq-definition_metric}
\D s^2 =  d ( \rho , \rho + t \dot{\rho} )^2 =  g_\rho ( \dot{\rho} , \dot{\rho} ) t^2 + \Oo ( t^3) \;.
\end{equation}

 In quantum mechanics, states are represented by non-negative operators $\rho$
 with unit trace from the Hilbert space $\Hh$ of the system into itself.
 We assume hereafter that $\Hh$ has finite dimension $n = \dim ( \Hh) < \infty$ and
 denote by $\states  $ the set of all quantum states of a given system.
The arccos Bures distance between two states $\rho$ and $\sigma$ is defined by~\cite{Nielsen}
\begin{equation} \label{eq-Bures_arccos_dist}
 d_{\rm B} (\rho ,\sigma ) = \arccos \sqrt{F(\rho,\sigma)} \;,
\end{equation}
where 
\begin{equation} \label{eq-fidelity}
  F(\rho ,\sigma) =  \Bigl( \tr |  \sqrt{\sigma} \sqrt{\rho} |  \Bigr)^2 =
   \Bigl( \tr ( \sqrt{\rho} \, \sigma \sqrt{\rho})^\onehalf  \Bigr)^2
\end{equation}
is the fidelity.
The set of all invertible states, 
\begin{equation}
\states^\inv   = \big\{ \rho : \Hh \to \Hh, \; ;\; \rho > 0 , \tr \rho =1 \big\}\;,
\end{equation}
equipped with the distance $d_{\rm B}$ forms a smooth open Riemannian manifold.
Its boundary $\partial \states  $ consists of density matrices $\rho$ having at least one vanishing
eigenvalue; for instance, pure states $\rho_\psi = \ketbra{\psi}{\psi}$ are on the boundary.

The tangent space at $\rho \in \states^\inv$ can be identified with the (real)  vector space
of self-adjoint traceless operators on $\Hh$,
\begin{equation}
  T_\rho \,\states
  =  \big\{ \drho : \Hh \to \Hh\;,\; \drho^\dagger = \drho \;,\; \tr  \drho  = 0 \big\}\;. 
\end{equation}
The metric $g_{\rm B}$ associated to the distance $ {d}_{\rm B}$ is given explicitly by~\cite{Hubner92,myreview}
\begin{equation} \label{eq-Bures_metric}
   ( g_{\rm B})_\rho ( \drho , \dsigma ) = \onehalf \re \!\sum_{k,l=1}^n \frac{\overline{\bra{k} \drho \ket{l}} \bra{k} \dsigma \ket{l}}{p_k + p_l}
\;,\; \drho , \dsigma \in T_\rho \,\states   \;,
\end{equation}
where $\{ \ket{k} \}_{k=1}^n$ is an orthonormal basis of eigenvectors of $\rho$ with eigenvalues $p_k$.

 \subsection{Purifications and smooth submersions}

Mixed quantum states of a given system can be described by introducing an auxiliary system  $\AAA$, called the 
ancilla, and viewing the system state $\rho$ as the reduced state of the system\,+\,ancilla.  
The dimension of the ancilla Hilbert space $\HhA$ is assumed to fulfill $n_\AAA \geq n$. We denote by
$\Kk = \Hh \otimes \HhA$ the Hilbert space of the composite system.
A purification of $\rho$  on $\Kk$
is a pure state $\ketbra{\Psi}{\Psi}$ such that $\ket{\Psi} \in \Kk$ and  $\rho =  \pi  ( \ket{\Psi})$, with
\begin{equation} \label{eq-quotient_map}
\pi  ( \ket{\Psi}) = \tr_\AAA  \ketbra{\Psi}{\Psi} \;,
\end{equation}
where $\tr_\AAA$ stands for the partial trace over the ancilla space $\Hh_\AAA$.
For our purpose, it is convenient to consider purifications as normalized vectors in $\ket{\Psi}\in \Kk$,
instead of pure states 
  (recall that a pure state is a normalized vector
  modulo a phase factor and can be represented as a rank-one projector $\ketbra{\Psi}{\Psi}$ in the projective space $P \Kk$).
The condition $\rho>0$ is satisfied if and only if 
$\ket{\Psi}$ has Schmidt decomposition 
\begin{equation} \label{eq-Schmidt_decomp}
  \ket{\Psi} = \sum_{k=1}^n \sqrt{p_k} \ket{k} \ket{\alpha_k}
\end{equation}
 with $n$ positive Schmidt coefficients $\sqrt{p_k} >0$, $k=1,\ldots,n$. Here,
{  $\{ \ket{k}\}_{k=1}^n$ is an orthonormal eigenbasis of $\rho$ and} $\{ \ket{\alpha_k}\}_{k=1}^{n_\AAA}$ is an arbitrary \ONB of $\Hh_\AAA$.
 The set of purifications of invertible states is thus the subset {   of the unit sphere in $\Kk$ given by}
\begin{eqnarray}
  \nn
  \Ss_{\Kk}^\inv & = & \big\{ \ket{\Psi} \in \Kk\; ; \; \| \Psi\| = 1 \; ,
  \;\ket{\Psi} \text{ has $n$  positive}
  \\
  & & \qquad \text{Schmidt coefficients} \big\}\;.
\end{eqnarray}  
The tangent space of $\Ss_\Kk^\inv$ at $\ket{\Psi} \in \Ss_\Kk^\inv$ is
\begin{equation} \label{eq-tangent_spaces_hypersphere}
T_{\ket{\Psi}}  \Ss_{\Kk} = \big\{ \ket{\dPsi} \in \Kk \;;\; \re \braket{\Psi}{\dPsi } = 0 \big\}\;.  
\end{equation}
A natural metric on $\Ss_{\Kk}^\inv$ is $g_\Ss ( \ket{\dPsi},\ket{\dPhi}) = \re \braket{\dPsi}{\dPhi}$.
Note that the scalar product is independent   of $\ket{\Psi}$.
This metric is that induced by  the euclidean metric on $\Kk$.
The Riemannian manifold $(\Ss_{\Kk}^\inv, g_\Ss)$ is isometric under the action 
$\ket{\Psi} \mapsto \identity \otimes U_\AAA \ket{\Psi}$ of the unitary group $U(n_\AAA )$  acting on the ancilla, that is,
\begin{equation}
  g_\Ss ( \identity \otimes U_\AAA \ket{\dot{\Psi}},\identity \otimes U_\AAA \ket{\dot{\Phi}}) = g_\Ss ( \ket{\dot{\Psi}},\ket{\dot{\Phi}})
\end{equation}
for any unitary $U_\AAA$ on $\HhA$.

We now argue that  $\states^\inv  $ can be viewed as the quotient of $\Ss_{\Kk}^\inv$ 
by the unitary group $U(n_\AAA)$. 
The map (\ref{eq-quotient_map}) defines a projection from $\Ss_{\Kk}^\inv$  onto $\states^\inv$.
The set  $\pi^{-1} (\rho)$ of all purifications of $\rho$ coincides with the orbit of $\rho$ under the group action,  
\begin{equation} \label{eq-orbits_of_rho}
  \pi^{-1} (\rho) = \big\{ \ket{\Psi}  = \identity \otimes U_\AAA \ket{\Psi_0}\, ;\, U_\AAA \text{ unitary on } \HhA \big\}\;,
\end{equation}
where $\ket{\Psi_0}\in \Ss_\Kk^\inv$ is some fixed purification. This can be proven by noting that 
any purification of $\rho$ has the form (\ref{eq-Schmidt_decomp}); thus it is obtained by applying
a local unitary $\identity \otimes U_\AAA$  to $\ket{\Psi_0}$.
Therefore, $ \states^\inv  $ can be identified with the quotient manifold $\Ss^\inv_{\Kk} / U(n_\AAA)$ and $\pi$ is the quotient map.

An important fact about $\pi$ is that its differentials
$\D \pi |_{\ket{\Psi}}: T_{\ket{\Psi}}  \Ss_{\Kk} \to  T_\rho \,\states  $ are surjective for any $\ket{\Psi} \in \Ss_\Kk^\inv$.
Such quotient maps with surjective differentials are called  smooth submersions.
The differentials of the map (\ref{eq-quotient_map}) are given by
\begin{equation} \label{eq-differential_pi}
  \D \pi |_{\ket{\Psi}} ( \ket{\dPsi} ) = \tr_\AAA ( \ketbra{\Psi}{\dPsi} + \ketbra{\dPsi}{\Psi} ) \;.
\end{equation}
To prove that $\pi : \Ss_\Kk^\inv \to \states^\inv$ is a smooth submersion, let us first note that for 
any fixed  $\drho \in T_\rho \states$, if $\ket{\dPsi} \in \Kk$ satisfies
\begin{equation} \label{eq-proof_D_pi_is_surjective}
  \D \pi |_{\ket{\Psi}} ( \ket{\dPsi} ) = \drho
\end{equation}
then $\ket{\dPsi} \in T_{\ket{\Psi}}  \Ss_{\Kk}$. In fact, by taking the trace of the right hand sides of (\ref{eq-differential_pi}) and
(\ref{eq-proof_D_pi_is_surjective}) one gets
$ 2 \re \braket{\Psi}{\dPsi} = \tr \drho =0$. Setting  $\rho = \pi ( \ket{\Psi})$, it is easy to check that
$\ket{\dPsi}= \onehalf \drho \rho^{-1} \otimes \identity \ket{\Psi}$ is a solution of (\ref{eq-proof_D_pi_is_surjective}).
Hence $\D \pi |_{\ket{\Psi}}$ is surjective.
Let us point out that this is not true if $\pi$ is defined on the whole unit sphere $\Ss_\Kk$ of $\Kk$, instead of $\Ss_\Kk^\inv$,
\ie, if one adds to $\states^\inv$ its boundary $\partial \states$.

\subsection{Riemannian submersions} \label{sec-Riemannian_submersions}

Our results in this paper rely on the notion of  Riemannian submersion.
In this subsection we review the properties of such submersions (see e.g.~\cite{Gallot_book} for  more details) and
show that the  partial trace (\ref{eq-quotient_map}) is an instance of Riemannian submersion $\Ss_\Kk^\inv \to \states^\inv$.

 Let  $\pi : \Xx \to \Ee$ be a smooth submersion, where $\Xx$ is a Riemannian manifold with metric $g_\Xx$. 
  The tangent space  $T_\psi \Xx$ at $\psi \in \Xx$ can be decomposed into a direct sum of two orthogonal subspaces
$\Vf_\psi = \ker ( \D \pi|_\psi )$ and $\Hf_\psi = \Vf_\psi^\bot$,  called respectively
  the vertical and horizontal subspaces (orthogonality is for the scalar product $(g_\Xx)_\psi$).
It can be shown that
there exists a unique Riemannian metric $g_\Ee$ on the quotient manifold $\Ee$
such that for all $\psi \in \Xx$, 
the restriction of $\D \pi |_{\psi}$ to the horizontal subspace $\mathfrak{h}_{\psi}$ is an isometry
from $(\Hf_\psi, (g_\Xx)_\psi)$ to $( T_{\rho} \Ee , (g_\Ee)_\rho)$, with $\rho = \pi ( \psi)$.
One then says that  $\pi: (\Xx,g_\Xx ) \rightarrow ( \Ee, g_\Ee )$ is a Riemannian submersion.  
It is not difficult to prove that $g_\Ee$ is associated to the distance $ d_\Ee $ on $\Ee$ defined by
\begin{equation} \label{eq-def_Riemannian_sub_dist}
  d_{\Ee} ( \rho , \sigma ) = \inf_{ \phi \in \pi^{-1} (\sigma )} d_\Xx ( \psi_1,\phi) \;,
\end{equation}
%
where $d_\Xx$ is a distance having the metric $g_\Xx$ and $\psi_1$ is an arbitrary (fixed) point on the orbit of $\rho$.

A nice property of Riemannian submersions is that the geodesics on the quotient space $\Ee$
can be obtained by projecting certain geodesics on  $\Xx$. More precisely, for any geodesic
$\Gamma: [t_0,t_1] \to \Xx$ on $( \Xx , g_\Xx)$ such that
$\dot{\Gamma} (0) \in \Hf_{\Gamma (0)}$, one  has~\cite{Gallot_book}:
\begin{itemize}
\item[(i)] $\dot{\Gamma}(t) \in {\Hf}_{\Gamma (t)}$ for any $t \in [t_0,t_1]$;
\item[(ii)] $\gamma= \pi \circ \Gamma$ is a geodesic on $(\Ee,g_\Ee)$. 
\end{itemize}
Conversely, any geodesic $\gamma$ on  $(\Ee,g_\Ee)$ with $\gamma (0)=\rho$ can be lifted locally to a geodesic 
$\Gamma$ on $( \Xx , g_\Xx)$ with horizontal tangent vectors such that $\Gamma (0) = \psi$, for any  $\psi \in \pi^{-1} ( \rho )$.
This property is illustrated in Fig.~\ref{fig-geodesics_Riemannian_submersion}.
We will call  horizontal geodesics the geodesics $\Gamma$ on $\Xx$ such that $\dot{\Gamma} (0) \in \Hf_{\Gamma (0)}$.

Let us apply this formalism to the smooth submersion $\pi$ given by (\ref{eq-quotient_map}).  
A natural distance on $\Ss_\Kk^\inv$ having the euclidean metric $g_\Ss$ is the norm distance $d_\Ss( \ket{\Psi},\ket{\Phi}) = \| \ket{\Psi} - \ket{\Phi}\|$.
The metric on $\states^\inv$ making $\pi$ a Riemannian submersion
turns out to be the Bures metric $g_B$.
This can be seen by invoking Uhlmann's theorem, which states that~\cite{Nielsen,Uhlmann76} 
\begin{equation} \label{eq-Uhmann_theo}
  d_{\rm B} ( \rho,\sigma) = \min_{\ket{\Psi} \in \pi^{-1}(\rho), \ket{\Phi} \in \pi^{-1} (\sigma)} \arccos \big| \braket{\Psi}{\Phi} \big|\;,
\end{equation}
where  in the \RHS  the arccos distance between pure states is minimized. Equivalently, the Bures
distance ${d}_{\rm{Bures}} (\rho,\sigma)= 2 \sin ( d_{\rm B}(\rho,\sigma)/2))$ is given by~\cite{Uhlmann86,myreview}
\begin{eqnarray} \label{eq-def_Bures_dist}
  \nn
& &      {d}_{\rm{Bures}} (\rho,\sigma)   =  \bigl( 2 - 2 \sqrt{F(\rho,\sigma )} \bigr)^{\frac{1}{2}}
     \\
& & \hspace*{5mm} =  
  \min_{\ket{\Psi} \in \pi^{-1}(\rho), \ket{\Phi} \in \pi^{-1} (\sigma)} \big\| \ket{\Psi} - \ket{\Phi} \big\|\;.
\end{eqnarray}
Note that the two distances $d_{\rm B}$ and $d_{\rm{Bures}}$ have the same metric $g_{\rm B}$, given by (\ref{eq-Bures_metric}).
Eqs. (\ref{eq-Uhmann_theo}) and (\ref{eq-def_Bures_dist}) tell us that the arccos and Bures distances between $\rho$ and $\sigma$ are 
the minimal distances
between the orbits of $\rho$ and $\sigma$.
By  (\ref{eq-orbits_of_rho}) and the unitary invariance of the scalar product in $\Kk$, the minima in these equations
can be carried out over all $\ket{\Phi} \in \pi^{-1}(\sigma)$ for some fixed $\ket{\Psi_1} \in \pi^{-1}(\rho)$.
Thus $d_{\rm{Bures}}$ has the form (\ref{eq-def_Riemannian_sub_dist}) with $d_\Xx=d_\Ss$.

\subsection{Vertical and horizontal subspaces}

Let us determine the vertical and horizontal subspaces in the case of the purification manifold  $\Xx=\Ss_\Kk^\inv$ and quotient map (\ref{eq-quotient_map}).
To determine
$\Vf_{\ket{\Psi}} = \ker ( \D \pi|_{\ket{\Psi}} )$, we observe that
curves contained in the orbit $\pi^{-1}(\rho)$ have by definition vertical tangent vectors. Thus, denoting by
$ T_{\ket{\Psi}} \pi^{-1} ( \rho)$ the tangent space of this orbit at $\ket{\Psi}$, it holds
$T_{\ket{\Psi}} \pi^{-1} ( \rho) \subset \Vf_{\ket{\Psi}}$. To show that the inclusion is an equality, we now prove that the two subspaces have the same dimension.
By the rank theorem and the surjectivity of $\D \pi|_{\ket{\Psi}}$, one has
\begin{eqnarray}
  \nn
& & \dim_\real ( \Vf_{\ket{\Psi}} )  =  \dim_\real ( T_{\ket{\Psi}} \Ss_\Kk) - \dim_\real ( T_\rho\, \states )
\\
& & \hspace*{2mm} =  (2 n n_\AAA - 1) - (n^2-1) = n ( 2 n_\AAA - n)\;.
\end{eqnarray}
On the other hand, by (\ref{eq-Schmidt_decomp}) and $\rho>0$, purifications of $\rho$ are in one-to-one correspondence with families $\{ \ket{\alpha_k}\}_{k=1}^n$ of $n$ orthonormal vectors in $\Hh_\AAA$. 
It is easy to show that these families form a manifold of real dimension $n(2 n_\AAA-n)$. Hence
$\Vf_{\ket{\Psi}} = T_{\ket{\Psi}} \pi^{-1} ( \rho)$. Thus, by (\ref{eq-orbits_of_rho}) one has
%
\begin{equation} \label{eq-vertical_subspace}
  \Vf_{\ket{\Psi}}  = \big\{ \identity \otimes K_\AAA \ket{\Psi} \; ;\; K_\AAA \text{ skew Hermitian} \big\}\;.
\end{equation}  
We point out that (\ref{eq-vertical_subspace}) is incorrect for non-invertible states $\rho\in \partial \states$. In fact, if $r=\rank (\rho) < n$ then $\dim_\real ( T_{\ket{\Psi}} \pi^{-1} ( \rho)) = r (2n_\AAA -r)$ is strictly smaller than $\dim_\real ( \Vf_{\ket{\Psi}}) = 2 n (n_\AAA - r) + r^2$.

In order to obtain the horizontal subspace $\Hf_{\ket{\Psi}} = \Vf_{\ket{\Psi}}^\bot$, we use
the Schmidt decomposition (\ref{eq-Schmidt_decomp}) and expand an arbitrary horizontal tangent vector
$\ket{\dPsi^\ho} = \sum_{k,l} c_{kl} \ket{k} \ket{\alpha_l} $ in the product basis $\{ \ket{k} \ket{\alpha_l} \}_{k,l=1}^{n,n_\AAA}$.
Since $\ket{\dPsi^\ho}$ is orthogonal to $\Vf_{\ket{\Psi}}$, one finds
\begin{eqnarray}
  0  & = & \re \bra{\dPsi^\ho} \identity \otimes K_\AAA \ket{\Psi}
  \\ \nn
  &  = &
  \sum_{k,l=1}^{n,n_\AAA}
  \big( \overline{c}_{kl} \sqrt{p_k} \bra{\alpha_l} K_\AAA \ket{\alpha_k}
  - c_{kl}\sqrt{p_k} \bra{\alpha_k} K_\AAA \ket{\alpha_l}  \big) 
\end{eqnarray}
for any skew Hermitian ancilla operator $K_\AAA$.
Choosing $K_\AAA = \I^{\,\nu} \ketbra{\alpha_l}{\alpha_k} - (-\I)^\nu \ketbra{\alpha_k}{\alpha_l}$ with $\nu=0$ or $1$, this gives
\begin{equation} \label{eq-proof_horizontal_space}
  c_{kl} \sqrt{p_k}
  = \begin{cases}
    \overline{c}_{lk} \sqrt{p_l} & \text{ if $1 \leq k,l \leq  n$}\\
    0                           & \text{ if $n < l \leq  n_\AAA$.}
  \end{cases} 
\end{equation}
Let us set $  L_\SSS =\sum_{k,l=1}^n \ell_{kl} \ketbra{k}{l}$ with $\ell_{kl} = c_{kl} p_l^{-\onehalf}$.
It follows from (\ref{eq-proof_horizontal_space}) that $  L_\SSS = L^\dagger_\SSS$. Furthermore,
\begin{equation}
  \ket{\dPsi^\ho} = \sum_{k,l=1}^n \ell_{kl} \sqrt{p_l} \ket{k} \ket{\alpha_l} 
   =  {  L_\SSS} \otimes \identity_\AAA \ket{\Psi}\;.
\end{equation}
Reciprocally, if $  \ket{\dPsi^\ho} = L_\SSS \otimes \identity_\AAA \ket{\Psi}$ with $  L_\SSS$ self-adjoint then 
$\re \bra{\dPsi^\ho} \identity \otimes K_\AAA \ket{\Psi}=0$. The condition $\re \braket{\dPsi^\ho}{\Psi}=0$
coming from the requirement that $\ket{\dPsi^\ho}$ is in the tangent space (\ref{eq-tangent_spaces_hypersphere}) yields the additional constraint
$  \langle L_\SSS \otimes \identity_\AAA \rangle_\Psi=0$.
Thus
\begin{equation} \label{eq-horizontal_space}
  \Hf_{\ket{\Psi}} \! = \! \big\{ {  L_\SSS} \otimes \identity_\AAA \ket{\Psi}  ;  L_\SSS \text{ self-adjoint,} \langle L_\SSS \otimes \identity_\AAA \rangle_\Psi\! =\! 0 \big\} .
\end{equation}  

In conclusion, we have shown that  the partial trace map (\ref{eq-quotient_map}) defines a Riemannian submersion from $\Ss_\Kk^\inv$ equipped with
the metric $g_\Ss$
to the manifold $\states^\inv$ equipped with the Bures metric $g_{\rm B}$.
This means that  $\D \pi$ is an isometry from $(\Hf_{\ket{\Psi}}, g_\Ss)$ to $(T_\rho\, \states , g_{\rm B})$, namely,
\begin{equation} \label{eq-metric_g_Ee_Riemannian_sub}
  (g_{\rm B})_\rho \big( \D \pi |_{\ket{\Psi}} ( \ket{\dPsi^\ho} ), \D \pi |_{\ket{\Psi}} (\ket{\dPhi^\ho}) \big)
  = \re\braket{\dPsi^\ho}{\dPhi^\ho}
\end{equation}
for any purification $\ket{\Psi}$ of $\rho$ and any horizontal tangent vectors $\ket{\dPsi^\ho},\ket{\dPhi^\ho}\in \Hf_{\ket{\Psi}}$.

\section{Bures geodesics} \label{sec_study_of_Bures_geodesics}

\subsection{Determination of the geodesics} \label{sec-Bures_geodesics}

We
determine in this subsection the Bures geodesics 
by  applying the mathematical framework of the preceding section
(see~\cite{Bhatia_17} for a similar approach in  the case of the space of
positive  definite matrices, \ie, unnormalized quantum states).

The shortest geodesic arc joining two vectors  $\ket{\Psi}$ and $\ket{\Phi}$ on the unit sphere $\Ss_\Kk^\inv$ equipped with the metric
$g_\Ss$ is the arc of great circle
\begin{equation} \label{eq-geodesics_sphere}
  \ket{\Psi_\geo (\tau)} = \cos \tau\, \ket{\Psi} + \sin \tau\,\ket{\dPsi} \quad  , \quad 0 \leq \tau \leq \theta
  \; , 
\end{equation}
where $\ket{\dPsi} \in  T_{\ket{\Psi}} \Ss_\Kk^\inv$ and
$\theta = \arccos ( \re \braket{\Psi}{\Phi})$ is the angle between $\ket{\Psi}$ and $\ket{\Phi}$.
We may assume without loss of generality that $\theta \not= 0,\pi$, since otherwise  $\ket{\Psi}$ and  $\ket{\Phi}=\pm \ket{\Psi}$ project
out to the same state $\rho=\sigma$.
The longest arc of great circle joining  $\ket{\Psi}$ and $\ket{\Phi}$ needs not be considered here, because it is the extension of the
shortest geodesic joining $\ket{\Psi}$ and $- \ket{\Phi}$ and the latter vector belongs to the same orbit as  $\ket{\Phi}$.
The geodesic tangent vector at $\tau=0$ is given by
%
\begin{equation} \label{eq-relation_between_dPsi_and_Phi}
  \ket{\dPsi_\geo (0)}= \ket{\dPsi} = \frac{1}{\sin\theta} \big( \ket{\Phi} - \cos \theta \ket{\Psi} \big)\;.
\end{equation}
It is easy to check on this formula that $\| \dPsi\|=1$, \ie, the geodesic (\ref{eq-geodesics_sphere}) has unit velocity.

According to the properties of Riemannian submersions (Sec.~\ref{sec-Riemannian_submersions}), the Bures geodesic arcs joining the invertible states
$\rho= \pi ( \ket{\Psi})$ and $\sigma = \pi ( \ket{\Phi})$ are  obtained by projecting
the arcs of great circle (\ref{eq-geodesics_sphere}) having horizontal tangent vectors $\ket{\dPsi} \in \Hf_{\Psi}$.
Let us consider the purification of $\rho$ given by
\begin{equation} \label{eq-Schmidt_decomp_Psi}
  \ket{\Psi} =  \sqrt{\rho} \otimes \identity_\AAA \sum_{k=1}^n \ket{k} \ket{\alpha_k}\;,
\end{equation}
where we have used the Schmidt decomposition (\ref{eq-Schmidt_decomp}). The last sum is an (unnormalized)  
maximally entangled state of the system and ancilla. 
Similarly, by (\ref{eq-orbits_of_rho}) any purification of $\sigma$
has the form
  \begin{equation} \label{eq-purification_sigma}  
    \ket{\Phi} =   \sqrt{\sigma} \, \rho^{-1/2} \otimes U_\AAA \ket{\Psi}\;, 
  \end{equation}
  where
   $U_\AAA$ is a unitary on  $\Hh_\AAA$.

It is convenient at this point to introduce the polar decomposition
\begin{equation} \label{eq-def_U_sigma_rho}
  \sqrt{\sigma}\sqrt{\rho} = U_{\sigma \rho} \Lambda_{\sigma \rho}\;\; , \;\;
  \Lambda_{\sigma \rho} =  |\sqrt{\sigma} \sqrt{\rho} | > 0\;,
\end{equation}  
where $U_{\sigma\rho}$ is unitary.

We have to determine the purifications $\ket{\Phi}$ of $\sigma$ such that the horizontality condition
$\ket{\dPsi} \in \Hf_{\ket{\Psi}}$ holds. We will show that:

\begin{lemma} \label{lemma-endpoint_horizontal_lift}
The purifications of $\sigma$ such that $\ket{\dPsi} \in \Hf_{\ket{\Psi}}$ are given by
\begin{equation} \label{eq-horizontality_cond_4}
  \ket{\Phi_V}  =  \MgeoV  \otimes \identity_\AAA \ket{\Psi}
\end{equation}
with
\begin{equation} \label{eq-def_M^(V)}
  \MgeoV  =  \sqrt{\sigma}\, U_{\sigma \rho} V \rho^{-1/2}
  =  \rho^{-1/2} \Lambda_{\sigma \rho} V \rho^{-1/2} \;,
\end{equation}
where $V$ is an arbitrary unitary and self-adjoint operator commuting with $\Lambda_{\sigma\rho}$. 
%
%
\end{lemma}

\Proof
In view of (\ref{eq-horizontal_space}),
(\ref{eq-relation_between_dPsi_and_Phi}), (\ref{eq-Schmidt_decomp_Psi}), and (\ref{eq-purification_sigma}), the horizontality  condition can be written as
\begin{eqnarray} \label{eq-horizontality_cond_2}
  \nn
& &   \frac{1}{\sin\theta} \bigg( \sqrt{\sigma} \, {  \identity }  \otimes  U_\AAA - \cos \theta \sqrt{\rho} \otimes \identity_\AAA \bigg) \sum_{k=1}^n \ket{k} \ket{\alpha_k}
  \\
  & & \qquad 
  = {  L_\SSS} \sqrt{\rho} \otimes \identity_\AAA \sum_{k=1}^n \ket{k} \ket{\alpha_k}
 \end{eqnarray} 
for some self-adjoint operator $  L_\SSS$ such that $\langle L_\SSS \otimes \identity_\AAA \rangle_\Psi = 0$.
In particular, one has $\bra{\alpha_l} U_\AAA \ket{\alpha_k} = 0$ for $n<l\leq n_\AAA$ and $1 \leq k \leq n$. We now use the identity
\begin{equation} \label{eq-action_local_op_on_max_entangled_state}
  \identity \otimes U_\AAA \sum_{k=1}^n \ket{k} \ket{\alpha_k} =  U_\AAA^{\rm T} \otimes \identity_\AAA \sum_{k=1}^n \ket{k} \ket{\alpha_k}\;,
\end{equation}
where $U_\AAA^{\rm T} = \sum_{k,l=1}^n \bra{\alpha_l} U_\AAA \ket{\alpha_k} \ketbra{k}{l}$ (when $n_\AAA > n$, (\ref{eq-action_local_op_on_max_entangled_state})~is true provided that  ${\mathrm{span}} \{ \ket{\alpha_1},\ldots ,\ket{\alpha_n}\}$ is invariant under $U_\AAA$, which is indeed the case here).
Observe that {    $U= U_\AAA^{\rm T}$ is unitary.} Multiplying both members of (\ref{eq-horizontality_cond_2}) by $\sqrt{\rho}$,
one deduces that this equation is equivalent to
\begin{equation}  \label{eq-horizontality_cond_3}
    \frac{1}{\sin\theta} \bigg( \sqrt{\rho} \sqrt{\sigma} \, U - \cos \theta\,\, \rho \bigg) = \sqrt{\rho}\, {  L_\SSS} \sqrt{\rho}\;.
\end{equation}
Therefore, (\ref{eq-horizontality_cond_2}) is equivalent to $\sqrt{\rho} \sqrt{\sigma}\, U$ being self-adjoint for some unitary $U$ on $\Hh$.
 (As we shall see below, this self-adjointness implies $\braket{\Psi}{\Phi} \in \real$, thus 
  $\braket{\Psi}{\dPsi} = \re \braket{\Psi}{\dPsi}=0$ by (\ref{eq-relation_between_dPsi_and_Phi}) and  $\ket{\dPsi} \in  T_{\ket{\Psi}} \Ss_\Kk^\inv$.
  Hence $\langle L_\SSS \otimes \identity_\AAA\rangle_\Psi= 0$ does not yield an additional condition.)

Let $V$ be the unitary operator given by $V=  U_{\sigma \rho}^\dagger U$.
Then by (\ref{eq-def_U_sigma_rho}), $\sqrt{\rho} \sqrt{\sigma} \,U$ is self-adjoint if and only if 
$\Lambda_{\sigma\rho}  V= V^\dagger  \Lambda_{\sigma \rho}$. This implies $V^\dagger \Lambda^2_{\sigma \rho} V = \Lambda^2_{\sigma \rho}$, \ie,
$V$ commutes with $\Lambda^2_{\sigma \rho}$
and thus with $\Lambda_{\sigma \rho}$. As a result,
  $V^\dagger \Lambda_{\sigma\rho} = V \Lambda_{\sigma\rho}$, which entails $V^\dagger= V$ (since $\Lambda_{\sigma\rho}>0$).
Thus the horizontality condition (\ref{eq-horizontality_cond_2}) holds if and only if $U = U_{\sigma \rho} V$
with $V$ unitary, self-adjoint, and $[V,\Lambda_{\sigma\rho}]=0$.
The corresponding purifications of $\sigma$ are given by (\ref{eq-horizontality_cond_4}).
\finpro

\vspace{2mm}

Note that $\braket{\Psi}{\Phi_V}$ is real.
By (\ref{eq-geodesics_sphere}) and (\ref{eq-relation_between_dPsi_and_Phi}), the horizontal geodesics on $\Ss_\Kk^\inv$ are given by
\begin{equation} \label{eq-horizontal_geodesics_sphere}
\ket{\Psi_{\geo,V} (\tau) } = \frac{1}{\sin \theta_V} \Big( \sin  (\theta_V - \tau) \ket{\Psi} + \, \sin \tau \ket{\Phi_V} \Big) 
\end{equation}
with $\cos \theta_V = \braket{\Psi}{\Phi_V} = \tr  \Lambda_{\sigma \rho} V$.

By using
the identity
\begin{equation} \label{eq-partial_trace_off_diag_term}
  \tr_\AAA \ketbra{\Phi_V}{\Psi} =  \sqrt{\sigma}\,U_{\sigma\rho} V \sqrt{\rho} =  \rho^{-1/2} \Lambda_{\sigma\rho} V  \sqrt{\rho}\;,
\end{equation}
we obtain the geodesics $\gamma_{\geo,V} (\tau) = \pi ( \ket{\Psi_{\geo,V} (\tau)})$ on $\states$, where $\pi$ is the
quotient map (\ref{eq-quotient_map}), 
\begin{eqnarray} \label{eq-Bures_geodesics_V}
  \nonumber
  && 
    \gamma_{\geo,V} ( \tau) =\frac{1}{\sin^2 \theta_V}
  \Big( \sin^2 ( \theta_V - \tau)\,  \rho  + \sin^2 ( \tau )\, \sigma +
  \\ 
  & & 
   +\sin ( \theta_V\! - \! \tau)  \sin \tau \Big(
   \rho^{-1/2} \Lambda_{\sigma\rho} V  \rho^{1/2} + {\mathrm{h.c.}} \Big)  \Big)
\end{eqnarray}
with $0 \leq \tau \leq \theta_V$.
Eq. (\ref{eq-Bures_geodesics_V})  generalizes formula (\ref{eq-shortest_Bures_geodesics}) of Sec.~\ref{main-results}. It coincides with this formula for $V = \identity$.
Thanks to  (\ref{eq-metric_g_Ee_Riemannian_sub}) and by the horizontality of $\ket{\dPsi_V}$, $\gamma_{\geo,V}$ has unit square velocity
$(g_{\rm B})_{\rho} (\dot{\gamma}_{\geo,V},\dot{\gamma}_{\geo,V}) =\| \dPsi_V\|^2=1$. Thus 
$\theta_V =\ell ( \gamma_{\geo,V} )$ is the geodesic length.

The geodesic with the smallest length is obtained for $V=\identity$.
In fact, for an arbitrary unitary and self-adjoint $V$ commuting with $\Lambda_{\sigma\rho}$, 
denoting by $\lambda_k >0$  and $v_k\in \{-1,1\}$ the eigenvalues of $\Lambda_{\rho\sigma}$ and $V$, one has

\begin{eqnarray} \label{eq-geodesic_length_V}
  & & 
  \cos \theta_V  =   \tr  \Lambda_{\sigma \rho} V  = \sum_{k=1}^n \lambda_k v_k
  \\ \nn
  & & \quad \leq  
  \sum_{k=1}^n \lambda_k = 
  \tr  \Lambda_{\sigma\rho} = \sqrt{F(\rho,\sigma)} = \cos \theta_1\;,
\end{eqnarray}
where we have set $\theta_1 = \theta_{V=\identity}$.
Similarly, the longest geodesic joining $\rho$ to $\sigma$ is obtained by choosing  $V=-\identity$ and has length
$\pi - \theta_1$.
In view of (\ref{eq-horizontality_cond_4}),
such a geodesic is the projection of the arc of great circle joining $\ket{\Psi}$ and the vector $\ket{\Phi_{-\identity}}=-\ket{\Phi_{\identity}}$
diametrically opposite to $\ket{\Phi_\identity}$ on the sphere $\Ss^\inv_\Kk$. The latter is obtained
by inverting time on the great circle through  
$\ket{\Psi}$ and $\ket{\Phi_\identity}$ 
 and replacing the arc length $\theta_1$ by $\pi - \theta_1$.
 Thus, by extending the shortest and longest geodesic arcs $\gamma_{\geo,\identity}$ and $\gamma_{\geo,-\identity}$ joining $\rho$ and $\sigma$
 to the interval $[0,\pi]$, one obtains the same closed curve albeit with opposite orientations.
More generally, the pair of geodesics $(\gamma_{\geo,V}, \gamma_{\geo,-V})$ enjoys the same property.

We have proven:

\begin{theorem} \label{prop-Bures_geodesics}
 The Bures geodesic arcs joining the two distinct invertible states $\rho$ and $\sigma \in \states^\inv$ 
  are given by
\begin{equation} \label{eq-Bures_geodesics}
  \gamma_{\geo,V} ( \tau) = X_{\rho\sigma,V} (\tau)\, \rho \, X_{\rho\sigma,V}(\tau)\; , \; 0 \leq \tau \leq \theta_V \;,
\end{equation}
where the geodesic length $\theta_V$ is  given by (\ref{eq-geodesic_length_V}) and $X_{\rho\sigma,V}(\tau)$ is the operator
defined by
\begin{equation} \label{eq-def_X(tau)}
X_{\rho\sigma,V} (\tau)
= \frac{1}{\sin \theta_V} \Big( \sin (\tau ) M_{\rho\sigma,V} + \sin ( \theta_V - \tau) \identity  \Big)
\end{equation}
with $\Mgeo$ given by (\ref{eq-def_M^(V)}). Here, 
$V$ is an arbitrary unitary self-adjoint operator commuting with 
 $\Lambda_{\sigma\rho}= |\sqrt{\sigma}\sqrt{\rho} |$.
Furthermore, the geodesic with the smallest length, denoted hereafter by  $ \gamma_\geo$,
 is obtained by choosing $V=\identity$ in (\ref{eq-def_M^(V)}), (\ref{eq-Bures_geodesics}) and (\ref{eq-def_X(tau)}) and has length
\begin{equation} \label{eq-shortest_geodesic_length}
   \ell ( \gamma_\geo ) = \theta_1 = {d}_{\rm B} ( \rho, \sigma) \; \in \; ( 0, \frac{\pi}{2}  ] \;.
\end{equation}
\end{theorem}

The explicit form (\ref{eq-Bures_geodesics}) of the Bures geodesics have been obtained  in Refs.~\cite{Barnum_thesis,Ericson05}
in the special case $V = \identity$. 
Our derivation shows that, in addition to this shortest geodesic, there are other geodesics having larger lengths $\theta_V$,
corresponding to $V \not= \identity$.
{  More precisely,} there are $2^n$ geodesic arcs joining two invertible states $\rho$ and $\sigma$ if
$\Lambda_{\sigma\rho}$ has a non-degenerate spectrum.
In contrast, if $\Lambda_{\sigma\rho}$ has a degenerate eigenvalue $\lambda_k$
there are infinitely many geodesics $\gamma_{\geo,V}$ joining $\rho$ and $\sigma$, in analogy with
what happens for diametrically opposite points on a sphere.
Actually, in the non--degenerate case  there are $2^n$ choices for $V$ because $V$ is diagonal in an
eigenbasis of $\Lambda_{\sigma\rho}$ (since $[\Lambda_{\sigma\rho},V]=0$) and thus is fully characterized by its eigenvalues  $v_k \in\{ 1,-1\}$. In the degenerate case, $V_k=\Pi_{k} V \Pi_k$
can be any $r_k \times r_k$ self-adjoint unitary matrix, where $\Pi_k$ and $r_k$ are the  eigenprojector and multiplicity
of $\lambda_k$, thus there are infinitely many choices for $V$.
Note that in all cases there are at most $2^n$ distinct geodesic lengths $\theta_V = \ell (\gamma_{\geo,V})$ since
$\theta_V $ only depends on the spectrum of $V$, see (\ref{eq-geodesic_length_V}). 
The geodesic with the shortest length 
is always unique and obtained for $V=\identity$.
The second shortest geodesic length $\theta_2$ is given by $\cos \theta_2 = \cos \theta_1 - 2 \lambda_{\rm{min}}$, $\lambda_{\rm{min}}$ being
the smallest eigenvalue of $\Lambda_{\sigma \rho}$. If $\lambda_{\rm{min}}$ is not degenerated, the corresponding geodesic 
is unique and obtained by choosing $V=V_{i_{\rm m}}$ such that it has a single negative eigenvalue $v_{i_{\rm m}}=-1$,
with $\lambda_{i_{\rm m}}= \lambda_{\rm{min}}$.

One infers from (\ref{eq-shortest_geodesic_length})  that the geodesic distance
(obtained as the minimal length of curves joining $\rho$ and $\sigma$)
is the arccos Bures distance  ${d}_{\rm B}$, see (\ref{eq-Bures_arccos_dist}).
This is the main reason for working with that distance, instead of the Bures distance (\ref{eq-def_Bures_dist}); $d_{\rm B}(\rho,\sigma)$ is the angle
$\theta_1 = \arccos \braket{\Psi}{\Phi_\identity}$ between the purification vectors
$\ket{\Psi}$ and $\ket{\Phi_\identity}$ of $\rho$ and $\sigma$ (recall that $\braket{\Psi}{\Phi_V} \in \real$).
Hence $\ket{\Phi}=\ket{\Phi_\identity}$ (and similarly $\ket{\Phi_{-\identity}} = - \ket{\Phi_\identity}$)
is a purification  of $\sigma$ maximizing the pure state fidelity $|\braket{\Psi}{\Phi}|$
in Uhlmann's theorem $ \sqrt{F(\rho, \sigma)}= \max_{\ket{\Phi}} |\braket{\Psi}{\Phi}|$~\cite{Uhlmann76,Nielsen}.
This gives a way to compute  $\ket{\Phi_{\identity}}$ numerically
using an optimization algorithm, instead of relying on the formula (\ref{eq-horizontality_cond_4})
(note that to compute  $\ket{\Phi_\identity}$ from (\ref{eq-horizontality_cond_4}) one needs to diagonalize $\rho$ and $\sqrt{\sigma}\sqrt{\rho}$).
 The purifications $\ket{\Phi_V}$ for $V \not= \pm \identity$ correspond to
relative maxima of $|\braket{\Psi}{\Phi}|$, which are smaller than the global maximum, see (\ref{eq-geodesic_length_V}).

The properties of the self-adjoint operators (\ref{eq-def_M^(V)}) are given in Appendix~\ref{sec-properties_M}.
As pointed out in~\cite{Ericson05}, for $V=\identity$, $  M_{\rho\sigma,\identity}$  is related to the optimal measurement to discriminate the distributions of measurement outcomes
in states $\rho$ and $\sigma$ (more precisely,  the Hellinger distance between these two distributions
is maximum for a projective measurement in the eigenbasis of $  M_{\rho \sigma, \identity}$~\cite{myreview}).

\subsection{Geodesics joining commuting states}

In the special case of commuting states $\rho$ and $\sigma$, the geodesics have the form 
\begin{eqnarray} \label{eq-geodesics_for_commuting_states}
  \gamma_{\geo,V} (\tau) & = & \sum_k p_{k,V} (\tau) \ketbra{k}{k}\\
  \nonumber
  p_{k,V}(\tau) & = & \bigg(\frac{\sin(\theta_V-\tau)}{\sin\theta_V} \sqrt{p_k} + v_k \frac{\sin \tau}{\sin \theta_V} \sqrt{q_k} \bigg)^2 \;,
\end{eqnarray}
where $\{ \ket{k}\}$ is an orthonormal basis of common eigenvectors of $\rho$ and $\sigma$ and $p_k$, $q_k$ are the corresponding eigenvalues.
In fact, then $ \MgeoV =\sum_k v_k \sqrt{q_k/p_k} \ketbra{k}{k}$ and the
purification (\ref{eq-horizontality_cond_4}) of $\sigma$ is given by
\begin{equation} \label{eq-purification_sigma_for_commuting_geodesics}
  \ket{\Phi_V} 
  = \sum_k v_k \sqrt{q_k}  \ket{k} \ket{\alpha_k}_\AAA\;.
\end{equation}
Replacing this expression into (\ref{eq-horizontal_geodesics_sphere})
yields the horizontal geodesic $\ket{\Psi_{\geo,V}}(\tau) = \sum_k \sqrt{p_{k,V}(\tau)} \ket{k} \ket{\alpha_k}_\AAA$, showing that $\gamma_{\geo,V} (\tau)$ commutes with $\rho$ and $\sigma$ at all times and is given by
(\ref{eq-geodesics_for_commuting_states}). Here, we have assumed that $\Lambda_{\sigma\rho}$ has 
a non-degenerated spectrum, \ie, $p_k q_k \not= p_l q_l$ if $k \not= l$.
Otherwise, there are infinitely many geodesics from $\rho$ to $\sigma$, which are not diagonal in the $\{ \ket{k}\}$-basis save for the one given by (\ref{eq-geodesics_for_commuting_states}).

The quantum
circuit of Fig.~\ref{fig-Q_circuit_geodesic_evol}(a) with the unitary $U_{\SSS \AAA}$ of Fig.~\ref{fig-Q_circuit_geodesic_evol}(b)
implements  a geodesic joining commuting states.
Indeed, using $\ket{\Psi}=  \sum_k \sqrt{p_k} \ket{w_k} \ket{k}_\AAA$ and $\ket{\dPsi} = U_{\SSS \AAA} \ket{1}\ket{0}_\AAA
= \sum_k \alpha_k \ket{w_k} \ket{k}_\AAA$ one finds that $\ket{\Phi_V}$ has the form
(\ref{eq-purification_sigma_for_commuting_geodesics}) with $\ket{k} \hookrightarrow \ket{w_k}$ and  $\ket{\alpha_k}_\AAA \hookrightarrow \ket{k}_\AAA$.

\subsection{Intersections with the boundary of quantum states}
\label{sec-int_Bures_geodesics_with_boundary}

We have so far determined the geodesics  on the open manifold $\states^\inv$ but have not discussed whether such geodesics can bounce on its
boundary  $\partial \states$.
Let us consider the  extension of the geodesic arc  (\ref{eq-Bures_geodesics}) joining the two  states $\rho$ and $\sigma$
to the time interval $[0,\pi]$. This extension is a closed curve, which we still denote by
$\gamma_{\geo,V}$.
Generalizing a result obtained in~\cite{Ericson05},
  we show in Appendix~\ref{sec-intersections_boundary} that this curve intersects $q_V$ times  the boundary,
  where $q_V$ is the number of distinct eigenvalues of the observable $\MgeoV$ in (\ref{eq-def_M^(V)}).
More precisely, let  $\rho_i$,  $i=1,\ldots, q_V$,  be the intersection points  of $\gamma_{\geo,V}$ with
$\partial \states$. Theorem~\ref{prop-intersection_Bures_geo_boundary} in Appendix~\ref{sec-intersections_boundary}
shows that the states $\rho_i$ have ranks $n-m_{i,V}$ and supports
$(\identity - P_{i,V})\Hh$, where $m_{i,V}$ and
  $P_{i,V}$ are the  multiplicity and  spectral projector of the $i$th eigenvalue of $\MgeoV$.
As a result, $\sum_{i=1}^{q_V} \dim ( \ker(\rho_i)) =n$. In particular, $\gamma_{\geo,V}$ intersects $\partial \states$ at a pure state if and only if
the spectrum of $\MgeoV$ has $q_V=2$ eigenvalues.
An interpretation of the  states $\rho_i$
and their kernel $P_{i,V} \Hh$  in quantum metrology is given in Sec~\ref{sec-geodesic_optimal_in_Qmetrology}  below.

Furthermore, it is shown in  Appendix~\ref{sec-intersections_boundary} that
the number of intersections on the part of $\gamma_{\geo,V}$ joining $\rho$ and $\sigma$
is equal to the multiplicity of
the eigenvalue $-1$ of $V$. In particular, the shortest geodesic $\gamma_{\geo}$ does not intersect  $\partial \states$ between $\rho$ and $\sigma$, while all other geodesics with $V \not= \identity$ do so at least once.  

\subsection{Geodesics passing through a pure state}

Consider an invertible state $\rho >0$ and
a pure state $\rho_1=\ketbra{\phi_1}{\phi_1}$ such that $ \langle \rho \rangle_{\phi_1}= \bra{\phi_1} \rho \ket{\phi_1} >0$.
As shown in Theorem~\ref{prop-intersection_Bures_geo_boundary}, there is up to time reversal only one geodesic joining $\rho$ and $\rho_1$. This geodesic is given by
\begin{eqnarray} \label{eq-geodesic_joing_mixed_and_pure_states}
  \nn
& &   \gamma_{\geo,\rho \to \rho_1} (\tau) =  \frac{1}{\sin^2 \theta_1} \bigg( \sin^2 ( \theta_1\! -\! \tau) \, \rho
  + \sin^2 (\tau) \times 
  \\
  & &
  \ketbra{\phi_1}{\phi_1} +  \frac{\sin(\theta_1-\tau) \sin (\tau)}{\cos \theta_1} \, \big\{ \rho ,  \ketbra{\phi_1}{\phi_1} \big\} \bigg)
\end{eqnarray}
with $\theta_1 = \arccos ( \langle \rho \rangle_{\phi_1}^{1/2} )$.
This geodesic intersects twice the boundary, at $\rho_1$ and at
another state $\rho_2$ of rank $n-1$ and support orthogonal to $\ket{\phi_1}$.
Eq. (\ref{eq-geodesic_joing_mixed_and_pure_states}) can be proven from (\ref{eq-Bures_geodesics_V}) by taking $\sigma = (1-\eps) \rho_1 + (\eps/n) \identity >0$ and
letting $\eps \to 0$.

If $\ket{\phi_1}$ is an eigenvector of $\rho$ with eigenvalue $p_1=\cos^2(\theta_1)>0$ then $\gamma_{\geo,\rho \to \rho_1}$
is a segment of straight line. In fact, in that case (\ref{eq-geodesic_joing_mixed_and_pure_states}) simplifies to
\begin{equation}
  \gamma_{\geo,\rho \to \rho_1} (\tau) = \sin^2 (\theta_1\! - \! \tau) \,\rho_\bot \! + \cos^2 (\theta_1\! -\! \tau)\, \ketbra{\phi_1}{\phi_1}\;,
\end{equation}
where $\rho_\bot = \Pi_\bot \rho\, \Pi_\bot/\sin^2(\theta_1)$ and $\Pi_\bot= 1 - \ketbra{\phi_1}{\phi_1}$ is the projector onto the
subspace orthogonal to $\ket{\phi_1}$. Note that this agrees with the general form (\ref{eq-geodesics_for_commuting_states}) of geodesics between commuting states. Recall 
 that the closest pure state to $\rho$, \ie, the state $\ket{\phi_1}$ maximizing the fidelity
$F(\rho,\rho_1)=\langle \rho \rangle_{\phi_1}$, is  an eigenvector of $\rho$
 associated to the maximal eigenvalue.
 Hence the shortest path joining $\rho$ to its closest pure state 
 is a segment of straight line intersecting $\partial \states$ transversally.

\section{Geodesics as physical evolutions} \label{sec-geodesics_physcial_evol}

We show in this section that the Bures geodesics correspond to
physical evolutions of the system coupled to an ancilla and that such evolutions are non-Markovian.

Let us recall that the dynamics of an open quantum system coupled to its environment  is obtained by letting the
total system (system and environment) evolve unitarily under some Hamiltonian $H$ 
and then tracing out over the environment (referred to as the ancilla $\AAA$ in what follows). The system state at time $t \geq 0$ is given by
\begin{equation} \label{eq-open_Q_system_dynamics}
  \rho(t) = \tr_\AAA  \E^{-\I t H} \ketbra{\Psi}{\Psi} \,\E^{\I t H} \;,
 \end{equation} 
where $\ket{\Psi}$ is the system--ancilla initial state.
Although one usually assumes that the system starts interacting with the ancilla
at $t=0$, so that
$\ket{\Psi} = \ket{\psi} \ket{\alpha}$ is a product state, in general the two subsystems can be initially entangled.

We have seen in Sec.~\ref{sec-Bures_geodesics} that the Bures geodesics $\gamma_{\geo,V}(\tau)$ joining two states $\rho$ and
$\sigma \in \states^\inv$
are the projection of horizontal pure--state geodesics $\ket{\Psi_{\geo,V} (\tau)}$
on an enlarged system with Hilbert space  $\Kk= \Hh \otimes \Hh_\AAA$, \ie, 
\begin{equation} \label{eq-Bures_geo_as_projecte_horiz_geo}
  \gamma_{\geo,V} (\tau)= \tr_\AAA  \ketbra{\Psi_{\geo,V}(\tau)}{\Psi_{\geo,V} (\tau)} \;,
\end{equation}
where $\ket{\Psi_{\geo,V} (\tau)}$ is given by (\ref{eq-horizontal_geodesics_sphere}).
The following theorem shows that one can associate   a
 Hamiltonian to the latter geodesic.

 \begin{theorem} \label{prop-geodesics_as-Q_evol}
Consider  the system-ancilla  Hamiltonian 
\begin{eqnarray} \label{eq-Hamiltonian_geodesic}
 \nn 
 \Hgeo & = & - \I \big( \ketbra{\Psi}{\dPsi_V} - \ketbra{\dPsi_V}{\Psi} \big)
 \\
   & = & \frac{- \I}{\sin \theta} \big( \ketbra{\Psi}{\Phi_V} - \ketbra{\Phi_V}{\Psi} \big)\;,
\end{eqnarray}
where $\ket{\Psi} \in \Kk$ is a fixed purification of $\rho$,
$\ket{\dPsi_V}$ is the horizontal tangent vector  to  $\ket{\Psi_{\geo,V}(\tau)}$
at $\tau=0$,
and $\ket{\Phi_V} \in \Kk$ is the corresponding purification of $\sigma$, see (\ref{eq-horizontality_cond_4}).   
Then
\begin{equation} \label{eq-psi_geo_Hamiltonian_evol}
  \ket{\Psi_{\geo , V} (\tau ) } = \E^{-\I \tau \Hgeo} \ket{\Psi}
\end{equation}
for any $\tau \geq 0$. As a result, the geodesic $\gamma_{\geo,V}$ coincides with the open quantum system time evolution 
\begin{equation} \label{eq-open_Q_system_geo_dynamics}
\gamma_{\geo,V} (\tau) = \tr_\AAA  \E^{-\I \tau \Hgeo} \ketbra{\Psi}{\Psi} \,\E^{\I \tau \Hgeo}\;.
\end{equation}
 \end{theorem}

\proof The horizontality condition  $\ket{\dPsi_V} \in \Hf_{\ket{\Psi}}$ entails $\braket{\Psi}{\dPsi_V} =0$,
see (\ref{eq-horizontal_space}) (note that general tangent vectors $\ket{\dPsi}$
at $\ket{\Psi}$ satisfy a weaker condition $\re  \braket{\Psi}{\dPsi} =0$). 
Furthermore, one has $\| \dPsi_V \| = 1$, see the statement following (\ref{eq-relation_between_dPsi_and_Phi}).
Let $\{ \ket{\Psi_k}\}_{k=0}^{n n_\AAA-1}$ be an \ONB  of $\Kk$ such that $\ket{\Psi_0}= \ket{\Psi}$ and $\ket{\Psi_1} = \ket{\dPsi_V}$.
The matrix of $\Hgeo$ in this basis has
a left upper corner given by the Pauli matrix $\sigma_y$, the other matrix elements being equal to zero. Such a matrix is
easy to exponentiate, yielding
\begin{eqnarray} \label{eq-exponenciation_geodesic_Hamiltonian}
\nn  
  \E^{- \I \tau \Hgeo}
  & = &
  \identity + ( \cos \tau -1 ) \big( \ketbra{\Psi}{\Psi} + \ketbra{\dPsi_V}{\dPsi_V} \big)
  \\
  & & - \sin \tau \big( \ketbra{\Psi}{\dPsi_V} - \ketbra{\dPsi_V}{\Psi} \big) \;.
\end{eqnarray}
Applying this operator to $\ket{\Psi}$ and comparing with (\ref{eq-geodesics_sphere})
yields the identity (\ref{eq-psi_geo_Hamiltonian_evol}).
The second equality in (\ref{eq-Hamiltonian_geodesic}) follows from the relation (\ref{eq-relation_between_dPsi_and_Phi}) 
between $\ket{\dPsi_V}$ and $\ket{\Phi_V}$.
\endproof

Note that the Hamiltonian $\Hgeo$ does not depend on the choice of the initial state $\ket{\Psi}$ on the horizontal geodesic $\ket{\Psi_{\geo,V} (\tau)}$.
Actually, let us fix another state
$\ket{\Psi_1}=\ket{\Psi_{\geo,V}(\tau_1)} =  \cos \tau_1 \ket{\Psi} + \sin \tau_1 \ket{\dPsi_V}$ on this
geodesic. Since $\ket{\dPsi_{\geo,V}(0)} \in \Hf_{\ket{\Psi}}$, by
property (i) of Sec.~\ref{sec-Riemannian_submersions}
the tangent vector   $\ket{\dPsi_{\geo,V} (\tau_1)}=  - \sin \tau_1 \ket{\Psi} + \cos \tau_1 \ket{\dPsi_V}$
is in the horizontal subspace at $\ket{\Psi_1}$.
One easily checks that the expression of $\Hgeo$ in the first line of  (\ref{eq-Hamiltonian_geodesic}) is invariant under the substitutions
$\ket{\Psi} \hookrightarrow \ket{\Psi_1}$ and $\ket{\dPsi_V} \hookrightarrow \ket{\dPsi_{\geo,V} (\tau_1)}$.

The geodesics $\gamma_{\geo,V}$ in Theorem~\ref{prop-Bures_geodesics} depend on 
two invertible states $\rho$ and $\sigma$. The purifications $\ket{\Psi}\in \Kk$ of $\rho$ are
entangled system-ancilla states.
However, since one can choose any initial state on  the geodesic,
$\rho$ can be taken to lie on the boundary $\partial \states$ of quantum states. Recall
that all geodesics intersect $\partial \states$, see Sec.~\ref{sec-int_Bures_geodesics_with_boundary}.
Let us discuss the special case for which $\gamma_{\geo,V}$ has an intersection with  $\partial \states$ given by a pure state.
For instance, all geodesics of a qubit satisfy this hypothesis (since $\partial \Ee_{\complex^2}$ is the set of pure qubit states).
One may then choose the initial  state $\rho$ on $\gamma_{\geo,V}$ to be a pure state $\rho_\psi=\ketbra{\psi}{\psi}$ 
having purifications
$
\ket{\Psi}= \ket{\psi} \ket{\alpha}
$
 given by product states, where
$\ket{\alpha}$ is an arbitrary ancilla pure state.
This means that the system and
ancilla are initially uncorrelated. 
In that   case one can extend the quantum evolution (\ref{eq-open_Q_system_geo_dynamics}) to arbitrary
(pure or mixed) initial states $\nu_\SSS \in \states$ of the system, by defining 
\begin{equation} \label{eq-Qevol}
  \Mm_{\geo,V,\tau}  ( \nu_\SSS ) = \tr_\AAA \E^{-\I \tau  \Hgeo} \,\nu_\SSS \otimes \ketbra{\alpha}{\alpha}\, \E^{\I \tau  \Hgeo}  \;.
\end{equation}
For all times $\tau$, $\Mm_{\geo ,V,\tau}$ is a quantum channel  (CPTP map).
The geodesic  $\gamma_{\geo,V} (\tau) = \Mm_{\geo,V,\tau} ( \rho_\psi )$ is obtained by taking the initial state 
$\nu_\SSS= \rho_\psi$.

It is clear from (\ref{eq-exponenciation_geodesic_Hamiltonian}) that the system-ancilla evolution operator $\E^{- \I \tau \Hgeo}$
is periodic in time with period $2\pi$. Thus the quantum  evolution  $\{ \Mm_{\geo,V,\tau} \}_{\tau \geq 0}$  is also periodic; more precisely, it
satisfies (\ref{eq-geo_evolutyion_is_periodic}). As a consequence, this evolution is  strongly non-Markovian.
 A quantitative study of this non-markovianity and a Kraus decomposition of  $\Mm_{\geo,V,\tau}$ will be presented in
 a forthcoming paper~\cite{Bures_geodesic_non_Markovian}.

 In conclusion, the Bures geodesics are not only mathematical objects but correspond to
 evolutions of the system coupled to an ancilla. This opens the route to the simulation of these geodesics on quantum computers and to
 their experimental observation.  Examples of quantum circuits simulating some geodesics have been given above
 (see Fig.~\ref{fig-Q_circuit_geodesic_evol}).

\section{Geodesics in quantum metrology} \label{sec-Q_metrology}

In this section we study quantum parameter estimation
for a quantum system coupled to an ancilla when measurements are not possible on the ancilla.
We show that the Bures geodesics features optimal state transformations for estimating a parameter
in a such situation.

\subsection{Quantum Fisher information and Bures metric} \label{sec-QFI_and_Bures_metric}

As explained in Sec.~\ref{sec_results_Q_metrology}, the best precision in the estimation of an unknown parameter $x$
using quantum probes in output states $\rho_x$ 
is given by the inverse square root of the QFI, see (\ref{eq-Q_Cramer_Rao_bound}).
The latter is by definition the maximum  over all POVMs
$\{ M_j\}$ of the CFI (\ref{eq-classical_Fisher_info}) with probabilities $p_{j|x}=\tr M_j \rho_x$. It is
given by~\cite{Helstrom68,Braunstein94} 
    \begin{equation}  \label{eq-QFI_and_semilogarithmic_derivative}
      \Ff_Q (x, \{ \rho_x \}_{x \in X} ) =   \tr [ \rho_x \SL_{x}^2  ]\;,
    \end{equation}  
where $\SL_x$ is a self-adjoint operator satisfying
    \begin{equation} \label{eq-def_SL}
      \onehalf \big\{ \SL_{x} \, , \, \rho_{x} \big\} = \dot{\rho}_{x}
    \end{equation}
 with $\dot{\rho}_x= \partial_x \rho_x$.  The operator $\SL_x$ is called the symmetric logarithmic derivative of $\rho_x$.
    A POVM  $ \{ M_j \}$ maximizes the CFI (\ie,
    $\Ff_\clas (x, \{ p_{j|x}) \}_{x \in X} ) = \Ff_Q (x,\{ \rho_x\}_{x \in X} )$) if and only if
    $M_j^{1/2} \rho_x^{1/2}$ is equal to $c_j M_j^{1/2} L_x \rho_x^{1/2}$ for any $j$, with $c_j \in \real$~\cite{Braunstein94}.
    If $\rho_x$ is invertible, this is equivalent to $M_j^{1/2} = c_j L_x  M_j^{1/2}$. Thus, the optimal POVMs $ \{ M_j^\opt \}$
    maximizing the CFI are such that  for any $j$, the support of $M_j^\opt$ is contained in an eigenspace of $L_x$.
    In particular, the projective measurement given by the spectral projectors
of $L_x$ is optimal. In general, the optimal measurements  $ \{ M_j^\opt \}$  depend on the estimated parameter $x$, as $L_x$ depends on $x$.

It is known that for invertible states $\rho_x \in \states^\inv$
the QFI is equal to the Bures metric (\ref{eq-Bures_metric}) up to a factor of four~\cite{Braunstein94,myreview},
  \begin{equation} \label{eq-QFI_and_Bures_metric}
  \Ff_Q (x, \{ \rho_x \}_{x \in X} ) =  4 
  ( g_{\rm B})_{\rho_x} ( \dot{\rho}_{x} , \dot{\rho}_{x} )\;.
  \end{equation}
      A similar relation holds for the CFI and the Fisher metric (see e.g.~\cite{Oberthaler14}).
        A way to derive the identity (\ref{eq-QFI_and_Bures_metric}) is as follows. Consider the geodesic  $\gamma_{\geo, V} ( \tau)$
        passing through $\rho_x$ at $\tau=0$ with tangent vector $\dot{\gamma}_{\geo, V} ( 0) = \dot{\rho}_x$.
        By (\ref{eq-Bures_geo_as_projecte_horiz_geo}), one has
        \begin{equation}
          \dot{\rho}_x   =  \tr_A \big( \ketbra{\dPsi_{\geo,V}}{\Psi_{\geo,V}} + \ketbra{\Psi_{\geo,V}}{\dPsi_{\geo,V}}\big)
          = \{ L_\SSS \, , \, \rho_x \} ,
        \end{equation}
        where the last equality comes from the  horizontality condition
        $\ket{\dPsi_{\geo,V}} = L_\SSS \otimes \identity_\AAA \ket{\Psi_{\geo,V}}$, see (\ref{eq-horizontality_cond}).
        Comparing with (\ref{eq-def_SL}), one sees that the self-adjoint operator $L_\SSS$ is nothing but the symmetric logarithmic
        derivative $\SL_x$ of $\rho_x$ up to a factor of $1/2$.
        Using the isometry property (\ref{eq-metric_g_Ee_Riemannian_sub})
        of the metric, one gets
        \begin{eqnarray} \nonumber
          & & 4 (g_B)_{\rho_x} (   \dot{\rho}_x \,,\,  \dot{\rho}_x  )  =  4 \| \dPsi_{\geo,V} \|^2
            \\ \nonumber
            & & \qquad =  \tr \big(  L_x \otimes \identity_\AAA \ket{\Psi_{\geo,V}}  \bra{\Psi_{\geo,V}}  L_x \otimes \identity_\AAA \big)
              \\ 
            &  & \qquad =  \tr ( \rho_x  L_x^2  \big) = \Ff_Q (x, \{ \rho_x \}_{x \in X} )\; . 
        \end{eqnarray}      
        We point out that the right-hand sides of (\ref{eq-QFI_and_semilogarithmic_derivative}) and (\ref{eq-QFI_and_Bures_metric}) are not always equal
  for non-invertible states  $\rho_x$.
  In fact, it is easy to show that the trace in (\ref{eq-QFI_and_semilogarithmic_derivative}) is given (up to a factor of four) by the \RHS of 
(\ref{eq-Bures_metric}) with $\dot{\sigma}=\dot{\rho}$ and a sum running over all indices $k,l$ such that $p_k + p_l >0$.
  On the other hand, as shown in~\cite{Safranek17}, the Bures metric (which is defined in terms of the infinitesimal distance
  $\D s^2$, see (\ref{eq-definition_metric})) is given by the same expression 
plus an additional term $2 \sum_{j,p_{j|x} =0} \partial^2_x {p}_{j|x}$ involving the second derivatives of the vanishing  eigenvalues of $\rho_x$
(note that this term is absent when $\rho_x >0$).  
As a result,
the QFI (\ref{eq-QFI_and_semilogarithmic_derivative}) is discontinuous at values $x$ at which
$\rank (\rho_{x}) $
is discontinuous, while the Bures metric remains continuous~\cite{Safranek17,Seveso2019}.
This can be illustrated by the  following example.
Let $\rho_x = \sum_j p_{j|x} \ketbra{j}{j}$ with $\{ \ket{j}\}$ a fixed orthonormal basis. Then
the QFI (\ref{eq-QFI_and_semilogarithmic_derivative}) equals the CFI 
(\ref{eq-classical_Fisher_info})  and has a jump for trajectories $x\mapsto \rho_x$ bouncing on $\partial \states$  at $x=x_0$.
More precise\-ly, each eigenvalue with a minimum at $x_0$ such that $p_{j|x_0}= \partial_x p_{j|x}(x_0)=0$  contributes to the jump amplitude by 
$-\lim_{x\to x_0} (\partial_x {p}_{j|x})^2 / p_{j|x}
= - 2 \partial^2_x{p}_{j|x}(x_0)$.
In contrast, $ ( g_{\rm B})_{\rho_x} ( \dot{\rho}_{x} , \dot{\rho}_{x} )$ is continuous at $x_0$ due to aforementioned
additional term canceling the discontinuity.

  \subsection{Pythagorean theorem and variational formula for the QFI} \label{sec-variational_formula_QFI}

  According to Stinespring's theorem~\cite{Stinespring55}, the action of an arbitrary quantum channel $\Mm_x$ on a state $\rho_\inp$ can be obtained
  by coupling the system to an ancilla and letting the composite
    system evolving unitarily, assuming an initial system-ancilla product state $\rho_\inp \otimes \ketbra{\alpha_0}{\alpha_0}$.
   We suppose in what follows that this composite system is in a pure state $\ket{\Psi_x}$ undergoing a 
    $x$-dependent unitary evolution of the form
    \begin{equation} \label{eq-unitary_evolution}
    \ket{\Psi_x} = \E^{-\I x H} \ket{\Psi_\inp}\;,  
    \end{equation}
    where $H$ is some {  $x$-independent} Hamiltonian on $\Kk=\Hh \otimes \Hh_\AAA$ and 
    the input system-ancilla state $\ket{\Psi_\inp}$ may be entangled or not.
    The output states of the system are given by 
    \begin{equation}
      \rho_x = \Mm_x ( \rho_\inp ) = \tr_\AAA \ketbra{\Psi_x}{\Psi_x}\;.
    \end{equation}
    For concreteness, we assume that $x$ belongs to an interval $X$ containing $0$.
  It is also convenient to replace $H$ by $\Delta H = H - \langle H \rangle_{\Psi_\inp} \identity$
  in (\ref{eq-unitary_evolution}). This amounts to multiply $\ket{\Psi_x}$ by an irrelevant
  phase factor  $\E^{\I x \langle H \rangle_{\Psi_\inp}}$.
  The QFI of the composite system reads  (see (\ref{eq-QFI_pure_states}))
    \begin{equation} \label{eq-QFI_pure_states_bis}
      \Ff_Q ( \{ \ket{\Psi_x} \}_{x \in X} ) =  4 \| \dPsi_x \|^2 = 4 \langle (\Delta H)^2\rangle_{\Psi_\inp}\;,
    \end{equation}
    where we have used that $\langle (\Delta H)^2\rangle_{\Psi_x}$ is independent of $x$.

    As explained in Sec.~\ref{sec_results_Q_metrology}, one can 
    decompose the tangent vector $\ket{\dPsi_x}$ into its horizontal and vertical
    parts,
    \begin{equation} \label{eq-decomposition_dot_Psi_horizontal_and_vertical}
      \ket{\dPsi_x}  = - \I \Delta H \ket{\Psi_x} =  \ket{\dPsi^{\ho}_{x}} - \I\, \identity \otimes B_{x}^{\ho} \ket{\Psi_x}
    \end{equation}
    with $ \ket{\dPsi^{\ho}_x} \in \Hf_{\ket{\Psi_x}}$ and $B_{x}^{\ho}$  a self-adjoint operator on $\Hh_\AAA$. Here, we have used the form (\ref{eq-vertical_subspace}) of the vertical subspace  $\Vf_{\ket{\Psi_x}}$. Note that  $\langle \identity \otimes B_{x}^{\ho} \rangle_{\Psi_x}=0$ since
    $\braket{\Psi_x}{\dPsi_x}=$ $\braket{\Psi_x}{\dPsi^{\ho}_x}=0$.
    As $\rho_x = \pi ( \ket{\Psi_x})$ and $\Vf_{\ket{\Psi}} = \ker ( \D \pi |_{\ket{\Psi_x}} )$, where $\pi$ is the projection (\ref{eq-quotient_map}),
    one has
    $\dot{\rho}_x = \D \pi|_{\ket{\Psi_x}} ( \ket{\dPsi_x^\ho} )$. 
    Thanks to (\ref{eq-metric_g_Ee_Riemannian_sub}) and (\ref{eq-QFI_and_Bures_metric}), the QFI is given by
    \begin{eqnarray} \label{eq-variational_formula_QFI}
     \nn
     & &  \Ff_Q ( x, \{ \rho_x \}_{x \in X} )
     \\ \nn
     & & \hspace*{1cm}
       =  4 ( g_{\rm B})_{\rho_x} \big(  \D \pi|_{\ket{\Psi_x}} ( \ket{\dPsi_x^\ho} ),  \D \pi|_{\ket{\Psi_x}} ( \ket{\dPsi_x^\ho} ) \big)
      \\ \nn
      & &\hspace*{1cm} =  4 \| \dPsi^{\ho}_x \|^2\\
      & &\hspace*{1cm}  =  4 \big( \| \dPsi_x \|^2 - \| \identity \otimes B_{x}^{\ho} \ket{\Psi_x} \|^2 \big)\;,
    \end{eqnarray}
    where  the third equality follows from (\ref{eq-decomposition_dot_Psi_horizontal_and_vertical}), the
    orthogonality of the horizontal and vertical subspaces, and the Pythagorean theorem.
    Using (\ref{eq-QFI_pure_states_bis}),
    one sees that (\ref{eq-variational_formula_QFI}) is equivalent to the formula (\ref{eq-formula_QFI_Pythagore}) of
    Sec.~\ref{sec_results_Q_metrology}. The skew Hermitian operator $\I B_x^{\ho}$ in (\ref{eq-variational_formula_QFI}) is the Uhlmann connection
    for the lift $x \in X \mapsto \ket{\Psi_x}$ of the curve  $x \mapsto \rho_x$~\cite{Uhlmann86}.

    It is instructive to derive from (\ref{eq-variational_formula_QFI}) the variational formula for the QFI from Ref.~\cite{Escher12}.
Consider the {  family of purifications of the $\rho_{y}$ given by
    $\ket{\widetilde{\Psi}_{y}} = \identity \otimes \E^{\I (y-x) B_x} \ket{\Psi_{y}}$,
    where $B_x$ is a self-adjoint operator on $\Hh_\AAA$. Its
    tangent vector at $y=x$ is $\partial_y \ket{ \widetilde{\Psi}_y} |_{x} = \ket{\dPsi_x} + \I \,\identity \otimes B_x \ket{\Psi_x}
    = -\I ( \Delta H - \identity \otimes B_x) \ket{\Psi_x}$. From  (\ref{eq-decomposition_dot_Psi_horizontal_and_vertical}) and 
    Pythagorean theorem again, the square norm of this tangent vector is}
\begin{eqnarray}
  \nn
  & &     \big\langle ( \Delta H  - \identity \otimes B_x )^2 \big\rangle_{\Psi_x}
  \\ \nn
   & & \hspace{1.4cm} =  
      \big\| \dPsi^{\ho}_x \big\|^2 + \big\| \identity \otimes ( B_x - B_x^\ho ) \ket{\Psi_x} \big\|^2
     \\
     & & \hspace{1.4cm} \geq \big\| \dPsi^{\ho}_x \big\|^2\;.
    \end{eqnarray}
    Hence the minimum of the {  \LHS}   over all $B_x$'s is equal to $\| \dPsi^{\ho}_x \|^2$ and the minimum is achieved for $B_x= B_{x}^{\ho}$.
    One deduces from  (\ref{eq-variational_formula_QFI}) that~\cite{Escher12}
\begin{equation} \label{eq-variational_formula_QFIbis}
      \Ff_Q (x, \{ \rho_x \}_{x \in X} )
       = 
    4 \min_{B_x = B_x^\dagger} \big\langle ( \Delta H  - \identity \otimes B_x )^2 \big\rangle_{\Psi_x} \;.
\end{equation}

Eq. (\ref{eq-variational_formula_QFI}) 
    tells us that the QFI of the system 
    is equal to the QFI (\ref{eq-QFI_pure_states_bis}) of the composite system 
    minus a non-negative quantity
    $4 \| \identity \otimes B_{x}^{\ho} \ket{\Psi_x} \|^2$
that can be interpreted as the amount of information on  $x$   in the ancilla.
Eq. (\ref{eq-variational_formula_QFIbis})
provides a variational formula for   the QFI.
Both expressions  (\ref{eq-variational_formula_QFI}) and (\ref{eq-variational_formula_QFIbis}) 
    have been derived in~\cite{Escher12} by using another method. We see here that they
have nice geometrical interpretations  in the framework of Riemannian submersions,
being simple consequences of the Pythagorean theorem.

\subsection{Optimal precision in parameter estimation in open quantum systems} \label{sec-geodesic_optimal_in_Qmetrology}

One deduces from (\ref{eq-QFI_pure_states_bis}) and (\ref{eq-variational_formula_QFI}) that the QFI of the probe is equal to the QFI
of the composite system when the tangent vector $\ket{\dPsi_x}$ is horizontal (\ie, $B_x^\ho=0$). In such a case, there is no information
loss on the parameter $x$ in the ancilla: joint measurements on the probe and ancilla do not lead to a better precision in the
estimation than local measurements on the probe. 
However, in general the state transformation does not preserve the horizontality of the tangent vector.
Let us assume {  for instance} that  $\ket{\dPsi_\inp} \in \Hf_{\ket{\Psi_\inp}}$, so that the probe QFI is equal to
 $ 4 ( \langle \Delta H )^2 \rangle_{\Psi_\inp}$  for $x=0$.
While  the QFI of the composite system remains the same for all values of $x$, 
the probe QFI (\ref{eq-variational_formula_QFI}) depends on $x$ and 
is strictly smaller than  $ 4 ( \langle \Delta H )^2 \rangle_{\Psi_\inp}$ for nonzero values of $x$ at which $\ket{\dPsi_x} \notin \Hf_{\ket{\Psi_x}}$, implying
a larger error $(\Delta x)_{\rm QCRB}$ for $x \not= 0$ than for $x=0$.
It is thus of interest to {  study} situations for which the tangent vector remains horizontal for all values of the parameter.
In such cases,  the minimal error $(\Delta x)_{\rm{QCRB}}$ is $x$-independent
and equal to the minimal error one would obtain from joint measurements on the probe and ancilla.

In the following, we assume that both the system-ancilla coupling Hamiltonian $H$ and the
 input state $\ket{\Psi_\inp}$ can be engineered at will. 
We look for Hamiltonians $H$ and input states $\ket{\Psi_\inp}$ satisfying:
\begin{itemize}
\item[(I)] the horizontality condition $\ket{\dPsi_x} \in \Hf_{\ket{\Psi_x}}$ holds for all $x \in X$;
\item[(II)] for a fixed Hamiltonian $H$ satisfying (I), $ \| \dPsi_x \|^2 =  \langle (\Delta H)^2\rangle_{\Psi_\inp}$ is maximum.
\end{itemize}
If  conditions (I) and (II) are satisfied then the QFI of the system is constant and maximal for all values of $x$,
\begin{equation} \label{eq-QFI_geodesic_evol}
\Ff_Q ( x, \{ \rho_x \}_{x \in X} ) = 4  \langle (\Delta H)^2\rangle_{\Psi_\inp} \;.
\end{equation}
We shall assume that the highest and smallest eigenvalues of $H$, $\epsilon_\mmax$ and $\epsilon_\mmin$, are non-degenerated. We set
\begin{equation}
{  \gap} =  \onehalf (\epsilon_\mmax - \epsilon_\mmin)\;.
\end{equation}
Our first result is:

\begin{theorem}  \label{prop-geodesics_and_parameter_estimation}
  Conditions {\rm (I)} and {\rm (II)} are fulfilled if and only if  $\ket{\dPsi_\inp} = - \I \Delta H \ket{\Psi_\inp} \in \Hf_{\ket{\Psi_\inp}}$
  and  one of the two equivalent conditions holds
  \begin{itemize}
\item[{\rm (a)}]  $\ket{\Psi_x}  = \E^{-\I  x {  \gap} \, H_{\geo, V}} \ket{\Psi_\inp}\;,\; x \in X$;
\item[{\rm (b)}]  $\Pi (\Delta H) \Pi = {  \gap} \, H_{\geo,V}$,
\end{itemize}  
where  $H_{\geo,V}$ is the geodesic Hamiltonian {\rm (\ref{eq-geodesic_Hamil})}
with $\ket{\Psi}$  and
$\ket{\dPsi}$ replaced by $\ket{\Psi_\inp}$ and $\ket{\dPsi_\inp}/\|\dPsi_\inp\|$, respectively,
and $\Pi$ is the projector onto the sum of the eigenspaces
  of $H$ with eigenvalues $\epsilon_\mmax$ and $\epsilon_\mmin$.
Note that {\rm (a)} implies
  \begin{equation} \label{eq-output_state-geo_evol}
    \rho_x = \gamma_{\geo,V} ( x {  \gap} )\;,\;x \in X
  \end{equation}
  where $\gamma_{\geo,V}$ is a geodesic starting at the state $\rho_\inp = \tr_\AAA \ketbra{\Psi_\inp}{\Psi_\inp}$.
\end{theorem}

This theorem tells us that among all probe-ancilla Hamiltonians $H$ having a fixed energy gap
  $2\gap$ and assuming that the input state $\ket{\Psi_\inp}$ maximizes the probe--ancilla QFI  for $H$, the Hamiltonians $H$ producing the
  highest QFI of the probe, for arbitrary parameter values $x$, satisfy property (b), \ie, they generate geodesic evolutions given by
  (\ref{eq-output_state-geo_evol}).
  
\vspace{0mm}

\proof Assume that (I) and (II) are fulfilled.
It has been pointed out in Sec.~\ref{sec_results_Q_metrology} that
the states $\ket{\Psi_\inp}$ maximizing the variance $\langle (\Delta H)^2\rangle_{\Psi_\inp}$ are
the superpositions
\begin{equation} \label{eq-Schrodinger_cat_states}
  \ket{\Psi_\inp} = \frac{1}{\sqrt{2}} \Big(   \ket{\epsilon_\mmax} + \E^{\I \varphi} \ket{\epsilon_\mmin} \Big)
\end{equation}  
with $\ket{\epsilon_\mmax}$ and $\ket{\epsilon_\mmin}$ two eigenstates of $H$ associated to  $\epsilon_\mmax$ and $\epsilon_\mmin$ and $\varphi \in \real$.
For such input states one has
\begin{equation} \label{eq-tangent_input_vector_state}
  \ket{\dPsi_\inp} = - \I \Delta H \ket{\Psi_\inp} = -\I \frac{  \gap}{\sqrt{2}} \Big(
  \ket{\epsilon_\mmax}  - \E^{\I \varphi}  \ket{\epsilon_\mmin}
  \Big)\;.
\end{equation}
Furthermore,
\begin{eqnarray} \label{eq-proof_theorem3}
  \nn
  \ket{\Psi_x} & =  & \E^{- \I x \Delta H} \ket{\Psi_\inp}
  \\ \nn
  & = & 
  \frac{1}{\sqrt{2}} \Big( \E^{-\I x \gap}  \ket{\epsilon_\mmax} + \E^{\I ( x \gap+ \varphi )} \ket{\epsilon_\mmin} \Big)
  \\
  & = &    \cos ( x \gap) \ket{\Psi_\inp} + \frac{1}{\gap} \sin ( x \gap) \ket{\dPsi_\inp}\;.
\end{eqnarray}
Since $\braket{\Psi_\inp}{\dPsi_\inp} = 0$ and $\| \dPsi_\inp \|^2 = \langle (\Delta H )^2 \rangle_{\Psi_\inp} =   \gap^2$,
  the last expression in (\ref{eq-proof_theorem3}) is nothing but  the arc of great circle (\ref{eq-geodesics_sphere})
  at time $\tau = x   \gap$, \ie, $\ket{\Psi_x} = \ket{\Psi_{\geo} ( x   \gap)}$ with $\ket{\Psi_{\geo}(\tau)}$  
  a geodesic on the unit sphere $\Ss_\Kk$ (see Sec.~\ref{sec-Bures_geodesics}).
By condition (I)  this geodesic is a horizontal geodesic, $\ket{\Psi_{\geo}(\tau)}= \ket{\Psi_{\geo,V} (\tau)}$.
In view also of (\ref{eq-psi_geo_Hamiltonian_evol}), one deduces that (a) is true.
Now, plugging
(\ref{eq-Schrodinger_cat_states}) and (\ref{eq-tangent_input_vector_state}) into the expression  of
the geodesic Hamiltonian yields
\begin{equation}
H_{\geo,V} = \ketbra{\epsilon_\mmax}{\epsilon_\mmax} - \ketbra{\epsilon_\mmin}{\epsilon_\mmin} =  {   \gap^{-1} }\Pi (\Delta H )  \Pi\, .
\end{equation}
This shows that (I) \& (II) $\Rightarrow$ (a) $\Rightarrow$ (b).

Reciprocally, assume that  $\ket{\dPsi_\inp} \in \Hf_{\ket{\Psi_\inp}}$ and
that (b) is true. 
Since $\epsilon_\mmax$ and $\epsilon_\mmin$ are not degenerated,
\begin{eqnarray} \label{eq-proof_reciprocal_theo3}
  \nonumber
 \Pi ( \Delta H) \Pi & = & \big( \epsilon_\mmax - \langle H\rangle_{\Psi_\inp} \big) \ketbra{\epsilon_\mmax}{\epsilon_\mmax}
  \\
   & &  - \big( \langle H\rangle_{\Psi_\inp} \!\! - \epsilon_\mmin  \big)  \ketbra{\epsilon_\mmin}{\epsilon_\mmin} .
\end{eqnarray}
Furthermore,
\begin{eqnarray} \label{eq-proof_reciprocal_theo3bis}
  \nonumber
  \epsilon H_{\geo,V} & = & -\I \epsilon \bigg( \ket{\Psi_\inp}\frac{\bra{\dPsi_\inp}}{\| \dPsi_\inp\|}
  - \frac{\ket{\dPsi_\inp}}{\| \dPsi_\inp\|}\bra{\Psi_\inp} \bigg)
  \\
  & = & 
 {  \gap}
   \big(
 \ketbra{\epsilon_+}{\epsilon_+} - \ketbra{\epsilon_{-}}{\epsilon_{-}} \big)\;,
\end{eqnarray}
where we have set
\begin{equation}
  \ket{\epsilon_\pm} = \frac{1}{\sqrt{2}} \bigg( \ket{\Psi_\inp} \pm \I \frac{\ket{\dPsi_\inp}}{\| \dPsi_\inp\|} \bigg)
  \; .
\end{equation}
Since $\braket{\Psi_\inp}{\dPsi_\inp} = 0$, the vectors $\ket{\epsilon_\pm}$ are normalized and orthogonal. Thus, by 
(\ref{eq-proof_reciprocal_theo3}) and (\ref{eq-proof_reciprocal_theo3bis}), the equality in (b) implies
$\ket{\epsilon_+} = \E^{-\I \varphi_+} \ket{\epsilon_\mmax}$ and $\ket{\epsilon_{-}} = \E^{-\I \varphi_{-}} \ket{\epsilon_\mmin}$, where 
$\varphi_\pm$ are real phases.
It follows that $\ket{\Psi_\inp}$ is given by (\ref{eq-Schrodinger_cat_states}) up to an irrelevant phase factor, with
$\varphi = \varphi_{+} - \varphi_{-}$, and
\begin{eqnarray}
  \nonumber
  \ket{\Psi_x} & = & \E^{-\I x \Delta H} \ket{\Psi_\inp} =  \E^{-\I x \, \Pi (\Delta H) \Pi} \ket{\Psi_\inp}    \\
  & = & \E^{-\I x {  \gap} \, H_{\geo,V}} \ket{\Psi_\inp} = \ket{\Psi_{\geo,V} (x   \gap )}
\end{eqnarray}    
(the second and last equalities follow from  (\ref{eq-Schrodinger_cat_states}) and (\ref{eq-psi_geo_Hamiltonian_evol}), respectively). 
Our hypothesis $\ket{\dPsi_\inp} \in \Hf_{\Psi_\inp}$ implies that $H_{\geo,V}$ is a geodesic Hamiltonian, \ie,
$\ket{\Psi_{\geo,V} (\tau)}$ is
a horizontal geodesic (see Property (i) of Sec.~\ref{sec-Riemannian_submersions}).
Hence condition (I) is fulfilled.
Furthermore, by (\ref{eq-Schrodinger_cat_states}) again,
$\| \dPsi_\inp\|^2 = \langle ( \Delta H )^2 \rangle_{\Psi_\inp} =   \gap^2$ is the maximal squared fluctuation of $H$.
Thus condition (II) also holds.
We have shown that (b) $\Rightarrow$ (a) $\Rightarrow$  (I) \& (II).
Finally,  by projecting the equality (a) onto $\states$, one obtains (\ref{eq-output_state-geo_evol}).
\endproof

It is worth noting that the probe-ancilla input state $\ket{\Psi_\inp}$ is not necessarily entangled. 
To see this, let us vary the  phase $\varphi$ in  (\ref{eq-Schrodinger_cat_states}) and observe that
$\ket{\Psi_\inp}$ then moves on the horizontal geodesic as 
$\ket{\Psi_\inp^0} \rightarrow$ $\ket{\Psi_\inp^\varphi} = \E^{-\I \varphi \Hgeo/2} \ket{\Psi_\inp^0}$ up to an irrelevant phase factor, where 
$\ket{\Psi_\inp^0}$ is the input state corresponding to $\varphi=0$.
Recall from Sec.~\ref{sec-int_Bures_geodesics_with_boundary} that all geodesics $\gamma_{\geo,V}$ intersect the boundary $\partial \states$ of
quantum states. Therefore, if for instance the probe is a qubit,  $\varphi$ can be chosen such that
$\rho_\inp=\ketbra{\psi_\inp}{\psi_\inp}$ is a pure state, that is, $\ket{\Psi_\inp}= \ket{\psi_\inp} \ket{\alpha_\inp}$ is a product state.
In other words, albeit the superposition (\ref{eq-Schrodinger_cat_states}) is in general entangled, an appropriate phase choice 
makes it separable.
Since preparing a probe and ancilla in an entangled state is challenging experimentally, this 
is a relevant observation. 
Let us stress that the aforementioned separability refers to a disentanglement between the probe and ancilla; if the probe
consists of $N$ qubits and $H$ acts independently on each qubit, we shall see in Sec.~\ref{sec-Heisenberg_limit} below that $\ket{\Psi_\inp}$ has maximal entanglement between the probe qubits. 
A phase choice such that the input state  $\ket{\Psi_\inp}$ is a product state
is also possible for higher-dimensional probes, but only for those geodesics such that the
observable $\MgeoV$ in (\ref{eq-def_M^(V)}) has two eigenvalues of multiplicities $n-1$ and $1$. In fact,
as argued in Sec.~\ref{sec-int_Bures_geodesics_with_boundary}, this condition
ensures that one of the intersection of $\gamma_{\geo,V}$ with $\partial \states$ is a pure state.

The next theorem characterizes all system-ancilla Hamiltonians $H$ generating horizontal geodesics, that is, 
coinciding (up to a numerical factor) with a geodesic Hamiltonian $H_{\geo,V}$ in a two-dimensional subspace.
It shows that such Hamiltonians have two eigenvectors related to each other by a local unitary acting on the system.

\begin{theorem}  \label{prop-geodesics_and_parameter_estimationIII}
  Let $\ket{\epsilon_1}$ and $\ket{\epsilon_2}$ be 
  two eigenstates of $H$ with distinct eigenvalues $\epsilon_1$ and $\epsilon_2$. 
  If
  \begin{equation} \label{eq-superposition_bis}
    \ket{\Psi_\inp} = \frac{1}{\sqrt{2}} \Big( \ket{\epsilon_{1}} + \E^{\I \varphi} \ket{\epsilon_{2}} \Big)
  \end{equation}
  then
  the unitary transformation $\ket{\Psi_x} = \E^{-\I x \Delta H} \ket{\Psi_\inp}$ is a horizontal geodesic if and only if
   $\ket{\epsilon_2} = U \otimes \identity_\AAA \ket{\epsilon_1}$ with $U$
  a local unitary acting on the system. In such a case  $\ket{\Psi_x} = \ket{\Psi_{\geo,V} ( x \gap)}$ with $\gap= ( \epsilon_{1}- \epsilon_{2})/2$.
\\
  In particular, conditions {\rm (I)} and {\rm (II)} hold if and only if 
  $\ket{\Psi_\inp}$ is given by {\rm (\ref{eq-Schrodinger_cat_states})} and $\ket{\epsilon_\mmin} = U \otimes \identity \ket{\epsilon_\mmax}$.
\end{theorem}

\proof
One deduces from
$\ket{\Psi_x} = \E^{-\I x \Delta H} \ket{\Psi_\inp}$ and  (\ref{eq-superposition_bis}) that $\ket{\dPsi_\inp}$ is given by (\ref{eq-tangent_input_vector_state}) upon substituting
$\ket{\epsilon_\mmax}$ and $\ket{\epsilon_\mmin}$  by $\ket{\epsilon_1}$ and $\ket{\epsilon_2}$.
The horizontality condition $ \ket{\dPsi_\inp} = L_\SSS \otimes \identity_\AAA  \ket{\Psi_\inp}$
for some self-adjoint operator $  L_\SSS$ such that $  \langle L_\SSS \otimes \identity_\AAA \rangle_{\Psi_\inp}=0$ can be rewritten as
\begin{equation}
    \ket{\epsilon_2} = \E^{-\I \varphi}  \frac{\gap- \I L_\SSS}{\gap+ \I L_\SSS}  \otimes \identity_\AAA \ket{\epsilon_1} = U \otimes \identity_\AAA\ket{\epsilon_1}\;,
\end{equation}
where $U$  is a unitary operator acting on the probe and $\epsilon= ( \epsilon_1-\epsilon_2)/2$. Reciprocally, if    
$\ket{\epsilon_2} = U \otimes \identity_\AAA \ket{\epsilon_1}$ then
\begin{equation}\label{eq-proof_teorem_5}
  \ket{\dPsi_\inp} = -\I \gap( \identity - \E^{\I \varphi}\, U ) ( \identity + \E^{\I \varphi}\,U   )^{-1}  \otimes \identity_\AAA \ket{\Psi_\inp}\;,
\end{equation}
where it is assumed that $-\E^{\I \varphi}$ is not an eigenvalue of $U$
(in such a way that $\identity + \E^{\I \varphi} U$ is invertible).
It is easy to show that the local operator in the \RHS of (\ref{eq-proof_teorem_5}) is self-adjoint.
Hence $\ket{\dPsi_\inp} \in \Hf_{\ket{\Psi_\inp}}$.
\endproof

\subsection{Optimal measurements} \label{sec-geodesic_optimal_measurements}

We now turn to the problem of determining the optimal measurement(s) on the probe maximizing the CFI.
As explained in Sec.~\ref{sec-QFI_and_Bures_metric}, these measurements are given in terms of  
the  symmetric logarithmic derivative $\SL_x$ of the output states $\rho_x = \gamma_{\geo,V} ( x \epsilon)$. 
Let us fix a state $\sigma$ on $\gamma_{\geo,V}$ such that
$\rho_x$ belongs to the geodesic arc between $\rho_\inp$ and $\sigma$.
Note that $\gamma_{\geo,V}$  is the same as the geodesic {  $\gamma_{\geo, V_x}^{\rho_x \to \sigma}$} starting at $\rho_x$ and passing through $\sigma$ 
translated in time by $- x   \epsilon$  (see Appendix~\ref{sec-properties_M} for an explicit proof). Thus,
by differentiating the latter geodesic at $\tau=0$ one gets the tangent vector
of  $\gamma_{\geo,V}$ at $\tau_x = x   \epsilon$. Making the substitutions
$\rho \hookrightarrow \rho_x$, {  $V \hookrightarrow V_x$}, and $\theta_V \hookrightarrow \theta_V - \tau_x$ in  (\ref{eq-Bures_geodesics}), this gives
\begin{equation} \label{eq-geo_tangent_vector}
  \dot{\gamma}_{\geo , V} (\tau_x)  =  \big\{\dot{X}_{\rho_x \sigma,{  V_x}} \, , \,  \rho_x \big\}
\end{equation}
with
\begin{equation} \label{eq-tangent_geodesic_metrology}
  \dot{X}_{\rho_x \sigma,{  V_x}}  =  \frac{1}{\sin ( \theta_V - \tau_x)} \Big( M_{\rho_x\sigma,{  V_x}} - \cos( \theta_V- \tau_x ) \identity \Big) .
\end{equation}
Plugging $\dot{\rho}_x=  \epsilon\, \dot{\gamma}_{\geo,V} ( \tau_x )$ into (\ref{eq-def_SL}) and comparing with
(\ref{eq-geo_tangent_vector}) yields
\begin{equation} \label{eq-semilogarithmic_deriv_metrology}
  \SL_x = {  2 \epsilon}\, \dot{X}_{\rho_x \sigma,{  V_x}}\;.
\end{equation}
In view of (\ref{eq-tangent_geodesic_metrology}) and (\ref{eq-semilogarithmic_deriv_metrology}), 
the eigenprojectors of $\SL_x$ are  the eigenprojectors  of $M_{\rho_x\sigma,{  V_x} }$. The latter eigenprojectors,
denoted hereafter by $P_{i,V_x}$,
are related to the intersection states $\rho_i$    of $\gamma_{\geo, V_x}^{\rho_x \to \sigma}$ with the boundary
of quantum states (see Sec.~\ref{sec-int_Bures_geodesics_with_boundary} and Appendix~\ref{sec-intersections_boundary}).
More precisely, one has $\ker ( \rho_i) = P_{i,V_x} \Hh$. But the $\rho_i$'s
are independent of the state $\rho_x$ on the geodesic $\gamma_{\geo,V}$   (in fact, $\gamma_{\geo, V_x}^{\rho_x \to \sigma}$ and  $\gamma_{\geo,V}$
  have the same intersections with $\partial \states$).
Therefore, the eigenprojectors $P_{i,V_x}$ do not depend on the estimated parameter $x$. A more explicit proof that  $P_{i,V_x}$ only depend on the geodesic $\gamma_{\geo,V}$ is given in Appendix~\ref{sec-properties_M}.
Recalling that the eigenprojectors of $L_x$ form an optimal POVM,  
we conclude that

\begin{theorem}  \label{prop-geodesics_and_parameter_estimationII}
 For output states given by (\ref{eq-output_state-geo_evol}), there exists a $x$-independent optimal
  POVM $\{ M_i^\opt \}$  given by
  the projective measurement with projectors $M_i^\opt = P_{i,V}$ onto the kernels of the intersection
  states $\rho_i$ of $\gamma_{\geo,V}$ with $\partial \states$.
\end{theorem}

More generally, thanks to the argument given in Sec.~\ref{sec-QFI_and_Bures_metric},
a POVM $\{ M_i^\opt \}$ is optimal if and only if 
  ${\rm{supp}} (M_i^\opt) \subset P_{i,V} \Hh$  up to permutations between the measurement operators.

As we have seen in  Sec.~\ref{sec-int_Bures_geodesics_with_boundary}, the number of eigenprojectors $P_{i,V}$
is equal to the number $q_V$ of distinct eigenvalues of $\MgeoV$.
In particular, if $\gamma_{\geo,V}$ intersects  $\partial \states$ at a pure state,
then the optimal measurement is a binary measurement consisting of $q_V=2$ projectors, the first one being of rank $n-1$ and the other of rank $1$.

By Theorems~\ref{prop-geodesics_and_parameter_estimation} and~\ref{prop-geodesics_and_parameter_estimationII}, the CFI for the measurement outcome probabilities $p_{i|x}^\opt = \tr P_{i,V} \rho_x$ is equal to the QFI $4 \langle ( \Delta H)^2\rangle_{\Psi_\inp}= 4 \epsilon^2$, being thus
 independent of the geodesic $\gamma_{\geo,V}$ in (\ref{eq-output_state-geo_evol}).
One may ask oneself about the dependence on $x$ and $\gamma_{\geo,V}$ of the distribution $\{ p_{i|x}^\opt\}_{i=1}^{q_V}$. Using
(\ref{eq-Bures_geodesics}) and the expression (\ref{eq-intersection_times_geodesic_boundary}) in Appendix~\ref{sec-intersections_boundary}
for the eigenvalues of $\MgeoV$, one finds
\begin{equation}
  p_{i|x}^\opt = \frac{\sin^2 ( x\epsilon  - \tau_i)}{\sin^2 (\tau_i)} p_{i|0}^\opt \quad , \quad i = 1,\ldots, q_V\;,
\end{equation}
where $p_{i|0}^\opt = \tr  P_{i,V} \rho$ and $\tau_i$ is the geodesic length between $\rho$ and the $i$th intersection state $\rho_i$.
For any geodesic $\gamma_{\geo,V}$ having a pure state intersection with the boundary ($q_V=2$), choosing  $\rho = \rho_1=\ketbra{\phi_1}{\phi_1}$ so that $\tau_1=0$, $\tau_2=\pi/2$, and
$p_{2|0}^\opt=1$ (see Theorem~\ref{prop-intersection_Bures_geo_boundary}), the binary optimal measurement yields a  geodesic--independent
distribution given by $\{ p_{1|x}^\opt=\sin^2(x\epsilon), p_{2|x}^\opt = \cos^2 (x\epsilon)\}$.
Furthermore, for any value of $q_V$ the conditional post-measurement states are independent of $x$, that is,
$(p_{i|x}^\opt )^{-1} P_{i,V} \rho_x P_{i,V} = (p_{i|0}^\opt)^{-1} P_{i,V} \rho P_{i,V}$ $\,\forall\,x \in X$.

\subsection{Heisenberg limit} \label{sec-Heisenberg_limit}

We show in this subsection that the estimation error of the open probe undergoing a geodesic evolution can reach the Heisenberg scaling.
To this end, we assume that a $N$-qubit probe is coupled to $N$ ancilla qubits $\AAA_1,\ldots, \AAA_N$.
The total Hilbert space $\Kk = \complex^{2^N} \otimes (\otimes_{\nu=1}^N \complex_{\AAA_\nu}^2 )$
has dimension $2^{4N}$.
The $\nu$th probe qubit $\SSS_\nu$ is coupled to the $\nu$th ancilla qubit $\AAA_\nu$ by a Hamiltonian $H_\nu$  {  such that 
its eigenvectors $\ket{e_{\nu,\pm}}$  with
maximal and minimal eigenvalues $e_{\nu,\pm}$} satisfy $\ket{e_{\nu,-}} = U_\nu \otimes \identity_{\AAA_\nu} \ket{e_{\nu,+}}$, where $U_\nu$ is a unitary acting on the $\nu$th probe qubit.
The total probe-ancilla Hamiltonian reads
\begin{equation} \label{eq-H_N}
H^{(N)} = \sum_{\nu=1}^N H_\nu \;,  
\end{equation}
where $H_\nu$ acts non-trivially on the $\nu$th probe and ancilla qubits only.
Then  $H^{(N)}$ has two eigenvectors $\ket{\epsilon_\mmax^{N}} = \otimes_{\nu=1}^N \ket{e_{\nu,+}}$ and
$\ket{\epsilon_\mmin^{N}} = \otimes_{\nu=1}^N \ket{e_{\nu,-}}$
associated to the highest and smallest eigenvalues
\begin{equation}
  \epsilon_\mmax^{N} = \sum_{\nu=1}^N e_{\nu,+} \;,\; \epsilon_\mmin^{N} = \sum_{\nu=1}^N e_{\nu,-}\;.
\end{equation}
Let us consider the  multipartite entangled input state
\begin{equation} \label{eq-input_state_maximally_ent_qubit_ancilla_pairs}
  \ket{\Psi_{\inp}^{(N)}}  = \frac{1}{\sqrt{2}} \Big( \tensorproduct_{\nu=1}^N   \ket{e_{\nu,+}} + \E^{\I \varphi}
  \tensorproduct_{\nu=1}^N \ket{e_{\nu,-}} \Big)\;,
\end{equation}
where  entanglement is between the $N$ qubit pairs $(\SSS_1 \AAA_1),\ldots, (\SSS_N \AAA_N)$.
Since $\ket{\epsilon_\mmax^{N}} =U^{(N)} \otimes \identity_\AAA \ket{\epsilon_\mmin^{N}}$ with $U^{(N)} = \otimes_{\nu=1}^N  U_\nu$, by Theorem~\ref{prop-geodesics_and_parameter_estimationIII}
the unitary transformation
\begin{equation}
\ket{\Psi_x^{(N)}}
    = \E^{-\I x H^{(N)}}
    \ket{\Psi_{\inp}^{(N)}}
    = \tensorproduct_{\nu=1}^N \E^{- \I x H_\nu}  \ket{\Psi_{\inp}^{(N)}}
\end{equation}
defines a horizontal geodesic on $\Ss_\Kk$. Thus  the probe  state
$\rho^{(N)}_x = \tr_\AAA  \ketbra{\Psi_x^{(N)}}{\Psi_x^{(N)}}$ follows a geodesic on ${\Ee}_{\complex^{2^N}}$,
\begin{equation}
\rho_x^{(N)} = \gamma_{\geo,V}^{(N)} ( x   \gap_N )  \quad , \quad {  \gap_N = \frac{\epsilon_\mmax^N - \epsilon_\mmin^N}{2}\;.}
\end{equation}
%
According to (\ref{eq-QFI_geodesic_evol}), 
the
QFI of the probe is given by
\begin{equation} \label{eq-QFI-Heisenberg_limit}
\Ff_Q ( \{ \rho_x^{(N)} \}_{x \in X} ) = 4 {   \gap_N^2}  =  \Bigg( \sum_{\nu=1}^N \Big( e_{\nu,+} - e_{\nu,-} \Big) \Bigg)^2 .
\end{equation}
If the eigenenergies $e_{\nu,\pm}$ of the Hamiltonians $H_\nu$ are independent of $\nu$, the QFI scales like $N^2$, implying
a minimal error 
$(\Delta x)_{\rm{QCRB}} \sim N_\meas^{-1/2} N^{-1}$ having the Heisenberg scaling.
Let us point out that if the energy gap $2 \epsilon=e_{\nu,+} - e_{\nu,-}$ of the $\nu$th qubit-ancilla pair is twice the 
  single qubit  energy, then the QFI (\ref{eq-QFI-Heisenberg_limit}) is  the same as the QFI obtained 
  by using the ancilla qubits as additional probes, the state transformation being generated by $2N$ Hamiltonians acting on single qubits and  the input state being the
  maximally entangled $2N$-qubit state. 
 Our setup has the advantage that, for the same precision, one has to measure only half of the $2N$ qubits.
An example of quantum circuit implementing this setup is shown in Fig.~\ref{fig-Q_circuit_metrology}.

By Theorem~\ref{prop-geodesics_and_parameter_estimationII}, an optimal measurement is a joint projective measurement on the $N$ probe qubits with projectors onto the kernels
of the intersection states $\rho_i^{(N)}$ of the geodesic $\gamma^{(N)}_{\geo,V}$ with the boundary
of the  $N$-qubit state manifold.

\section{Conclusions and perspectives} \label{sec-Conclusion}

In this work we have studied the geodesics on the manifold of quantum states for the Bures distance.
We have determined these geodesics  and have shown that they are physical, as they correspond to quantum evolutions of an open system coupled to an ancilla. The
corresponding system-ancilla coupling Hamiltonian has been derived explicitly. Examples of quantum circuits implementing some geodesics have been given. Furthermore, we have proven that the geodesics are optimal
for single-parameter estimation in open quantum systems, where the unknown parameter is a phase shift 
  multiplying a parameter-independent system--environment Hamiltonian.
Actually, among all such Hamiltonians $H$ with a fixed energy gap
  $2 \gap=\eps_\mmax-\epsilon_\mmin$ between the maximal and minimal
  eigenvalues and all 
  input state $\ket{\Psi_\inp}$ maximizing
  the QFI, \ie, such that  $4 \langle (\Delta H)^2\rangle_{\Psi_\inp}= 4 \gap^2$,  if one cannot measure the ancilla then the best precision for all parameter values is obtained when $H$ and $\ket{\Psi_\inp}$  generate a geodesic evolution of the probe.

These results open the route to experimental observations of geodesics in  multi-qubit quantum information platforms
offering a high degree of control on the Hamiltonian. Such experimental realizations would be of interest for  high-precision estimations
in situations where only a part of these qubits can be measured. 

The methods developed in this paper, which are borrowed from Riemannian geometry, are expected to be
applicable as well to non ideal quantum metrology setups. For instance, when the coupling with the environment provokes energy losses and dephasing,
additional couplings with engineered reservoirs could be tailored to modify the state transformation so that it becomes
closer to a geodesic. This would increase the precision of the estimation by reducing the amount of information on the parameter lost in the environment.
Alternatively, one could investigate whether $H$ can be steered to a geodesic Hamiltonian by  
 using external controls acting either on the probe or on its environment.

Another potential field of application of the Bures geodesics is incoherent quantum control. In order 
to efficiently steer a quantum mixed state $\rho$ to a given desired state $\sigma$, an idea is to
adjust the control parameters in such a way as to follow as closely as possible
the shortest geodesic joining $\rho$ and $\sigma$~\cite{Mauro-thesis}.
Among other  directions worth exploring is the relation between the geodesic evolutions and the quantum speed limit in open  systems~\cite{Davidovich2012}.

The present work can be contextualized as belonging to an emerging broader research topic.
Information geometry has been developed in the last decades by Amari and coworkers~\cite{Amari_book00,Amari_book16}  in an attempt to use concepts and methods from
Riemannian geometry in information theory. It has been successfully applied to many fields, such as machine learning, signal processing, optimization,
statistics, and neurosciences. The application of this approach to quantum information processing  remains largely
unexplored. It will hopefully open new challenging perspectives.

\vspace{0.3cm}
\noindent
    {\bf Acknowledgments}
    The author acknowledges support from the ANID Fondecyt Grant No 1190134 and is grateful
    to Fethi Mahmoudi and G\'erard Besson for useful discussions. 
\vspace{0.3cm}



\newpage

\appendix
\renewcommand{\theequation}{\Alph{section}\arabic{equation}}
\setcounter{equation}{0}

\section{Properties of the operator  $M_{\rho\sigma, V}$}
\label{sec-properties_M}

The operator $\MgeoV=\rho^{-1/2} |\sqrt{\sigma} \sqrt{\rho}\,| \rho^{-1/2}$ in the expression (\ref{eq-Bures_geodesics}) of
the Bures geodesics has the following properties:  
\begin{itemize}
\item[(a)] $\MgeoV \,\rho \, M_{\rho\sigma, V}= \sigma$;
\item[(b)] $\tr \rho  \MgeoV  = \cos \theta_V$,  $\;\tr  \rho M^2_{\rho\sigma, V}  = 1$;
\item[(c)] $M_{\sigma \rho,\widetilde{V}} = M_{\rho\sigma,V}^{-1}$ and $\;M_{\rho\sigma,V} \, \rho  =  \sigma M_{\sigma \rho,\widetilde{V}}$, where
  $\widetilde{V} = U_{\sigma\rho} V U_{\sigma\rho}^\dagger$.
\end{itemize}
Properties (a) and (b) follow from the definition of $\MgeoV$ and from (\ref{eq-geodesic_length_V}). The first identity in (c)
follows from the equality $ \MgeoV = \sqrt{\sigma} U_{\sigma \rho} V \rho^{-1/2}$, see (\ref{eq-def_M^(V)}),
and the fact that the unitary $U_{\rho \sigma}$ in the polar decomposition
of $\sqrt{\rho}\sqrt{\sigma}$ is equal to the adjoint of $U_{\sigma \rho}$
(recall that if $O = U | O|$ then $O^\dagger = U^\dagger | O^\dagger |$). The second identity is then deduced from (a).
Observe that (a) is equivalent to $\gamma_{\geo,V} ( \theta_V)= \sigma$, see (\ref{eq-Bures_geodesics}).
Property~(b) can be used to show that $\tr  \gamma_{\geo,V} ( \tau) =1$ for all $\tau$ (as it should be since 
$\gamma_{\geo,V} (\tau)$ is a quantum state).
Property (c)~insures that  the geodesic joining $\sigma$ to $\rho$, obtained by exchanging $\rho$ and $\sigma$
in (\ref{eq-Bures_geodesics}), coincides with the time-reversed geodesic
$\gamma_{\geo,\widetilde{V}} (\theta_V-\tau)$. 
By using the self-adjointness and unitarity of $V$ and $[\Lambda_{\sigma\rho},V]=0$, it is easy to show that
$\widetilde{V}$ enjoys the same properties, the commutation being  with  
$\Lambda_{\rho\sigma}=|\sqrt{\rho}\sqrt{\sigma}|= U_{\sigma\rho} \Lambda_{\sigma\rho} U_{\sigma\rho}^\dagger$.

The next property tells us how $M_{\rho\sigma, V}$ is transformed as one moves $\rho$ along the geodesic $\gamma_{\geo,V}$, keeping $\sigma$ fixed. 
For any invertible state $\rho_t = \gamma_{\geo,V}(t)$ on  $\gamma_{\geo,V}$, with $0 \leq t \leq \theta_V$, one has
\begin{itemize}  %
\item[(d)]
   $ \MgeoV = M_{\rho_t \sigma,V_t} X_{\rho\sigma,V} (t)$, 
\end{itemize}
where $X_{\rho\sigma,V}(t)$ is given by (\ref{eq-def_X(tau)}) and $V_t$ is some self-adjoint unitary operator commuting with $\Lambda_{\sigma \rho_t} = | \sqrt{\sigma} \sqrt{\rho_t} |$.
This formula is related to the fact that the geodesics joining $\rho_t$ and $\sigma$ are the
geodesics joining $\rho$ and $\sigma$ shifted in time, 
\begin{equation} \label{eq-shifted_geodesics}
  \gamma_{\geo,V_t}^{(t)} ( \tau)  = \gamma_{\geo, V} (t+ \tau) \;,\;0 \leq \tau \leq \theta_V - t\;.
\end{equation}
We will prove in Appendix~\ref{sec-intersections_boundary} that the spectrum of $V_t$ is constant in time save at the intersection
times of $\gamma_{\geo, V}$ with the boundary $\partial \states$, where some eigenvalues of $V_t$ may jump from
$-1$ to $+1$. In particular, if $V=\identity$ then $V_t=\identity$ for $0 \leq t \leq \theta_1$.

Formula (d) can be proven directly from  (\ref{eq-def_M^(V)}) and (\ref{eq-Bures_geodesics}), but it is simpler to derive it from the properties
of horizontal geodesics on the hypersphere $\Ss_\Kk^\inv$. In fact, it is clear geometrically that the arc of great circle joining $\ket{\Psi_t} = \ket{\Psi_{\geo} (t)}$ to
  $\ket{\Phi_V}$ has length $\theta_V-t$ and is contained in the arc of great circle joining $\ket{\Psi}$ to $\ket{\Phi_V}$, so that it is parametrized by
\begin{equation} \label{eq-shifted_arc_of_great_circle}
  \ket{\Psi_{\geo}^{(t)} (\tau)} = \ket{\Psi_{\geo} (t + \tau)}\;,\;0 \leq \tau \leq \theta_V - t\;.
\end{equation}
Since $\ket{\Psi_{\geo}(\tau)}$ is a horizontal geodesic, its tangent vector $\ket{\dPsi_t}$  at $\ket{\Psi_t}$ is horizontal
(property (i) of Sec.~\ref{sec-Riemannian_submersions}).
  According to the result of Sec.~\ref{sec-Bures_geodesics}, this is equivalent to
  \begin{equation} \label{eq-horizontality_cond_for_Psi_0}
    \ket{\Phi_V} = M_{\rho_t \sigma, V_t} \otimes \identity_\AAA \ket{\Psi_t}
  \end{equation}
  for some  self-adjoint unitary $V_t$ commuting with $\Lambda_{\sigma \rho_t }$.
  By (\ref{eq-horizontality_cond_4}), (\ref{eq-horizontal_geodesics_sphere}), and (\ref{eq-def_X(tau)}), one has
\begin{equation} \label{eq-formula_Psi_0}
  \ket{\Psi_t} = \ket{\Psi_{\geo} (t)} = X_{\rho\sigma,V} (t) \otimes \identity_\AAA \ket{\Psi}\;.
\end{equation}
Plugging (\ref{eq-formula_Psi_0}) into (\ref{eq-horizontality_cond_for_Psi_0}) one gets
\begin{equation}
  \ket{\Phi_V}=
    M_{\rho_t \sigma, V_t} X_{\rho\sigma,V} (t) \otimes \identity_\AAA \ket{\Psi} = \MgeoV \otimes \identity_\AAA \ket{\Psi} .
\end{equation}
One easily shows that the second equality is equivalent to (d) (for instance, one may rely on (\ref{eq-Schmidt_decomp_Psi})).
Furthermore,  (\ref{eq-shifted_arc_of_great_circle}) implies
(\ref{eq-shifted_geodesics}) since  the projection on $\states^\inv$ of the horizontal geodesic
$ \ket{\Psi_{\geo}^{(t)} (\tau)}$ is the Bures geodesic
joining $\rho_t$ and $\sigma$ with unitary $V_t$.

An important consequence of (d) for applications to quantum metrology is the following.
As shown in Appendix~\ref{sec-intersections_boundary}, $X_{\rho\sigma,V}(t)$ is
invertible when $\rho_t$ is invertible, \ie, when $t$ is not an intersection time of $\gamma_{\geo,V}$ with $\partial \states$.
In such a case $M_{\rho_t \sigma,V_t} = \MgeoV X_{\rho\sigma,V} (t)^{-1}$ is a function of the self-adjoint operator
$\MgeoV$ (recall that $X_{\rho\sigma,V} (t)$ is a function of $\MgeoV$, see (\ref{eq-def_X(tau)})).
Thus the eigenprojectors of $M_{\rho_t \sigma,V_t}$ are $t$-independent and coincide with the eigenprojectors of $\MgeoV$.

\section{Intersections of the geodesics  with the boundary of quantum states} \label{sec-intersections_boundary}

In this appendix we study the intersections of the Bures geodesics with  the boundary of quantum states  $\partial \states$.
As explained in
the main text, we consider the  extensions of the geodesic arcs $\gamma_{\geo,V}$ joining two states $\rho$ and
$\sigma \in \states^\inv$ to the time interval $[0,\pi]$, given by (\ref{eq-Bures_geodesics}) with $0 \leq \tau \leq \pi$.
These extensions  are closed geodesic curves, which are denoted by the same symbol $\gamma_{\geo,V}$.
Recall that these curves depend on a self-adjoint unitary operator $V$ commuting with $\Lambda_{\sigma\rho}^2 = \sqrt{\rho} \,\sigma \sqrt{\rho}$. The arc length of $\gamma_{\geo,V}$ between $\rho$ and $\sigma$ is denoted by $\theta_V$
(see Theorem~\ref{prop-Bures_geodesics}).

\begin{theorem}  \label{prop-intersection_Bures_geo_boundary}
One has 
\begin{itemize}
\item[(i)] $\gamma_{\geo,V}$  intersects $q_V$ times $\partial \states$, where $q_V$
  is the number of distinct eigenvalues of the operator $\MgeoV$ in (\ref{eq-def_M^(V)}).
  While the shortest geodesic $\gamma_{\geo}$ does not intersect $\partial \states$  between $\rho$ and $\sigma$,
  \ie, $\gamma_{\geo} ( [0,\theta_1] ) \subset \states^\inv$,
  the other geodesics  with $V \not= \identity$ do so at least once.
  More precisely, the number of intersections of $\gamma_{\geo,V} ([0,\theta_V])$ with $\partial \states$ is equal to
  the multiplicity of the eigenvalue $-1$ of $V$.
\item[(ii)]  The intersection points  $\rho_i$ of $\gamma_{\geo,V}$ with
$\partial \states$ have  ranks $n-m_{i,V}$ and supports $(\identity - P_{i,V})\Hh$, where $m_{i,V}$ and
  $P_{i,V}$ are the  multiplicities of the eigenvalues and the spectral projectors of $\MgeoV$, respectively.
  In particular,
\begin{equation}
  \sum_{i=1}^{q_V} \dim ( \ker \rho_i ) = n\;.
\end{equation}
\item[(iii)] Given an invertible state $\rho \in \states^\inv$ and a pure state $\ket{\phi_1}$ such that $\bra{\phi_1}\rho \ket{\phi_1}>0$,
  there are exactly two geodesics
  passing through $\rho$ and intersecting $\partial \states$ at $\rho_1=\ketbra{\phi_1}{\phi_1}$, namely the
  shortest geodesic $\gamma_{\geo,\rho \to \rho_1} (\tau)$ joining $\rho$ and $\rho_1$ and its time reversal $\gamma_{\geo, \rho \to \rho_1} (\pi - \tau)$.
  Moreover, $\gamma_{\geo, \rho \to \rho_1}$ intersects twice $\partial \states$; the other intersection  point $\rho_2$ has rank $n-1$ and support orthogonal to
  $\ket{\phi_1}$,  being therefore separated from $\rho_{1}$ by a geodesic distance $\pi/2$.
\end{itemize}
\end{theorem}

The results (i) and (ii) have been proven in Ref.~\cite{Ericson05} in the particular case $V=\identity$.
It follows from (ii) that $\gamma_{\geo,V}$
intersects  $\partial \states$ at a pure state 
if and only if $\MgeoV$ has two eigenvalues of multiplicities $n-1$ and $1$.
Note that this is always the case for $n=2$ (for a qubit, $\partial \states$ is the set of pure states).

For a qutrit ($n=3$), there are up to time-reversal four geodesics passing through two generic states $\rho>0$ and $\sigma>0$, where by generic we mean that $\Lambda_{\sigma\rho} = | \sqrt{\sigma} \sqrt{\rho}|$ has a non-degenerate spectrum,  see the discussion after  Theorem~\ref{prop-Bures_geodesics}.
The shortest geodesic $\gamma_\geo$ joining $\rho$ and $\sigma$, obtained for $V = \identity$, does not intersect $\partial \states$
between these two states. Its time-reversal is the geodesic obtained for $V = -\identity$.
The three other geodesics correspond to $V$ having spectrum $\{ 1,1,-1\}$
(or  $\{ -1, -1,1\}$ for their time-reversal). They intersect the boundary once (twice for the time-reversal) between
$\rho$ and $\sigma$. If $\MgeoV$ has non-degenerated eigenvalues then  $\gamma_{\geo,V}$  has $q_V=3$ intersections $\rho_i$ with $\partial \states$, which have rank $2$.

\vspace{2mm}

\noindent {\it Proof}.
(i) To simplify the notation
we do not write explicitly the dependence  on $\rho$, $\sigma$, and $V$ of the operators $\MgeoV$, $X_{\rho\sigma,V}$, etc.
Following the arguments
of~\cite{Ericson05}, we observe that  in view of (\ref{eq-Bures_geodesics}),
$\gamma_{\geo,V} (\tau) \in \partial \states$ if and only if
$\det  \gamma_{\geo,V} (\tau) = \det X (\tau)^2 \det  \rho =0$, that is, $\det  X ( \tau)  =0$.
The last determinant is the characteristic polynomial of $M$, see (\ref{eq-def_X(tau)}).
Thus $\gamma_{\geo,V}$ intersects  $q$ times $\partial \states$, at times
$\tau_1 < \ldots < \tau_q$ given by
\begin{equation} \label{eq-intersection_times_geodesic_boundary}
  \frac{\sin ( \theta_V - \tau_i)}{\sin \tau_i} = - \mu_i
  \; \Leftrightarrow \;
  {\rm{cotan}}\, \tau_i  =  \frac{\cos \theta_V-\mu_i}{\sin\theta_V} 
  \;,
\end{equation}
where $\mu_1 < \ldots < \mu_q$ are the distinct eigenvalues of $M$.
If $V=\identity$ then $M >0$ and thus
$\mu_1>0$. Hence ${\rm{cotan}}\,\tau_1 < {\rm{cotan}} \,\theta_1$, so that the first intersection time satisfies $\tau_1 > \theta_1$. This
tells us that the shortest geodesic arc  $\gamma_\geo ([0,\theta_1])$
starting at $\rho$ and ending at $\sigma$  is contained in $\states^\inv$, \ie, it does not intersect $\partial \states$.
In contrast, let us show that if $V \not= \identity$ then $M$
has at least one negative eigenvalue $\mu_1 <0$.
In fact,
$V$ has at least one eigenvalue $v_{k}=-1$. Denote by  $\ket{\varphi_k}$ a common eigenvector
of $V$ and $\Lambda= |\sqrt{\sigma} \sqrt{\rho}|$ for the eigenvalues $v_k$ and $\lambda_k$, respectively. Then
\begin{equation}
  \bra{\varphi_k} \sqrt{\rho} \, M \sqrt{\rho} \ket{\varphi_k}=
  \bra{\varphi_k} \Lambda  V \ket{\varphi_k} = - \lambda_k <0\;.
\end{equation}
By the variational principle  it follows that $\mu_1 <0$.
Hence $M$ has $s \geq 1$ negative eigenvalues $\mu_1<\ldots < \mu_s <0$. One deduces from (\ref{eq-intersection_times_geodesic_boundary}) that 
${\rm{cotan}} \, \tau_{i} >  {\rm{cotan}} \,\theta_V$ and thus $\tau_{i} < \theta_V$ for $i=1,\ldots,s$. A reversed inequality holds for $i>s$. This shows that
$\gamma_{\geo,V}$  intersects the boundary $s$ times on its part between $\rho$ and $\sigma$.
The fact that $s$ is equal to the multiplicity of the eigenvalue $-1$ of $V$ follows from a similar argument, using the min-max theorem
for self-adjoint operators.

(ii)
Let us now prove that the intersection states $\rho_i= \gamma_\geo ( \tau_i) \in \partial \states$
have  ranks
$r_i  =  n- m_i$ and supports
\begin{equation} \label{eq-support_intersection_state}
  \supp ( \rho_i) =  Q_i \Hh  = [\ker( M - \mu_i )]^\bot\;,
\end{equation}
where $m_i$ and $P_i$ are the multiplicity and spectral projector of $M$ for the eigenvalue $\mu_i$ and $Q_i=\identity -P_i$.
We first note that 
\begin{equation} \label{eq-def_X_i}
 X_i = X (\tau_i) = \frac{\sin \tau_i}{\sin\theta}  ( M - \mu_i )
\end{equation}
has rank $r_i$ and support $Q_i \Hh $, so  that $X_i=Q_i X_i Q_i$. But
\begin{equation}
\rho_i=X_i \,\rho \, X_i\;,
\end{equation}
hence
$\ker ( \rho_i) \supset \ker ( X_i )$. Reciprocally, let $\ket{\varphi} \in \ker ( \rho_i)$.
Then $\rho \, X_i \ket{\varphi}\in P_i \Hh$, that is, $Q_i \,\rho \, Q_i X_i \ket{\varphi} = 0$. Since  $Q_i\, \rho \,Q_i$ is invertible on $Q_i \Hh$ (recall that $\rho>0$), one has
$\ket{\varphi} \in \ker ( X_i)$. This implies that $\ker ( \rho_i) = \ker ( X_i )=P_i \Hh$ and thus
$\supp ( \rho_i) = Q_i \Hh$, as announced above.

(iii)
Let $\gamma_{\geo,V}$ be a geodesic starting at $\rho$ and intersecting the boundary
at $\rho_1=\ketbra{\phi_1}{\phi_1}$. 
Thanks to (\ref{eq-support_intersection_state}) one has 
\begin{equation} \label{eq-M-mu_1}
  M - \mu_1 = \langle  M - \mu_1 \rangle_{\phi_1} \ketbra{\phi_1}{\phi_1}\;.
\end{equation}
Using  (\ref{eq-intersection_times_geodesic_boundary}) and (\ref{eq-M-mu_1}) one obtains
\begin{eqnarray} \label{eq-expectation_M-mu_1}
  \nonumber
& &   \langle  M - \mu_1 \rangle_{\phi_1}  \langle \rho \rangle_{\phi_1}  
   =  \tr  (M - \mu_1)\rho  
   \\
\nonumber   
&  & \quad = \tr M \rho -  \cos \theta  +\sin \theta \,{\rm{cotan}} \tau_1
\\
&  & \quad =  \sin \theta \,{\rm{cotan}} \tau_1 \;,
\end{eqnarray}
where the last equality follows from
property (b) of Appendix~\ref{sec-properties_M}.
Furthermore, equating $X_1 \rho X_1$ with $\rho_1=\ketbra{\phi_1}{\phi_1}$
and using (\ref{eq-def_X_i}) and (\ref{eq-M-mu_1}),  one gets
$\cos \tau_1 = \pm \langle \rho \rangle_{\phi_1}^{1/2}$ .
For any choice of the operator $V$,
$\tau_1$ is thus either equal to ${d}_{\rm B}(\rho,\rho_1)= \arccos (\langle \rho \rangle_{\phi_1}^{1/2})$ 
or to $\pi$ minus this distance. 
We can now express $\mu_1$ and $\langle  M - \mu_1 \rangle_{\phi_1}$ in terms of $\theta$ and 
$\tau_1$, replace these expressions
into (\ref{eq-M-mu_1}), and use (\ref{eq-def_X(tau)}) in the main text to obtain 
\begin{equation} \label{eq-X_for_geo_between_rho_and_pure_state}
  X ( \tau)
= \frac{1}{\sin \tau_1} \bigg( \sin ( \tau_1 - \tau) \identity + \frac{\sin \tau}{\cos\tau_1} \ketbra{\phi_1}{\phi_1} \bigg) 
  .
\end{equation}
Observe that $ X ( \tau)$ depends on $\ket{\phi_1}$ and $\tau_1$  but not on $\theta$. Therefore, there are exactly two geodesic arcs
$\gamma_{\geo,V}(\tau)= X(\tau) \rho X(\tau)$ starting at $\rho$ and intersecting $\partial \states$ at $\rho_1$.
The first one has length $\tau_1= {d}_{\rm B}(\rho,\rho_1)$ (shortest geodesic
$\rho \to \rho_1$), the second one has length $\tau_1=\pi - {d}_{\rm B}(\rho,\rho_1)$ (time-reversed geodesic).
The other affirmations in (iii) are direct consequences of (ii).
Note that if $\langle \rho \rangle_{\phi_1}=0$ then there are no geodesic joining $\rho$ to  $\rho_i$ because in such a case
(\ref{eq-def_X_i}) and (\ref{eq-M-mu_1}) imply that $X_1 \rho X_1$ vanish, in contradiction with $X_1 \rho X_1=\rho_1$.
\finpro

\vspace{3mm}

Let us now apply Theorem~\ref{prop-intersection_Bures_geo_boundary} 
and a continuity argument to
 determine the self-adjoint unitary operators $V_t$ appearing in property (d)
 and Eq. (\ref{eq-shifted_geodesics}) of Appendix.~\ref{sec-properties_M}.
 Recall that $V_t$ is associated to
 the  time-shifted geodesic $\gamma_{\geo,V_t}^{(t)}(\tau) = \gamma_{\geo,V}(t+\tau)$ joining $\rho_t$ and $\sigma$ and that 
 $V_t$ commutes with $\Lambda_{\sigma \rho_t} = | \sqrt{\sigma} \sqrt{ \rho_t} |$.
 Denoting as above by $\tau_1 < \tau_2 < \cdots < \tau_q$ the intersection times of $\gamma_{\geo,V}$ with $\partial \states$,
we first assume that $0 \leq t < \tau_1$.
 One deduces from property (d) that
 \begin{equation} \label{eq-continuity_argument_V_t}
    V_t =  \Lambda_{\sigma \rho_t}^{-1} \sqrt{\rho_t} \MgeoV X_{\rho\sigma,V} (t)^{-1} \sqrt{\rho_t}\;.
 \end{equation}
 Here, we have used that $\Lambda_{\sigma \rho_t}$ and $X_{\rho\sigma,V}(t)$ are invertible for  
 $0 \leq t < \tau_1$
 (in fact, for  $t\not= \tau_i$ one has $\det \rho_t = (\det X_{\rho\sigma,V}(t ) )^2 \det \rho \not= 0$, 
 see the proof of Theorem~\ref{prop-intersection_Bures_geo_boundary}).
 Furthermore,  $\Lambda_{\sigma \rho_t}^{-1}$ and $X_{\rho\sigma,V}(t)^{-1}$ are continuous in time in view of the continuity of
  $\rho_t$ and of (\ref{eq-def_X(tau)}), respectively. It follows that
  $V_t$ is continuous in time on $[0, \tau_1)$.
   Thus its eigenvalues $v_k(t) \in \{-1,1\}$ are time-independent on this interval and
$V_t = \sum_{k} v_k \ketbra{\varphi_k (t)}{\varphi_k (t)}$ for any $t \in [0, \tau_1)$,
where $v_k$ are the eigenvalues of $V=V_0$ and $\{ \ket{\varphi_k (t)} \}_{k=1}^n$ is a time--continuous \ONB diagonalizing $\Lambda_{\sigma \rho_t}$.
 In particular, if $V= \identity$ then $V_t = \identity$ for all $t \in [0,\theta_1]$    (in fact, in such a case $\tau_1 > \theta_1$ by
 Theorem~\ref{prop-intersection_Bures_geo_boundary}(i)).  Eq.~(\ref{eq-shifted_geodesics}) then ensures that
  $\gamma_\geo^{(t)}(\tau) = \gamma_\geo (t+ \tau)$ is the shortest geodesic arc joining $\rho_t$ and $\sigma$ and has length
 $d_B(\rho_t,\sigma) = \theta_1 - t$ with $t=d_B (\rho,\rho_t)$.
 This is consistent with the additivity property of the distance,
\begin{equation} \label{eq-saturation_triangle_inequality}
       d_B ( \rho, \sigma ) = d_B ( \rho, \rho_t ) + d_B ( \rho_t, \sigma)\;,\; \rho_t \in \gamma_\geo ( [0,\theta_1])\, .
\end{equation}
On the other hand, if $\tau_i < t < \tau_{i+1}<\theta_V$ then the number of intersections  with $\partial \states$ of the time-shifted geodesic arc
$\gamma_{\geo,V_t}^{(t)} ( [0,\theta_V-t])$ is reduced by $i$ as compared to the number of
intersections of $\gamma_{\geo,V} ( [0,\theta_V]) $ with $\partial \states$. According to Theorem~\ref{prop-intersection_Bures_geo_boundary}(i),
the multiplicity of the  eigenvalue $-1$ of $V_t$ equals  $s-i$, where $s$ is the multiplicity for $V$.
By the same argument as above, the eigenvalues of $V_t$ and their multiplicities  are constant between $\tau_i$ and $\tau_{i+1}$, but
the multiplicities jump by one at the intersection times. In particular, the identity (\ref{eq-saturation_triangle_inequality}) does not hold
for $t> \tau_1$.

\vspace{5mm}



\end{document}

%% file: geGriemann_submersion.pstex_t
\begin{picture}(0,0)%
\includegraphics{geGriemann_submersion.pstex}%
\end{picture}%
\setlength{\unitlength}{4144sp}%
\begingroup\makeatletter\ifx\SetFigFont\undefined%
\gdef\SetFigFont#1#2#3#4#5{%
  \reset@font\fontsize{#1}{#2pt}%
  \fontfamily{#3}\fontseries{#4}\fontshape{#5}%
  \selectfont}%
\fi\endgroup%
\begin{picture}(12107,8452)(972,-8220)
\put(11845,-6449){\makebox(0,0)[lb]{\smash{{\SetFigFont{34}{40.8}{\familydefault}{\mddefault}{\updefault}{\color[rgb]{.69,0,.69}$\dot{\sigma}$}%
}}}}
\put(12726,-2940){\makebox(0,0)[lb]{\smash{{\SetFigFont{34}{40.8}{\familydefault}{\mddefault}{\updefault}{\color[rgb]{0,0,0}$|\dot{\Phi}_V\rangle$}%
}}}}
\put(3855,-4547){\makebox(0,0)[lb]{\smash{{\SetFigFont{34}{40.8}{\familydefault}{\mddefault}{\updefault}{\color[rgb]{.69,0,.69}${\rm{d}}\pi |_{\Psi}$}%
}}}}
\put(9690,-3481){\makebox(0,0)[lb]{\smash{{\SetFigFont{34}{40.8}{\ttdefault}{\mddefault}{\updefault}{\color[rgb]{0,0,1}$|\Phi_V\rangle$}%
}}}}
\put(3499,-425){\makebox(0,0)[lb]{\smash{{\SetFigFont{34}{40.8}{\familydefault}{\mddefault}{\updefault}{\color[rgb]{0,0,1}$|\dot{\Psi}\rangle$}%
}}}}
\put(2801,-3819){\makebox(0,0)[lb]{\smash{{\SetFigFont{34}{40.8}{\familydefault}{\mddefault}{\updefault}{\color[rgb]{0,0,0}${\mathfrak{H}}_{\Psi}$}%
}}}}
\put(11100,-3904){\makebox(0,0)[lb]{\smash{{\SetFigFont{34}{40.8}{\familydefault}{\mddefault}{\updefault}{\color[rgb]{0,0,0}${\mathfrak{H}}_{\Phi_V}$}%
}}}}
\put(1679,-3386){\makebox(0,0)[lb]{\smash{{\SetFigFont{34}{40.8}{\ttdefault}{\mddefault}{\updefault}{\color[rgb]{0,0,1}$|\Psi\rangle$}%
}}}}
\put(3865,-2330){\makebox(0,0)[lb]{\smash{{\SetFigFont{34}{40.8}{\familydefault}{\mddefault}{\updefault}{\color[rgb]{0,0,0}$|\dot{\Psi}_h \rangle$}%
}}}}
\put(3723,-5836){\makebox(0,0)[lb]{\smash{{\SetFigFont{34}{40.8}{\familydefault}{\mddefault}{\updefault}{\color[rgb]{.69,0,.69}$\dot{\rho}$}%
}}}}
\put(1353,-7856){\makebox(0,0)[lb]{\smash{{\SetFigFont{34}{40.8}{\familydefault}{\mddefault}{\updefault}{\color[rgb]{0,.82,0}$\mathcal{E}$}%
}}}}
\put(10759,-239){\makebox(0,0)[lb]{\smash{{\SetFigFont{34}{40.8}{\ttdefault}{\mddefault}{\updefault}{\color[rgb]{0,0,1}$|\Phi\rangle$}%
}}}}
\put(5667,-6635){\makebox(0,0)[lb]{\smash{{\SetFigFont{34}{40.8}{\familydefault}{\mddefault}{\updefault}{\color[rgb]{0,.82,0}$\gamma_{g,V}$}%
}}}}
\end{picture}%